\documentclass[sigconf,nonacm]{acmart}
\usepackage{amsmath}
\usepackage{mathtools}
\usepackage{tikz}
\usepackage{multirow}
\usepackage{float}
\usepackage{subcaption}
\usepackage{microtype}
\usepackage{verbatim}
\usepackage{paralist}
\usepackage{graphicx}
\usepackage{pgffor}
\usepackage{makecell}
\usepackage[most]{tcolorbox}
\usetikzlibrary{trees}
\theoremstyle{definition}
\newtheorem{definition}{Definition}

\newif\ifversionExtented
\versionExtentedtrue


\AtBeginDocument{%
  }

\newcommand\vldbavailabilityurl{https://doi.org/10.5281/zenodo.20486310}

\definecolor{tatr}{HTML}{DC050C}
\definecolor{xy}{HTML}{E8601C}
\definecolor{vgt}{HTML}{F4A736}
\definecolor{docling}{HTML}{F7F056}
\definecolor{monkey}{HTML}{882E72}
\definecolor{pedia}{HTML}{D1BBD7}
\definecolor{got}{HTML}{AE76A3}
\definecolor{camelot}{HTML}{CAE0AB}
\definecolor{pdfplumber}{HTML}{CAE0AB}
\definecolor{pymupdf}{HTML}{CAE0AB}
\definecolor{grobid}{HTML}{4EB265}
\definecolor{qwen}{HTML}{7BAFDE}
\definecolor{gpt}{HTML}{5289C7}
\definecolor{gemini}{HTML}{0D3E80}
\definecolor{mathpix}{HTML}{44AA99}

\colorlet{camelot}{camelot!80!black}
\colorlet{pymupdf}{pymupdf!80!black}
\colorlet{pdfplumber}{pdfplumber!80!black}
\colorlet{docling}{docling!85!black}
\colorlet{pedia}{pedia!90!black}

\newcommand{\Camelot}{\textcolor{camelot}{Camelot}}
\newcommand{\PdfPlumber}{\textcolor{pdfplumber}{PDFPlumber}}
\newcommand{\Pymupdf}{\textcolor{pymupdf}{PyMuPDF}}
\newcommand{\Grobid}{\textcolor{grobid}{Grobid}}
\newcommand{\Docling}{\textcolor{docling}{Docling}}
\newcommand{\TATR}{\textcolor{tatr}{TATR}}
\newcommand{\TATRs}{\textcolor{tatr}{TATR}}
\newcommand{\VGT}{\textcolor{vgt}{VGT+TATR-struct}}
\newcommand{\VGTs}{\textcolor{vgt}{VGT}}
\newcommand{\XY}{\textcolor{xy}{XY+TATR}}
\newcommand{\XYs}{\textcolor{xy}{XY}}
\newcommand{\pedia}{\textcolor{pedia}{TabPedia}}
\newcommand{\got}{\textcolor{got}{GOT}}
\newcommand{\monkey}{\textcolor{monkey}{MonkeyOCR}}
\newcommand{\qwen}{\textcolor{qwen}{Qwen2.5-VL}}
\newcommand{\gpt}{\textcolor{gpt}{GPT-4.1}}
\newcommand{\gemini}{\textcolor{gemini}{Gemini 2.5 Pro}}
\newcommand{\mathpix}{\textcolor{mathpix}{Mathpix}}

\newcommand{\Fbbox}{\(\mathbf{F_1}^{\!\!\textit{bbox}}\)}
\newcommand{\Ftxt}{\(\mathbf{F_1}^{\!\!\textit{txt}}\)}
\newcommand{\FTop}{\(\mathbf{F_1}^{\!\!\text{Top}}\)}
\newcommand{\FCon}{\(\mathbf{F_1}^{\!\!\text{Con}}\)}
\newcommand{\FTEDS}{\(\mathbf{F_1}^{\!\!\text{TEDS}}\)}
\newcommand{\AP}{\(\mathbf{AP}\)}
\newcommand{\APTop}{\(\mathbf{AP}^\text{Top}\)}
\newcommand{\APCon}{\(\mathbf{AP}^\text{Con}\)}
\newcommand{\APTEDS}{\(\mathbf{AP}^\text{TEDS}\)}

\begin{document}
\title{Benchmarking Table Extraction from~Heterogeneous~%
  Scientific~PDF~Documents
  \ifversionExtented{[Extended~Version]}\else\fi}

\author{Marijan Soric}
\orcid{0009-0001-6058-8550}
\affiliation{%
  \institution{DI ENS, ENS, PSL University, CNRS, Inria}
  \city{Paris}
  \country{France}
}
\email{marijan.soric@inria.fr}

\author{Cécile Gracianne}
\orcid{0000-0003-4232-5359}
\affiliation{%
  \institution{BRGM}
  \city{Orléans}
  \country{France}
}
\email{c.gracianne@brgm.fr}

\author{Ioana Manolescu}
\orcid{0000-0002-0425-2462}
\affiliation{%
  \institution{Inria, Institut Polytechnique de Paris}
  \city{Palaiseau}
  \country{France}
}
\email{ioana.manolescu@inria.fr}

\author{Pierre Senellart}
\orcid{0000-0001-5109-3700}
\affiliation{%
  \institution{DI ENS, ENS, PSL University, CNRS, Inria}
  \city{Paris}
  \country{France}
}
\email{pierre@senellart.com}

\renewcommand\leq\leqslant
\renewcommand\geq\geqslant

\renewcommand{\shortauthors}{Marijan Soric, Cécile Gracianne, Ioana Manolescu, and Pierre Senellart}
\begin{abstract}
  Table Extraction (TE) consists in extracting tables from PDF documents,
  in a structured format enabling automatic processing.
  While numerous TE tools exist, the variety of methods and techniques
  makes it difficult for users to choose the most appropriate one.
  We propose a novel benchmark for assessing end-to-end TE methods (from PDF to the final table) over 86k pages.
  We contribute an analysis of TE evaluation metrics, and a novel,
  rigorous evaluation process, which
  allows scoring each TE sub-task as well as end-to-end TE, and captures model uncertainty.
  Along with prior datasets, our benchmark comprises two new heterogeneous datasets of 39k samples.
  We run our benchmark on diverse models, including off-the-shelf libraries,
  tools, computer vision-based models and modern approaches using general and specialized vision language models.
  The results demonstrate that TE remains challenging:
  current methods suffer from a lack of generalizability when facing heterogeneous data,
  and from limitations in robustness and interpretability.
\end{abstract}


\maketitle

\begingroup\small\noindent\raggedright
This is the extended version, including appendices, of a paper published in the
\emph{Proceedings of the 32nd ACM SIGKDD Conference on Knowledge Discovery and
Data Mining V.2 (KDD~'26)}, \url{https://doi.org/10.1145/3770855.3817462}.\\
\textbf{Availability:}
The source code of the implementation accompanying this paper is available at
\url{\vldbavailabilityurl}.
\endgroup

\begin{figure}[thbp]
  \centering
  \includegraphics[width=.9\linewidth]{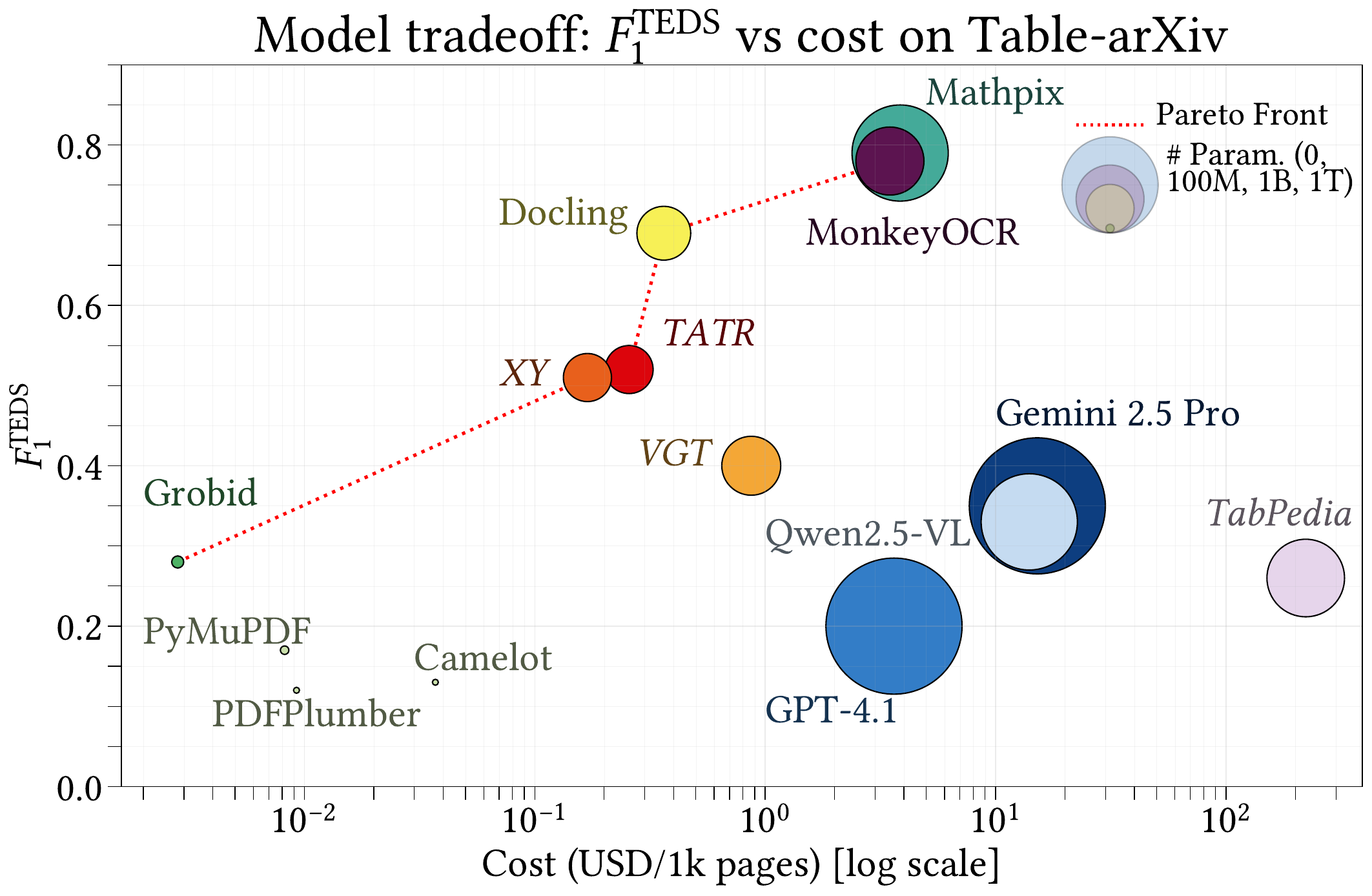}
  \caption{Performance vs. computational cost} \label{fig:tradeoff}
\end{figure}
\section{Introduction}
In PDF documents, such as scientific publications, business or
administrative reports, etc., tables are often used to represent data
in a way that is easy to read by humans. Such tables often contain
valuable information, which cannot
be easily found in other sources. Thus, retrieving the set of all tables from
these documents is crucial in order to exploit such data. %
This is the goal of \textbf{table extraction}
(\textbf{TE}, in short).
TE is often a first step of data
analysis pipelines, e.g., to answer questions over
tables~\cite{guo2025birdie,imam2025pneuma} or
combine them within a data lake~\cite{DBLP:conf/edbt/KhatiwadaSM26,DBLP:journals/csur/PatonCW24,DBLP:journals/pvldb/KoutrasZQLIFKK25}.

\begin{figure}[thbp]
  \centering
  \includegraphics[width=.9\linewidth]{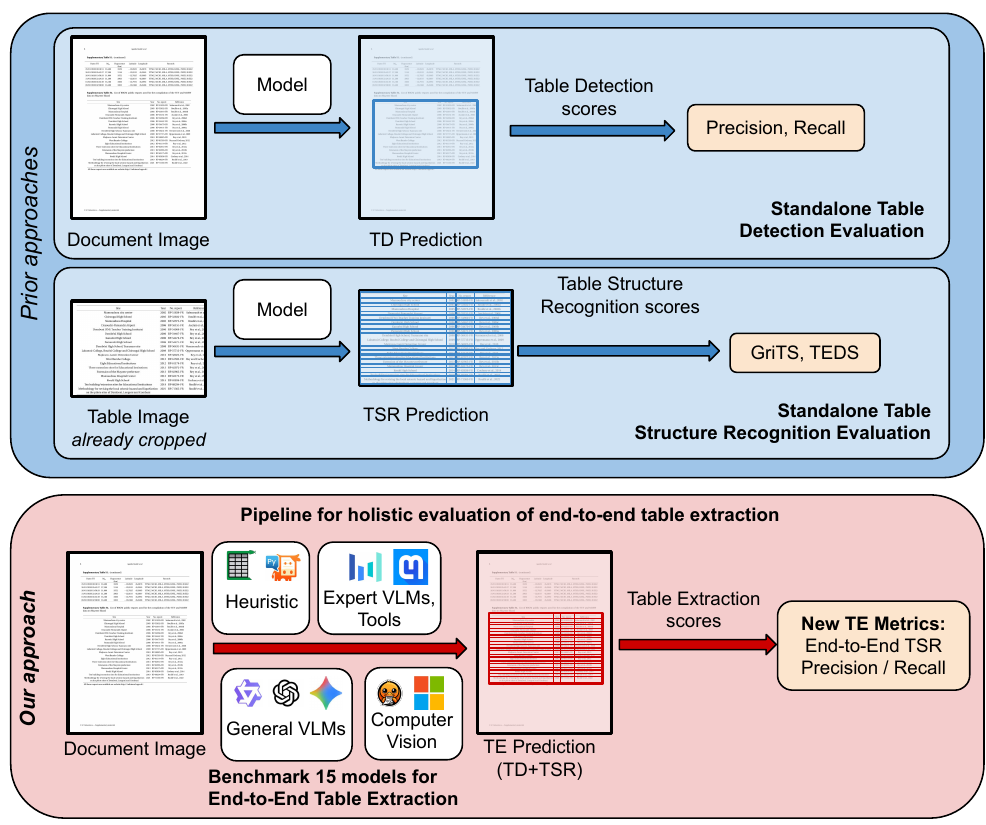}
  \caption{Prior approaches evaluate Table Detection and Table Structure Recognition in isolation,
    each on its own dataset and with TSR starting from gold-cropped table images.
    Our pipeline instead evaluates 15 end-to-end models on the same documents,
    feeding the TD output into TSR and scoring the result with dedicated end-to-end TE metrics.} \label{fig:pipeline}
\end{figure}

Prior research has identified two core TE sub-tasks:  \textbf{Table
  Detection} (\textbf{TD}) and \textbf{Table Structure Recognition}
(\textbf{TSR}), often studied and evaluated independently, each on its own dataset~\cite{6628853, li2019tablebank, chi2019complicated, smock2023aligning}.
However, in any real pipeline TD feeds TSR, so detection errors propagate downstream.
Evaluating the two in isolation, on separate datasets (\autoref{fig:pipeline}), therefore fails to capture what we actually care about,
which is why an \emph{end-to-end} evaluation is needed. Moreover,
new methods are typically compared only against others of the same family on homogeneous data,
rather than across diverse approaches and document layouts.

In this paper, we address this gap by proposing carefully justified metrics for TE
methods and, based on these metrics,
comparing a large set of methods
in an \emph{end-to-end} fashion, on scientific PDF documents.
We review existing methods on
heterogeneous datasets (diverse in terms of layout and table styles),
ranging from simple rule-based libraries to object detection systems
based on transformers, general and task-specialized Vision Language Models (VLMs),
and a commercial tool.
As modern approaches (based on vision language models) require heavy computation,
we also investigate the tradeoff between performance and computational cost, as illustrated in
\autoref{fig:tradeoff}.

Our main contributions are as follows:
\begin{inparaenum}
  \item We compare academic methods to those widely used
  in practice, on four benchmark datasets. While prior work assessed TE subtasks in isolation,
  this paper enables end-to-end quality evaluation of TE methods.
  \item We propose a new framework for
  TE evaluation, with formally justified, new end-to-end metrics;
  \item We create and
  publicly share two novel datasets for TD, TSR, and TE, with high-quality
  ground-truth data.
\end{inparaenum}

We distinguish \emph{text-based} PDFs, i.e., files produced via tools such as Office or \LaTeX,
from \emph{non-text} ones, which are mostly results of scan or photography.
While the latter include historical (legacy) documents, the former are continuously created, in increasing numbers, e.g.,
\href{https://arxiv.org/stats/monthly_submissions}{on arXiv}\footnote{\url{https://arxiv.org/}}.
Thus, in this work, \emph{we focus on text-based PDFs only}.
Non-text PDFs could be converted to text-based ones via Optical Character Recognition, however,
this may introduce extra errors, whose impact we seek to avoid in our TE benchmark.

We define the TE, TD, and TSR tasks in Section~\ref{sec:background}, also
discussing the relevant literature. We describe our
datasets and the methods we compare in
Sections~\ref{sec:datasets} and~\ref{sec:methods}, respectively.
We present evaluation metrics and our evaluation methodology in
Section~\ref{sec:metrics}.
Experimental results are in Section~\ref{sec:experiments}.
Our benchmark is archived at \href{\vldbavailabilityurl}{\url{\vldbavailabilityurl}}, together with additional material and an extended
version with appendix.

\section{Table Extraction}
\label{sec:background}

We view a \textbf{table} extracted from a document
as a list of \emph{rows}, each of which consists of a list of \emph{cells}.
Further, each cell has a \emph{content}, that may be textual, numerical, mixed, etc.
Note that different rows may have different numbers of cells; this is
due to \emph{merged cells}, i.e.,
cells spanning several rows and/or columns.

\subsection{Table Extraction Tasks}
Table Extraction (TE) involves detecting and recognizing a table's logical structure and content from its
unstructured presentation in a document. TE can be decomposed in subtasks, as follows.

\paragraph{Table Detection (TD)}
This consists in \emph{detecting} tables in an input document (PDF document,
or page given as a bitmap image). We assume tables are \emph{aligned with the axes} of
the pages, and might have been \emph{rotated} by \(\pm 90\) degrees.
Typically, a table is detected as a \textbf{rectangular bounding box}.
Note that some TE models do not output bounding boxes, but only HTML tables.

\paragraph{Table Structure Recognition (TSR)}
This consists in \emph{recognizing the structure
  of a table in terms of rows, columns, and cells}.
The input can be a cropped table image or a page image with detected
table bounding box.
Table Content Recognition (TCR) \emph{identifies the content (characters) within each cell}.
Clearly, TCR depends on TSR for the identification of cells; the two tasks are often performed together.
Thus, by a small abuse of language, we will denote \emph{both tasks} by the TSR acronym.
At the end of TSR, the table is fully captured in a structured format,
and can be rendered in HTML with its rows,
cells, spans, and content.
We give a complete formal definition of an extracted table's structure
(cells, spans, and merged cells) in \autoref{app:table-formalism}.

\subsection{Prior datasets and benchmarks}

The literature comprises several benchmarks for TD and TSR, the latter with or without TCR.
Some datasets focus \emph{solely on TD}, such as PubLayNet~\cite{zhong2019publaynet} and
TNCR~\cite{ABDALLAH2021}.
Others target \emph{TSR only}, e.g.:
SciTSR~\cite{chi2019complicated}, focusing on scientific papers;
PubTabNet~\cite{zhong2020imagebasedtablerecognitiondata}, from biomedical articles;
TabLeX~\cite{10.1007/978-3-030-86331-9_36}, which uses \LaTeX{} for dataset generation;
SynthTabNet~\cite{TableFormer2022}, a synthetically generated dataset;
WTW~\cite{Long_2021_ICCV} and iFLYTAB~\cite{zhang2023semv2}, from photos and scanned documents.
Other datasets address \emph{TD and TSR, but not TCR}, including:
ICDAR-2019~\cite{ICDAR19}, the dataset of a TD competition;
TableBank~\cite{li2019tablebank}, a larger dataset built
using weak supervision from Word and \LaTeX{} documents on the internet;
PubTables-1M~\cite{TATR}, the largest dataset available (one million tables).
Some datasets support TE evaluation but \emph{consist of only images (scans) of documents}, %
as opposed to PDFs with text. Most recent examples include
TabRecSet~\cite{yang2023large} (real-world scanned documents), and
OmniDocBench~\cite{11091844} (manuscript annotations, magazines).
Among the few text-based PDF datasets %
is that from the ICDAR-2013~\cite{6628853} TD competition.
While FinTabNet~\cite{9423317, smock2023aligning} (financial reports) meets our criteria,
its poor data quality led us to exclude it from our study.
A comprehensive dataset overview is provided in \autoref{app:datasets}.

Most prior benchmarks evaluate only a \emph{limited set of TE methods},
often focusing on introducing a new dataset rather than studying existing approaches~\cite{huang2024detection}.
In contrast, our benchmark evaluates a \emph{diverse range of methods},
fifteen in total, from classical to very recent ones. %
In particular, we provide the first thorough comparison of recent VLMs, proposed for document
layout analysis, with specialized TE methods, on common datasets, with
a unified comparison framework.

Existing benchmarks tackle \emph{TD and TSR as two independent problems} and \emph{do not evaluate the quality of end-to-end TE methods}.
Consider a two-step TE method involving two models where the output of TD becomes the input of TSR model.
TSR \emph{benchmarks} start from detected (cropped) table images.
In contrast, in \emph{real-world deployments},
the quality of TSR output can strongly degrade if its input (TD output) is of poor quality.
This impact is missed when evaluating TD and TSR separately; or, in reality, what matters is %
\emph{end-to-end} TE result quality.
This is why we need an \textbf{end-to-end benchmark}, and thus
\textbf{new metrics} for evaluating the end-to-end model synergy. %

Another limitation of existing benchmarks is that \emph{they  only
  consist of positive instances}, i.e., each sample includes at least
one table. As we will show, this \emph{inflates precision scores} and
\emph{compromises their relevance}  for real-world performance.

\paragraph{Existing TE methods}
An end-to-end TE model processes a document (or a page) to extract tables.
Early heuristics methods relying on hard-coded
patterns were limited to specific table types~\cite{pyreddy1997tinti,wangt2001automatic,10.1007/11551188_67,jahan2014locating,itonori1993table,chandran1993structural}.
Machine Learning (ML) later improved generalization~\cite{trecs, 10.1145/2425333.2425395,article, kasar2013learning} but
required heavy feature engineering, which was eventually surpassed by Deep Learning (DL) models~\cite{DLA, 10.1145/3657281}.
Thanks to larger training datasets, deeper networks
and more efficient architectures, these now achieve superior accuracy and versatility.
Notably, \cite{hao2016table} used Convolutional Neural Networks (CNNs) for TD, and DeepDeSRT~\cite{8270123} was the first to use Faster R-CNN-based
models~\cite{7410526} for TD and TSR.

Some models are focused on certain data types, for
example scans of paper documents with distortion~\cite{Ma_2023}, others
on robustness. More recently, expert VLMs specialized for TE or document parsing have emerged.
We detail state-of-the-art methods in
Section~\ref{sec:methods}.

\section{Our benchmark} \label{sec:datasets}

\begin{table}[htbp]
  \caption{Datasets used in our benchmark (L: \LaTeX, W: Word, P: Professional publishing system, O: Other).}
  \label{tab:datasets}
  \resizebox{\columnwidth}{!}{
    \begin{tabular}{@{}l rrrr rrr@{}}
      \toprule
      \multirow{2}[1]{*}{\textbf{Dataset}} & \multicolumn{4}{c}{\% \textbf{Type}} & \multicolumn{3}{c}{\textbf{Count}}                                                      \\
      \cmidrule(lr){2-5}\cmidrule(lr){6-8}
                                           & L                                    & W                                  & P       & O  & \# Pages & \# Positives & \# Tables \\
      \midrule
      PubTables   %
                                           & 0                                    & 1                                  & 85      & 14
                                           & 46\,942                              & 46\,942                            & 55\,990                                            \\
      \textbf{Table-arXiv} %
                                           & 100                                  & 0                                  & 0       & 0
                                           & 36\,869                              & 5\,214                             & 6\,308                                             \\
      \textbf{Table-BRGM}  %
                                           & 11                                   & 81                                 & 0       & 8
                                           & 2\,003                               & 256                                & 361                                                \\
      ICDAR-2013  %
                                           & ?                                    & ?                                  & ?       & ?
                                           & 238                                  & 135                                & 163                                                \\
      \bottomrule
    \end{tabular}}
\end{table}

Our benchmark for evaluating the quality of end-to-end TE methods
is based on four datasets of \emph{scientific} text-based PDF
documents.
These are structured documents
primarily used in academic, research, or technical fields.
Their tables vary widely in design (depending on the publisher), and
domain, e.g.,
biomedical,
mathematical, physical, geological, governmental (administrative), etc.
Each dataset contains:
\textbf{as input}, document pages (both as PDF files and as individual
rasterized images)
with coordinates of each word appearing within the page (such
coordinates can be obtained by standard PDF or image analysis libraries such as
PDFAlto and PDFMiner\footnote{\url{https://github.com/kermitt2/pdfalto}, \url{https://github.com/pdfminer/pdfminer.six}});
\textbf{as output}, every table identified within each page with
bounding box coordinates, as well as the table content and structure
formatted as HTML markup.
\autoref{tab:datasets} shows dataset statistics with novel contributions in bold.
We use the PDF metadata, extracted via \verb|pdfinfo|, to classify document types.
For ICDAR-2013, we only have access to processed PDFs (documents with deleted pages),
which prevents inference of the document type.

\paragraph{PubTables-Test}
First, we used a subset of the largest prior TD/TSR test dataset,
PubTables-1M, from scientific biomedical articles in PubMed Central Open Access~\cite{pcoa},
enriched with the PDF files and ground truth tables as HTML markup.
The documents
tend to be fairly homogeneous in the way tables are typeset;
the PDF metadata indicates that most documents
were produced using professional publishing systems such as Adobe
InDesign.

\paragraph{Table-arXiv}
To ensure \textbf{heterogeneity} within our data,
we have also automatically generated ``Table-arXiv'',
a new dataset of 37k samples built from the \LaTeX{} source
files of  arXiv preprints.
These sources were sampled from \emph{the last 20 years} and
\emph{across all arXiv domains}. %
Using the \LaTeX{} source code, we automatically generated TD and TSR (with TCR) annotations:
for TD, by instrumenting the commands starting and ending a
\verb|tabular| environment,
to add anchors that allow tracking their locations
within the generated PDF; for TSR, we used
LaTeXML\footnote{\url{https://github.com/brucemiller/LaTeXML}}
to generate HTML markup tables from the source file.

\paragraph{Table-BRGM}
We have manually constructed and annotated the dataset ``Table-BRGM'',
ensuring high annotation quality,
that is domain-specific and comprises French and English
geological reports sourced from a selection of BRGM
documents; %
BRGM
is France's reference public institution in Earth sciences.
The documents contain geological survey reports and records.
Its interest is that \emph{its layout significantly differs from the
  others}, mainly because these were mostly produced with a Word
processing software rather than a typesetting system such as \LaTeX.

\paragraph{ICDAR-2013}
Finally, we also considered the standard ICDAR-2013 dataset, a
well-known TE benchmark.
This dataset, introduced at the 2013 ICDAR Table Competition,
contains European Union and US Government PDF reports.

\smallskip

These datasets contain \textbf{heterogeneous} tables: small and large tables, tables with or
without borders, empty cells, merged cells, etc., in English and
French.

\section{End-to-end methods}
\label{sec:methods}

\begin{table}[htbp]
  \centering
  \caption{End-to-end methods for table extraction} \label{tab:all-model}
  \resizebox{\columnwidth}{!}{\begin{tabular}{@{}l ccc ccc c r c @{}}
      \toprule
      \multirow{2}[1]{*}{\textbf{Method}} & \multirow{2}[1]{*}{\textbf{Type}}                             & \multirow{2}[1]{*}{\textbf{Input}} & \multirow{2}[1]{*}{\textbf{TCR}} & \multicolumn{3}{c}{\textbf{Output}} & \multirow{2}[1]{*}{\textbf{Steps}} & \multirow{2}[1]{*}{\#\textbf{Param.}} & \multirow{2}[1]{*}{\textbf{Usage}}                               \\
      \cmidrule(lr){5-7}
                                          &                                                               &                                    &                                  & BBox                                & Conf.                              & HTML                                  &                                                                  \\
      \midrule
      \Camelot                            & \multirow{4}{*}{\makecell[l]{Baseline \S\ref{sec:libraries}}} & PDF                                & PDFMiner                         & $\checkmark$                        &                                    & $\checkmark$                          & 1                                  & 0M   & \multirow{4}{*}{CPU} \\
      \Pymupdf                            &                                                               & PDF                                & PyMuPDF                          & $\checkmark$                        &                                    & $\checkmark$                          & 1                                  & 0M   &                      \\
      \PdfPlumber                         &                                                               & PDF                                & PDFMiner                         & $\checkmark$                        &                                    & $\checkmark$                          & 1                                  & 0M   &                      \\
      \Grobid                             &                                                               & PDF                                & PDFAlto                          & $\checkmark$                        &                                    & $\checkmark$                          & 1                                  & 0.3M &                      \\ \cmidrule(lr){1-1} \cmidrule(lr){2-10}
      \TATR                               & \multirow{4}{*}{\makecell[l]{Computer                                                                                                                                                                                                                                                                                       \\ Vision \S\ref{sec:cv}}} & Image                              & PDFAlto                          & $\checkmark$                          & $\checkmark$                         & $\checkmark$                            & 2                                  & 60M  & \multirow{4}{*}{CPU} \\
      \XY                                 &                                                               & Image                              & PDFAlto                          & $\checkmark$                        & $\checkmark$                       & $\checkmark$                          & 2                                  & 60M  &                      \\
      \VGT                                &                                                               & Image                              & PDFAlto                          & $\checkmark$                        & $\checkmark$                       & $\checkmark$                          & 2                                  & 280M &                      \\
      \Docling                            &                                                               & PDF                                & EasyOCR                          & $\checkmark$                        &                                    & $\checkmark$                          & 2                                  & 130M &                      \\ \cmidrule(lr){1-1} \cmidrule(lr){2-10}
      \qwen                               & \multirow{3}{*}{\makecell[l]{General                                                                                                                                                                                                                                                                                        \\ VLMs \S\ref{sec:lvlm}}}  & Image                              & VLM                              & $\checkmark$                          &                                    & $\checkmark$                            & 1                                  & 32B  & GPU                  \\
      \gpt                                &                                                               & Image                              & VLM                              & ?                                   &                                    & $\checkmark$                          & 1                                  & ?    & API                  \\
      \gemini                             &                                                               & Image                              & VLM                              & $\checkmark$                        &                                    & $\checkmark$                          & 1                                  & ?    & API                  \\ \cmidrule(lr){1-1} \cmidrule(lr){2-10}
      \pedia                              & \multirow{3}{*}{\makecell[l]{Expert                                                                                                                                                                                                                                                                                         \\ VLMs \S\ref{sec:evlm}}}   & Image                              & PDFAlto                          & $\checkmark$                          &                                    & $\checkmark$                            & 2                                  & 8B   & \multirow{3}{*}{GPU} \\
      \got                                &                                                               & Image                              & VLM                              &                                     &                                    & $\checkmark$                          & 1                                  & 0.7B &                      \\
      \monkey                             &                                                               & Image                              & VLM                              & $\checkmark$                        & $\checkmark$                       & $\checkmark$                          & 1                                  & 3B   &                      \\ \cmidrule(lr){1-1} \cmidrule(lr){2-10}
      \mathpix                            & ? \S \ref{sec:mathpix}                                        & Both                               & ?                                & $\checkmark$                        &                                    & $\checkmark$                          & ?                                  & ?    & API                  \\
      \bottomrule                                                                                                                                                                                                                                                                                                                                                       \\
    \end{tabular}}
\end{table}

We tested various methods, ranging from simple Python
libraries, to well-established extraction platforms, computer vision-based methods,
general VLMs and expert VLMs. \autoref{tab:all-model} summarizes
the methods, grouped by type. Some methods (noted with ``Conf.'' in
\autoref{tab:all-model}) output a confidence score (between 0 and~1) along with their predictions,
indicating how confident the model is that a table exists in a given area.
We refer to models that return a confidence as ``probabilistic models'' (even though
the scores are not necessarily interpretable as probabilities).
``BBox'' and ``HTML'' state whether the tool outputs a table's coordinates
(bounding box, \textbf{BBox}) and/or its structure in HTML.
``\#Param.'' is the number
of model parameters; ``Usage'' states how the
tool is run (on CPU, GPU, or through an API).

\subsection{Baseline Models}
\paragraph{Commonly used Python libraries}\label{sec:libraries}

We selected the most popular TE Python libraries on GitHub: PDFPlumber~\cite{pdfplumber},
PyMuPDF\footnote{\url{https://github.com/pymupdf/PyMuPDF}} and
Camelot\footnote{\url{https://github.com/camelot-dev/camelot}}, with
respectively (as of June~2026) 10.4k, 9.9k, and 3.7k stars.
Taking text-based PDF as input, they have rule-based heuristics
using the \emph{lines} that are explicitly drawn or implied by the word alignment on the page
for table detection and extraction.

\paragraph{Grobid}\label{sec:grobid}
(GeneRation Of BIbliographic
Data\footnote{\url{https://github.com/grobidOrg/grobid}}~\cite{GROBID})
is a machine-learning library for extracting, parsing, and
re-structuring raw documents such as PDF into structured XML/TEI-encoded documents, with a particular focus on
technical and scientific publications; it is used in production, e.g., on
France's national preprint server HAL. %
Grobid structures the document's text body into paragraph, section titles, reference, figures, \emph{tables},
etc., which it represents as XML.

\subsection{Computer Vision-based methods}\label{sec:cv}
More advanced methods for TE rely on computer vision techniques, taking images as input
(not PDF). We study four end-to-end TE methods, %
with different architectures.
Confidence scores are typically obtained from the final softmax layer of the neural network
and are used to classify bounding boxes.

\paragraph{TATR}
TAble TRansformer (TATR)~\cite{TATR} is a model released by authors of PubTables-1M, pretrained on their dataset,
using the DEtection TRansformer (DETR)~\cite{DETR} architecture.
It consists of %
two parts: a backbone model (CNN) which extracts image features,
fed into an encoder--decoder transformer model for object detection tasks.
In contrast with prior approaches, e.g.~\cite{lin2018focallossdenseobject,zhou2019objectspoints,tian2019fcosfullyconvolutionalonestage},
DETR does not need to be given ``anchors'' (regions where to look for candidate objects to extract).
Instead, DETR learns a small, fixed number of position encodings, so-called \emph{object queries}, and uses them for extraction.
TATR is composed of two models:
one for TD, \emph{TATR-detect}, whose output feeds the \emph{TATR-struct} model which performs TSR.
Given a page as an image, TATR-detect performs object detection
to locate tables.
Then, in the cropped images, TATR-struct
detects table structure elements.
TATR-struct expects as input the locations of tokens (letters, digits, symbols, etc.) in the document as vertical and horizontal coordinates in pages; we use PDFAlto
to extract them. Finally, we obtain a complete table, including textual cell content,
and export it as HTML markup.

\paragraph{XY-cut+TATR}
Next, we introduce a new probabilistic model we devised: XY-cut+TATR.
We noted that TATR-detect was sometimes wrong on pages containing multiple tables: either some were missed, or, when recognized, their confidence scores were low.
Object queries (decoder input) are positional embeddings learned during training,
which contain abstract information about spatial
locations (``where should the model look?'').
Thus, each output, which is directly associated to an object query, is roughly specialized in an area.
To improve TATR results, we pre-process each page given as input, by separating layout components that are isolated by white space;
for this, we rely on the algorithm XY-cut~\cite{XY}. XY-cut works as follows.
By counting the number of black pixels\footnote{Assuming PDF reports
  are written in black on a white background.} along each axis, $X$ and $Y$, we
obtain \emph{profiles of pixel distributions}, which are used to isolate
rectangle components to form sub-images.
Each sub-image is then fed as input to
TATR-detect. This allows to help the model focus on individual areas.
We will designate this model as \emph{XY}.

\paragraph{VGT+TATR-struct}
\label{sec:vgt-pipeline}
This model follows a pipeline similar to that of TATR,
where TATR-detect is replaced with a multimodal vision component,
Vision Grid Transformer~\cite{da2023visiongridtransformerdocument}
(VGT). This is a Document Layout Analysis tool; it produces structured representations of documents.
In contrast with TATR, which only uses visual information, VGT also relies on textual information, which it views as a 2D grid.
This model achieves state-of-the-art results
on TD over the PubLayNet~\cite{zhong2019publaynetlargestdatasetdocument} dataset.
VGT has two parts:
a Vision Transformer, %
which models visual information, inspired by ViT~\cite{vit} and DiT~\cite{dit};
and a Grid Transformer (GiT), that models 2D language information with 2D token-level grid.
Overall, the framework aims to generate better embeddings from the input image page, in order to feed the detection framework
that performs object
detections using 6 classes, including table.
The detection framework is the Cascade R-CNN~\cite{cai2017cascadercnndelvinghigh}
detector, which is implemented based on the detection library Detectron2~\cite{wu2019detectron2}.

\paragraph{Docling} \label{sec:docling}
Docling~\cite{auer2024doclingtechnicalreport} is an advanced Python
library for PDF document processing and understanding (60.8k stars on GitHub).
It extracts tables in two steps: it leverages
RT-DETR~\cite{10657220} for TD, and TableFormer~\cite{TableFormer2022} for TSR,
while relying on EasyOCR\footnote{\url{https://github.com/JaidedAI/EasyOCR}},
a third-party OCR library, for TCR.
Docling does not provide confidence scores for TD.
RT-DETR, based on DETR, finds various layout components (table, text, title, etc.) in each page.
RT-DETR adds an efficient hybrid encoder that processes multi-scale features
(inspired by Deformable-DETR~\cite{zhu2021deformabledetrdeformabletransformers});
and the uncertainty-minimal query selection (inspired by DINO~\cite{zhang2022dinodetrimproveddenoising})
that improves the quality of initial object queries.
This model has been re-trained on the DocLayNet dataset~\cite{10.1145/3534678.3539043},
comprising diverse data sources with varied layouts.
TableFormer~\cite{TableFormer2022} is a vision-transformer model
that identifies the structure of a table by using a custom structure token
language~\cite{10.1007/978-3-031-41679-8_3}, starting from an image of the table.
Its architecture consists of an encoder (CNN backbone network and transformer encoder)
producing features that represent the input image, and two transformer decoders:
Structure Decoder and Cell BBox Decoder. The former generates the logical table structure
as a sequence of tokens (HTML tags), while the latter
predicts simultaneously bounding boxes of table cells.

\subsection{General VLMs}\label{sec:lvlm}
General VLMs are large-scale multimodal models designed for broad applications,
including document parsing.
By integrating large OCR datasets during pretraining, models such as
Qwen2.5-VL~\cite{bai2025qwen25vltechnicalreport} have shown strong performance in
document content extraction tasks.

We selected Qwen2.5-VL 32B, an open source model able to localize objects using bounding boxes,
document parsing and ``robust structured data extraction from tables''.
We also use OpenAI's GPT-4.1~\cite{gpt4.1} and
Google's Gemini 2.5 Pro~\cite{comanici2025gemini25pushingfrontier} as proprietary models,
since they are recent, popular, and reported to perform very well on multimodal tasks.
Qwen2.5-VL and Gemini 2.5 Pro are expected to have visual capacities
and spatial understanding because of their pretraining.
It remains unclear whether GPT-4.1 supports object localization, but we
maintained the same prompt across all VLMs for consistency.
We use these general VLMs in zero-shot prompting:
given an image and a prompt, the VLM outputs normalized table coordinates and HTML tables.
See \autoref{app:vlm-prompt} for the full prompt.
As we will discuss in \autoref{sec:eval-html}, the model's limited visual capacity poses
a challenge for evaluation, since TD analysis typically requires table coordinates.
API inference cost a total of 320 USD for GPT-4.1 and 1\,200 USD for Gemini 2.5 Pro.
Also, we cannot guarantee the absence of
contamination: these VLMs may have been trained on the datasets from the benchmark.

\subsection{Expert VLMs}\label{sec:evlm}
Expert VLMs are large multimodal models specifically trained for document parsing tasks.
\paragraph{TabPedia}
TabPedia~\cite{10.5555/3737916.3738146} is a specialized VLM in visual table understanding, able to perform
TD, TSR, table querying and table question answering.
TabPedia is composed of High and Low Resolution Vision Encoders, using the Swin-B~\cite{liu2021Swin} and
CLIP visual encoder ViT~\cite{pmlr-v139-radford21a} architecture respectively.
The pre-training stage aims to align the visual features to the large language
model (LLM) \mbox{Vicuna-7B}~\cite{vicuna}, and the fine-tuning stage focuses on visual table-aware understanding.
These pre-trained models have been finetuned on PubTables-1M, FinTabNet and PubTabNet for TD and TSR.

\paragraph{GOT}
GOT-OCR 2.0~\cite{wei2024general} is an ``OCR 2.0'' model, based on VLMs.
GOT relies on three modules: an image encoder (VitDet~\cite{10.1007/978-3-031-20077-9_17}), a linear layer, and
an output decoder (OPT-125M~\cite{zhang2022optopenpretrainedtransformer} or Qwen-0.5B).
It follows three training stages: a pre-training for the vision encoder,
then a joint-training for the encoder-decoder
and finally a post-training for the language decoder.
The authors curated a dataset of tables by crawling a large collection of \LaTeX{} source files from arXiv.
GOT outputs a \verb|.tex| file describing the whole page, including the tables,
but without coordinates.

\paragraph{Monkey-OCR}
MonkeyOCR-pro-3B~\cite{li2025monkeyocrdocumentparsingstructurerecognitionrelation}
is a unified vision and language framework for robust document parsing.
MonkeyOCR adopts a Structure-Recognition-Relation triplet paradigm,
which simplifies the multi-tool pipeline of modular approaches while
avoiding the inefficiency of using general VLM for full-page document processing.
The first stage employs a YOLO-based~\cite{zhao2024doclayoutyoloenhancingdocumentlayout}
document layout detector; the
second stage performs content recognition within detected regions through a LLM.
In the final stage, the logical reading order of detected elements is inferred through relation prediction.
The training dataset has not been disclosed.

\subsection{Commercial tool}\label{sec:mathpix}
\paragraph{Mathpix}
We evaluated Mathpix\footnote{\url{https://mathpix.com/}}, a closed-source commercial tool accessed via API.
While its model architecture is not publicly disclosed,
Mathpix is a STEM-specialized system rather than a general-purpose model.
It focuses on scientific and technical document
understanding (math, chemistry, tables, figures, and multi-column layouts).
API inference cost a total of 310 USD.
\section{Metrics}
\label{sec:metrics}
In this section, we present metrics used for TD, TSR, and TE, as well as our evaluation methodology.

We first introduce some notations.
From now on, \emph{we highlight the first occurrence of each notation encasing it in a box, for quick reference}.
Let \fbox{\(\mathcal P\)} be the set of TE model outputs: each of its elements is a
tuple that includes a bounding box, a predicted HTML table $\hat y$,
and, if available, a confidence score $c_{\text{table}}$.
When the models lack a confidence score, we consider  \(c_\text{table}=1\) in all  $\mathcal{P}$ entries.
We define  \fbox{$\mathcal P^+$} \(\subseteq \mathcal{P}\) as the set of
positive predictions: those that we consider to be  actual tables.
For probabilistic models, where confidence scores can be interpreted as table
probabilities, we use a threshold \fbox{\(\theta_c\)} to define
positive predictions. This threshold draws a decision boundary: only
predictions with scores above it are counted as positive. In
this case, we set \fbox{\(\mathcal P_{\theta_c}\)} to be \(\{(\hat
y,c_{\mathrm{table}}) \in \mathcal P
\mid c_\text{table}>\theta_c \}\), with confidence score threshold
\(\theta_c \in [0,1]\).
For the non-probabilistic models, \(c_\text{table}=1\), and thus
we have \(\mathcal P^+=\mathcal{P}\).
More details about the metrics (threshold-free TD metric and the calibration error)
are provided in \autoref{app:metrics}.

\subsection{Table detection} \label{sec:metrics-td}
TD performance can be evaluated using traditional metrics inspired from Information Retrieval.
Specifically, a True Positive (TP) is a table correctly
recognized, a False Positive (FP) is a table found by the model where a
human user would not consider a table exists, and a False Negative (FN)
is a table recognized by human users but missed by the model.
Given a table predicted by the model, and a table in the gold standard, to identify TPs, we rely on the
Intersection-over-Union (IoU) metric between the areas of the prediction, and the  gold standard tables. If IoU is above a given
threshold \fbox{\(\theta_J\)}\(\in [0,1]\), the prediction is counted as a TP.
For a given TD method, \fbox{\(\mathcal P^{++} \)}\(\subseteq \mathcal{P}^+\) denotes the set of true positive predictions.

\paragraph{Average precision} \label{sec:ap}
For probabilistic models, the set of positive predictions
\({\mathcal P^+ \in \left\{ \mathcal P_{\theta_c}\right\}_{\theta_c}}\) depends on the choice of a confidence
score threshold \(\theta_c\) and a IoU threshold $\theta_J$, thus we obtain a set of corresponding
scores \( \{X_{\theta_c,\theta_J}\}_{\theta_c,\theta_J}\), where $X$ may denote Precision,
Recall, etc.
To aggregate over several \(\theta_c\)  values, in the object detection literature,
the Average Precision metric is used~\cite{DBLP:journals/corr/LinMBHPRDZ14}.
The Average Precision at IoU threshold \(\theta_J\)  is the area under
the \emph{Precision--Recall curve} for a fixed $\theta_J$, varying $\theta_c$.
By ranking predictions in the increasing order of their confidence score, we can compute
Precision and Recall, at confidence score~\(\theta_c\) (where the positive predictions are elements from \(\mathcal P_{\theta_c}^+ \))
and at matching threshold \(\theta_J\), which leads to pairs of
Precision--Recall (the curve).
The benefit of the  \(\mathit{AP}_{\theta_J}\) metric is to account for the entire
recall range, not just a single recall threshold. We compute it using
Scikit-learn~\cite{Scikit}.

\paragraph{Estimating model confidence via its (mis)calibration}
\label{sec:miscalibration}
The metrics used above ($X_{\theta_c,\theta_J}$, and \(AP_{\theta_J}\)) rely
on binary decisions, where each prediction is either positive or negative.
We seek a finer-grain evaluation,
accounting not only for these binary choices, but also
for the associated confidence scores.
A probabilistic model is said to be \textbf{calibrated} when its probabilities
(confidence scores) are aligned with the real-world outcomes~\cite{guo2017calibrationmodernneuralnetworks}.
A calibrated model is desirable because its predictions can be seen as more trustworthy. Formally, model calibration is defined by:
\(
\mathbb P[ \widehat Y=Y \mid \widehat P = p]=p
\)
where \([\widehat{Y}=Y]\) is the event ``the model prediction is right''
and \([\widehat{P}=p]\) is the event ``the confidence score (seen as a RV)
is equal to \(p\)'',  for \(p \in [0,1]\). If this is the case, the random
variable \(\widehat P\) perfectly reflects  the prediction trustworthiness.
Based on this definition, the miscalibration of a classifier can be measured by the \emph{expected calibration error (ECE)}~\cite{ece}.

However, in the case of object detection, we cannot enumerate \([\widehat{Y}=Y]\), but only
\([\widehat{Y}= Y=1]\), because there is a multitude of negative locations (possible bounding boxes that do not contain a table).
Thus, for an object detection model, calibration is measured with respect to the
precision~\cite{Kuppers_2020}, leading to \emph{the object detection version of ECE}, or
\fbox{\(\operatorname{D-ECE}\)}, in short:
\(
\operatorname{D-ECE} = \mathbb E_{\widehat{P}} \left[
  \left\lvert \mathbb P\left[\widehat Y=Y=1 \,\middle|\, \widehat P = p\right]-p \right\rvert \right].
\)
To compute D-ECE efficiently,
we approximate (discretize) %
by partitioning predictions into equal
bins, and taking a weighted average of the bins' (Precision--Confidence) gap.

\begin{figure*}[!t]
  \centering
  \includegraphics[width=0.9\linewidth]{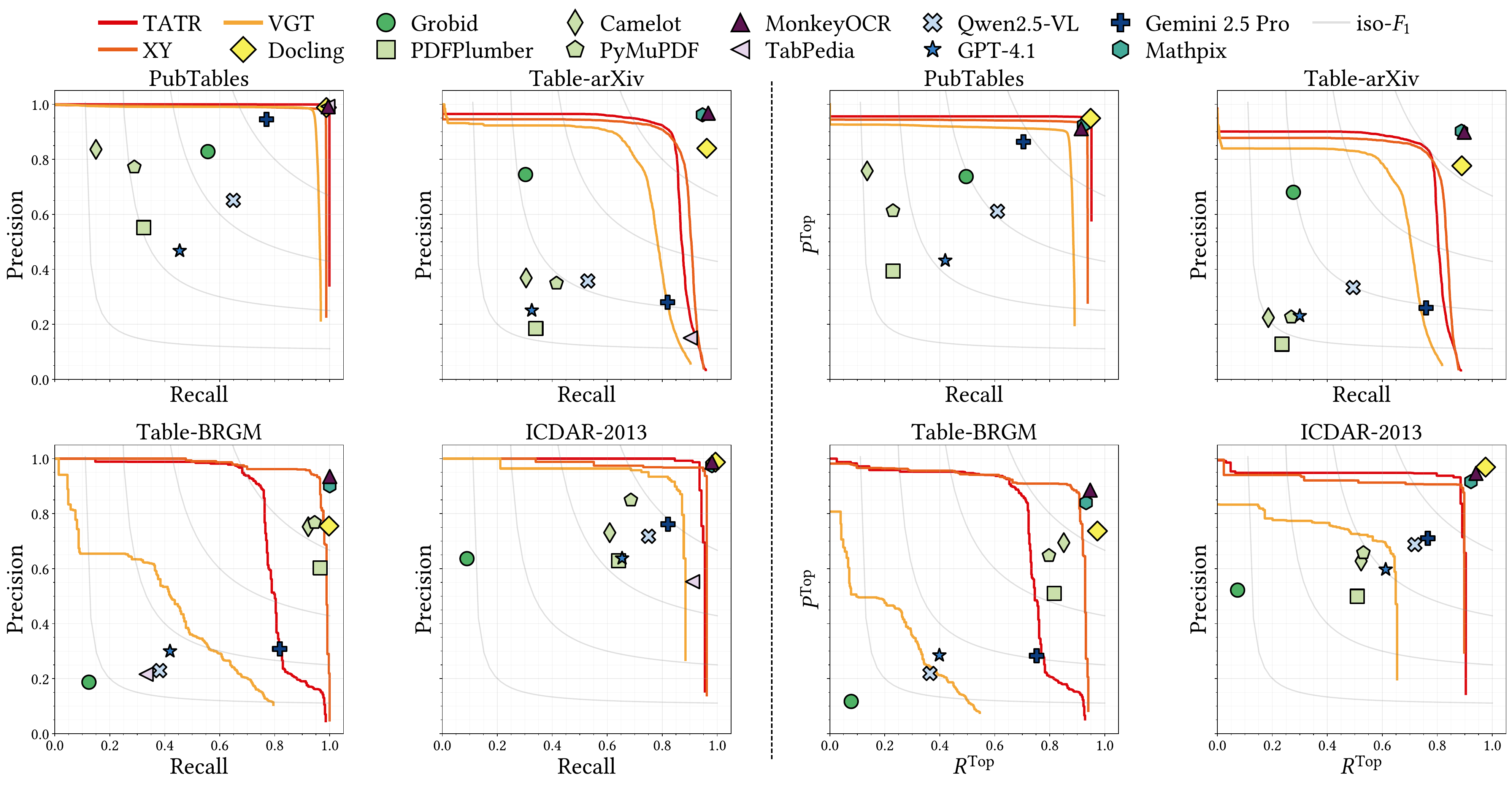}
  \caption{(left) Precision--Recall for \emph{bbox} TD; (right) \(P^{\mathrm{Top}}-R^{\mathrm{Top}}\) for \emph{bbox} TE}\label{fig:td-te}
\end{figure*}

\subsection{Table structure recognition} \label{sec:metrics-tsr}
Once a table has been correctly detected (TD yielded a true positive,
TP),
we want to evaluate TSR result quality,
by checking that the predicted table has the correct table shape (topology, structure),
and that cells have the expected content.

A na\"ive approach for TSR evaluation could be to tackle it as an object detection task
by comparing predicted bounding box cells with Ground Truth (GT) cells by their absolute positions.
This approach is vulnerable to \emph{shifts}: a slight shift along one or
both coordinates of a predicted bounding box cell would disproportionately penalize the models.
Therefore, we prefer metrics that rely on the structure, not on absolute coordinates.
The 4-gram BLEU score~\cite{bleu} (BLEU-4) metric compares predictions and GT
HTML tables as strings with markup structure (tags) and content but
lacks spatial structure evaluation,
and needs exact matches for cells. %
The DAR metric~\cite{DAR} (Directed Adjacency Relation)
is based on immediate neighbors, i.e., the relative positions of non-empty cells. %
As shown in~\cite{zhong2020imagebasedtablerecognitiondata}, DAR cannot detect
errors caused by empty cells and misalignment of cells beyond
immediate neighbors (it does not capture the global 2D structure) and uses exact match.
Thus, we chose to use two metrics: GriTS~\cite{grits},
which evaluates tables directly in their matrix form and computes a similarity between these matrices,
and TEDS~\cite{zhong2020imagebasedtablerecognitiondata} which compares
tables seen as trees of HTML tags. We describe these metrics below.

\paragraph{TEDS}
\label{sec:metrics-teds}
The metric Tree-Edit-Distance-Based Similarity (TEDS) measure views each HTML table as follows.
There is a root \verb|<table>| with children \verb|<tr>| (nodes are table rows).
The leaves of the tree are cells \verb|<td>| with attributes \verb|colspan|, \verb|rowspan| and \verb|content|.

TEDS builds upon the previous Tree-Edit Distance~\cite{PAWLIK2016157},
in turn inspired by the Levenshtein Distance~\cite{Levenshtein}. The latter counts the minimum number of insertions, deletions, and substitutions to
transform a string into another string. Similarly, TEDS counts
the node deletions, insertions, relabeling and content modifications needed to transform a tree (HTML
table) into another, as follows: the cost to \emph{insert or delete} and \emph{relabel} a node is 1;
the cost of \emph{relabeling a cell content} is the normalized Levenshtein similarity
between the two strings. Formally, for trees \(T_P,T_{GT}\):
\(
{\rm TEDS}(T_P,T_{GT})=1-\frac{{\rm Edit Dist}(T_P,T_{GT})}{\max(|T_P|,|T_{GT}|)}
\)
where \(|\cdot|\) is the number of nodes.

\paragraph{GriTS}
\label{sec:metrics-grits}
The Grid Table Similarity (GriTS) family of metrics considers table
structure from three perspectives:
\emph{Topology}, describing the rows and columns each cell spans over, in a two-dimensional grid;
\emph{Content}, which refers to the textual content within each cell;
and \emph{Location}, denoting the rectangular span in a pixel matrix that is occupied by each cell.
This leads to \emph{three GriTS similarity metrics}. %
We ignore the one based on Location (coordinates) since it suffers from the shift problem described above.
Each metric takes as input two matrices of cells, the prediction and the GT, and computes
their \emph{2-dimensional Longest Common Subsequence} (2D-LCS, in short).
First, tables are represented as matrices \(P, G\) (prediction, GT) entries in \(E\), which is
\(\mathbb Z^4\) and  \(V^{\mathbb N}\) for the Topology metric and Content metric, respectively.
From $(P,G)$, we extract \emph{the two substructure matrices} \(\widetilde P\) and \(\widetilde G\)
(of the same shape between themselves) \emph{that are the most similar according to \(f\)},
the penalty function between the grid cells' properties.
The \emph{penalty function} \(f\) depends on the subtask (Topology, Content), as explained below.
With \(|\cdot|\), the matrix size (number of grid cells):
\(
\mathrm{GriTS}_f(P,G)=2\frac{\sum_{i,j} f ( \widetilde P_{i,j}, \widetilde G_{i,j} )}{|P|+|G|}.
\)
The \emph{entry-wise penalty function} \(f \colon E \times E \rightarrow [0,1]\)
measures the \emph{partial correctness} between two entries with the same coordinates;
it replaces the 0 or 1 value used for the vanilla 2D-LCS computation~\cite{GLCS}.
Concretely, \(f\) is IoU for Topology, and LCS for Content.

\paragraph{HTML markup vocabulary to be used in our evaluation}
One last aspect needs to be settled with respect to the metrics based on HTML markup (TEDS and GriTS).
To avoid being penalized by different sets of HTML tags used by different models, we normalize each output
preserving only three tags: \verb|<table>|, \verb|<tr>|, \verb|<td>|.
From now on,
we implicitly consider that each TSR metric applies over such normalized outputs (and normalized GT) only.

\subsection{Table extraction} \label{sec:metrics-te}
We aim at evaluating TE to compare directly end-to-end methods as a whole.
To this end, we propose a metric similar in spirit with the well-known Precision and Recall metrics, because these are easily interpretable.
Specifically, we replace \emph{binary} values (1 for TP and 0 for FP) by
\emph{continuous scores, parameterized by the TSR score},
which measures how accurately table structure and content are extracted.
Recall that  \(\theta_J\) is  the IoU threshold above which table detection is considered to have found a table.

\begin{definition}[TSR Precision and Recall]
  Denoting for each  detected (positive) table \(i\) its IoU \(J_i\), given a TSR metric which associates to each positive $i$ the score  \(s^\mathrm{TSR}_i\), we define the TSR precision and recall as:
  \[
    P^{\mathrm{TSR}}\hspace{-.5em}=\frac{1}{\lvert \mathcal{P}^+ \rvert}\sum_{i \in \mathcal{P}^+} s^\mathrm{TSR}_i \cdot \mathbf{1}_{[J_i>\theta_J]}
    \qquad
    R^{\mathrm{TSR}}\hspace{-.5em}=\frac{1}{\lvert \mathcal{G} \rvert}\sum_{i \in \mathcal{P}^+} s^\mathrm{TSR}_i \cdot \mathbf{1}_{[J_i>\theta_J]}
  \]
  where \(\mathbf{1}_{[J_i>\theta_J]}\) is the indicator function whose value is $0$
  for tables whose IoU is below \(\theta_J\) and $1$ for the others and
  $|\mathcal{G}|$ is the number of ground-truth tables. %
\end{definition}

Intuitively, in the metrics defined above, we propagate the strength (or confidence) of the table detection, in order to also reflect it in the ``downstream'' task of TSR.

\begin{table*}[!t]
  \caption{\AP{} scores for \emph{bbox} TD and Average \(\mathbf{AP}^{\text{TSR}}\) for \emph{bbox} TE across datasets for probabilistic models.} \label{tab:te-ap}
  \resizebox{0.9\textwidth}{!}{\begin{tabular}{@{}l cccc cccc cccc  cccc@{}} \toprule
      \small
      \multirow{3}[4]{*}{\textbf{Method}} & \multicolumn{4}{c}{\textbf{PubTables}} & \multicolumn{4}{c}{\textbf{Table-arXiv}}                & \multicolumn{4}{c}{\textbf{Table-BRGM}} & \multicolumn{4}{c}{\textbf{ICDAR-2013}}                                                                                                                                                                                                                                                                                                                          \\
      \cmidrule(lr){2-5} \cmidrule(lr){6-9} \cmidrule(lr){10-13}  \cmidrule(lr){14-17}
                                          & \textbf{TD}                            & \multicolumn{3}{c}{\textbf{TE metrics \textit{(bbox)}}} & \textbf{TD}                             & \multicolumn{3}{c}{\textbf{TE metrics \textit{(bbox)}}} & \textbf{TD}   & \multicolumn{3}{c}{\textbf{TE metrics \textit{(bbox)}}} & \textbf{TD}   & \multicolumn{3}{c}{\textbf{TE metrics \textit{(bbox)}}}                                                                                                                                                      \\
      \cmidrule(lr){2-2} \cmidrule(lr){3-5} \cmidrule(lr){6-6} \cmidrule(lr){7-9} \cmidrule(lr){10-10} \cmidrule(lr){11-13} \cmidrule(lr){14-14} \cmidrule(lr){15-17}
                                          & \AP                                    & \APTop                                                  & \APCon                                  & \APTEDS                                                 & \AP           & \APTop                                                  & \APCon        & \APTEDS                                                 & \AP              & \APTop           & \APCon           & \APTEDS          & \AP           & \APTop           & \APCon           & \APTEDS          \\
      \midrule
      \TATR                               & \textbf{1.00}                          & \textbf{0.91}                                           & \textbf{0.80}                           & \textbf{0.80}                                           & \textbf{0.86} & \textbf{0.73}                                           & \textbf{0.56} & \textbf{0.52}                                           & \underline{0.82} & \underline{0.73} & \underline{0.67} & \underline{0.64} & \textbf{0.96} & \textbf{0.81}    & \textbf{0.67}    & \textbf{0.66}    \\
      \XY                                 & \underline{0.98}                       & \underline{0.88}                                        & \underline{0.78}                        & \underline{0.78}                                        & \textbf{0.86} & \textbf{0.73}                                           & \textbf{0.56} & \underline{0.51}                                        & \textbf{0.95}    & \textbf{0.87}    & \textbf{0.78}    & \textbf{0.73}    & \textbf{0.96} & \underline{0.79} & \underline{0.64} & \underline{0.63} \\
      \VGT                                & 0.96                                   & 0.81                                                    & 0.69                                    & 0.70                                                    & 0.75          & 0.60                                                    & 0.44          & 0.40                                                    & 0.46             & 0.21             & 0.15             & 0.14             & 0.88          & 0.51             & 0.33             & 0.34             \\
      \bottomrule
    \end{tabular}}
\end{table*}
\begin{table*}[!t]
  \caption{\(\mathbf{F_1}\)-scores for \emph{bbox} and \emph{txt} TD and \(\mathbf{F_1}^{\!\!\text{TSR}}\)-scores for \emph{bbox} TE across datasets for models.} \label{tab:te-f1}
  \small
  \resizebox{\textwidth}{!}{
    \begin{tabular}{@{} l cc ccc  cc ccc  cc ccc  cc ccc @{}} \toprule
      \multirow{3}[4]{*}{\textbf{Method}} & \multicolumn{5}{c}{\textbf{PubTables}} & \multicolumn{5}{c}{\textbf{Table-arXiv}}                & \multicolumn{5}{c}{\textbf{Table-BRGM}} & \multicolumn{5}{c}{\textbf{ICDAR-2013}}                                                                                                                                                                                                                                                                                                                                                                                                                                             \\
      \cmidrule(lr){2-6} \cmidrule(lr){7-11} \cmidrule(lr){12-16} \cmidrule(lr){17-21}
                                          & \multicolumn{2}{c}{\textbf{TD}}        & \multicolumn{3}{c}{\textbf{TE metrics \textit{(bbox)}}} & \multicolumn{2}{c}{\textbf{TD}}         & \multicolumn{3}{c}{\textbf{TE metrics \textit{(bbox)}}} & \multicolumn{2}{c}{\textbf{TD}} & \multicolumn{3}{c}{\textbf{TE metrics \textit{(bbox)}}} & \multicolumn{2}{c}{\textbf{TD}} & \multicolumn{3}{c}{\textbf{TE metrics \textit{(bbox)}}}                                                                                                                                                                                                                                     \\
      \cmidrule(lr){2-3} \cmidrule(lr){4-6} \cmidrule(lr){7-8} \cmidrule(lr){9-11} \cmidrule(lr){12-13} \cmidrule(lr){14-16} \cmidrule(lr){17-18} \cmidrule(lr){19-21}
                                          & \Ftxt                                  & \Fbbox                                                  & \FTop                                   & \FCon                                                   & \FTEDS                          & \Ftxt                                                   & \Fbbox                          & \FTop                                                   & \FCon            & \FTEDS           & \Ftxt            & \Fbbox           & \FTop            & \FCon            & \FTEDS           & \Ftxt            & \Fbbox           & \FTop            & \FCon            & \FTEDS           \\
      \midrule
      \Camelot                            & 0.23                                   & 0.25                                                    & 0.23                                    & 0.20                                                    & 0.20                            & 0.11                                                    & 0.33                            & 0.20                                                    & 0.15             & 0.13             & 0.75             & 0.83             & 0.76             & 0.68             & 0.68             & 0.54             & 0.66             & 0.57             & 0.51             & 0.50             \\
      \Pymupdf                            & 0.30                                   & 0.42                                                    & 0.33                                    & 0.29                                                    & 0.27                            & 0.13                                                    & 0.38                            & 0.25                                                    & 0.19             & 0.17             & 0.63             & 0.85             & 0.71             & 0.64             & 0.62             & 0.55             & 0.76             & 0.59             & 0.51             & 0.49             \\
      \PdfPlumber                         & 0.30                                   & 0.41                                                    & 0.29                                    & 0.25                                                    & 0.22                            & 0.09                                                    & 0.24                            & 0.17                                                    & 0.13             & 0.12             & 0.52             & 0.74             & 0.63             & 0.56             & 0.54             & 0.46             & 0.63             & 0.50             & 0.43             & 0.41             \\
      \Grobid                             & 0.61                                   & 0.67                                                    & 0.59                                    & 0.51                                                    & 0.47                            & 0.33                                                    & 0.43                            & 0.39                                                    & 0.31             & 0.28             & 0.06             & 0.15             & 0.09             & 0.07             & 0.06             & 0.12             & 0.16             & 0.13             & 0.12             & 0.11             \\
      \Docling                            & \textbf{0.96}                          & \textbf{0.99}                                           & \textbf{0.95}                           & \textbf{0.86}                                           & 0.86                            & \textbf{0.70}                                           & 0.90                            & \underline{0.83}                                        & \underline{0.72} & 0.69             & 0.78             & 0.86             & 0.84             & \underline{0.80} & 0.79             & \textbf{0.99}    & \textbf{0.99}    & \textbf{0.97}    & \textbf{0.90}    & \textbf{0.90}    \\
      \qwen                               & 0.90                                   & 0.65                                                    & 0.61                                    & 0.52                                                    & 0.54                            & 0.50                                                    & 0.43                            & 0.40                                                    & 0.33             & 0.33             & 0.68             & 0.29             & 0.27             & 0.24             & 0.24             & 0.92             & 0.73             & 0.70             & 0.64             & 0.64             \\
      \gpt                                & 0.68*                                  & 0.46*                                                   & 0.43*                                   & 0.35*                                                   & 0.36*                           & 0.29                                                    & 0.28                            & 0.26                                                    & 0.20             & 0.20             & 0.43             & 0.35             & 0.33             & 0.27             & 0.27             & 0.86             & 0.65             & 0.60             & 0.53             & 0.53             \\
      \gemini                             & 0.85*                                  & \underline{0.85}*                                       & 0.78*                                   & 0.72*                                                   & 0.74*                           & 0.27                                                    & 0.42                            & 0.39                                                    & 0.34             & 0.35             & 0.45             & 0.45             & 0.41             & 0.38             & 0.39             & 0.80             & 0.79             & 0.74             & 0.70             & 0.71             \\
      \pedia                              & --                                     & \textbf{0.99}                                           & TO                                      & TO                                                      & TO                              & --                                                      & 0.26                            & TO                                                      & TO               & TO               & --               & 0.26             & TO               & TO               & TO               & --               & 0.69             & TO               & TO               & TO               \\
      \got                                & 0.39                                   & --                                                      & --                                      & --                                                      & --                              & 0.14                                                    & --                              & --                                                      & --               & --               & 0.07             & --               & --               & --               & --               & 0.28             & --               & --               & --               & --               \\
      \monkey                             & \textbf{0.96}                          & \textbf{0.99}                                           & 0.91                                    & \underline{0.85}                                        & \underline{0.87}                & 0.67                                                    & \textbf{0.97}                   & \textbf{0.90}                                           & \textbf{0.79}    & \underline{0.78} & \underline{0.82} & \textbf{0.97}    & \textbf{0.92}    & \textbf{0.86}    & \textbf{0.84}    & \underline{0.98} & \underline{0.98} & \underline{0.94} & \underline{0.88} & 0.87             \\
      \mathpix                            & \underline{0.93}                       & \textbf{0.99}                                           & \underline{0.92}                        & 0.84                                                    & \textbf{0.88}                   & \underline{0.68}                                        & \underline{0.95}                & \textbf{0.90}                                           & \textbf{0.79}    & \textbf{0.79}    & \textbf{0.91}    & \underline{0.95} & \underline{0.88} & 0.78             & \underline{0.82} & 0.95             & \underline{0.98} & 0.92             & 0.84             & \underline{0.89} \\
      \bottomrule
    \end{tabular}
  }
\end{table*}

\subsection{Evaluation Methodology}
\label{sec:methodology}
We now explain how we apply these metrics to ensure a fair comparison of all models
and to evaluate end-to-end TE methods.

\paragraph{Text-based Table Detection} \label{sec:eval-html}
For \textbf{methods that output table locations} (bounding boxes), we use the metrics
presented in Section~\ref{sec:metrics-td}, based on IoU.
For models which \textbf{do not output bounding boxes} (\got{}, \autoref{tab:all-model}),
we replace the IoU by a Jaccard index (over tables content) to determine
whether a table has been detected, based only on the predicted and
GT tables HTML markup.
Given the \emph{multisets of 2-grams} obtained from the text
content of a predicted table $(S_P)$, respectively, from the GT $(S_{GT})$,
we define:
\(
\text{Jaccard}(S_P, S_{GT})= \frac{|S_P \Cap S_{GT} |}{|S_P \Cup S_{GT}|}
\)
where
\(\Cup\) is the multiset union (retaining each multiset element with its maximum multiplicity), and \(\Cap\)
the multiset intersection (based on minimum multiplicity).
Accordingly, we call ``\emph{bbox} TD''
and ``\emph{txt} TD'' the TD evaluations based on
bounding boxes, respectively, table content. After the TD evaluation,
TSR evaluation %
is carried over the true positives \(\mathcal{P}^{++}\).

\paragraph{Evaluating TSR impacted by TD}
\label{sec:tsr-evaluation}
Traditionally, TD and TSR are evaluated independently.
However, in downstream tasks, models are used sequentially to perform
end-to-end TE, thus TSR evaluation depends on TD.
Note that most recent models tested are one-step TE methods,
and we may not be able to separate (discern) the two sub-models.  %
Aiming for an end-to-end evaluation, we choose not to use a predefined dataset as input for TSR.
Instead, we use the set of true positives \(\mathcal{P}^{++} \subseteq \mathcal{P}^{+}\) obtained from
the TD inference part; we use \(\theta_J=0.5\) to distinguish TP from FP.
In this way, the TSR score better reflects what can be observed in the use of the end-to-end model.
The final TSR score is then obtained by averaging the scores over the set of TP.
This has two consequences:
($i$)~\emph{TSR evaluation is not independent of TD}; %
($ii$)~We can \emph{no longer directly compare TSR results between models},
since the input dataset \(\mathcal{P}^{++}\) is different for each model
(determined by the model's performance on TD).
Consequently, if TD fails to correctly\footnote{For instance, a TP with IoU at 51\% may miss part of the table's content.}
detect tables, the structure model will be penalized. %
While this may seem ``unfair'' to TSR models, it reflects \emph{the reality of end-to-end TE result quality}, which are measured on the final results.
Our end-to-end evaluation protocol,
including the \emph{text}-Jaccard matching and how TSR is evaluated conditionally on TD,
is detailed in \autoref{app:methodology}; the relationship between \emph{bbox} and \emph{txt} TD evaluation is further analyzed in \autoref{sec:jaccard-bbox-content}.

\section{Experimental evaluation}
\label{sec:experiments}

We begin by analyzing TD results (\autoref{sec:td-results}), comparing results obtained
with \emph{bbox} and \emph{txt} TD, model calibration,
and the tradeoff between performance and computational cost.
Next, we analyze TE results (\autoref{sec:te-results}) using our new end-to-end evaluation framework.
Our code is archived \href{\vldbavailabilityurl}{on Zenodo}.
The configurations, pipelines, and implementation details for every model under
test are reported in \autoref{sec:setting}.
Experimental results will answer the following questions raised previously:
\begin{inparaenum}
  \item[(Q1)] Which models demonstrate the most consistent performance across all datasets?
  \item[(Q2)] Do low-cost models perform well in some cases?
  \item[(Q3)] Are VLMs reliable and competitive?
  \item[(Q4)] Which models offer the best tradeoff between performance and computational cost across datasets?
  \item[(Q5)] Are confidence scores trustworthy?
\end{inparaenum}
Additional per-dataset results, reliability diagrams,
and performance/cost tradeoff plots are presented in \autoref{app:more-results}.

\subsection{Table detection performance}\label{sec:td-results}
\autoref{fig:td-te} (left) plots Precision--Recall curves from \emph{bbox} TD over our four datasets.
Models are distinguished by colors (same as in~\autoref{tab:all-model}).
We plot \emph{probabilistic model scores as curves} and
represent \emph{the other models as dots}, using IoU threshold of \(\theta_J = 0.5\).
Moreover, we draw iso-\(F_1\) curves in gray (\(F_1\) values: 0.2, 0.4, 0.6, 0.8).
\emph{The closer a curve is to the top-right corner, the better
  its trade-off between Precision and Recall}.
Finally, we compute Average Precision scores, denoted as \fbox{\(AP\)}, which is the area under the
Precision--Recall curve (see \autoref{tab:te-ap}) to easily compare curves.
\autoref{tab:te-f1} gathers \(F_1\)-scores for other models with \emph{txt}
and \emph{bbox} TD (denoted \fbox{\Ftxt}, \fbox{\Fbbox}).
We use the former to compare GOT with other models.
For Gemini and GPT, scores with an asterisk (*) are for results
obtained using a subset of the data (40k over 47k images), as inference was stopped after 1\,500 USD due to budget constraints.
The best results are shown in bold, and the second best are
underlined.

Having four datasets allows us to observe different behaviors of models
depending on the document layout and table styles.
Indeed, models' performance varies when changing the datasets.
For example, Grobid, trained on scientific documents, has low
\(F_1\)-score on Table-BRGM and ICDAR-2013 but significantly higher on PubTables.
In contrast, heuristic methods perform well on Word-typeset tables with borders (Table-BRGM),
but struggle on \LaTeX{}-typeset tables (Table-arXiv).
Among probabilistic models, XY achieves better performance on
Table-BRGM,
but does not maintain its lead over the other three datasets.
On our PubTables dataset, TATR obtains an almost perfect score because it has been pretrained
on PubTables-train, and our test set has similar inputs (in terms of layout or table style).
VGT always outperforms TATR and XY.
The Docling, MonkeyOCR and Mathpix methods, relying on distinct approaches (object detection and expert VLM for the former two)
share the best performance on our benchmark with almost perfect \(F_1^{\textit{bbox}}\)-scores
(Q1).
Since we noticed that MonkeyOCR outputs only perfect confidence scores, we decided to use \(\mathcal P^+=\mathcal P\),
which is why it is compared to other non-probabilistic models.

GOT generates unnecessary nested tables, and its table content is
misaligned with the ground truth,
resulting in poor \(F_1^{\textit{txt}}\)-scores;
its heuristics are unreliable.
TabPedia performs well on PubTables (it has been pre-trained on its train set),
detects most tables in Table-arXiv and ICDAR-2013 (high recall),
but generates many false positives (low precision).
We interrupted TabPedia experiments during the TSR phase due to timeout (TO), exceeding allocated GPU hours.
A full inference would have resulted in computational costs an order of
magnitude higher than the second most expensive model (\autoref{fig:tradeoff}).

On our two new datasets, Table-arXiv and Table-BRGM,
the precision of general VLMs falls significantly.
These datasets include negative samples, as they comprise real scientific documents
which do not contain tables on every page (see \autoref{tab:datasets}).
Since \emph{VLMs almost always generate a prediction for each page}, this results in high number of FP,
while their recall remains high (Q3).
Such behavior is not observable on a dataset like PubTables, which consists
solely of pages containing tables.

Depending on the layout and table styles, for example on Table-BRGM,
\emph{heuristic methods} such as PyMuPDF, Camelot \emph{compete with computer vision-based approaches
  and even outperform VLMs} (GPT, Gemini, TabPedia, GOT)
\emph{for a much lower computational cost} (\autoref{fig:tradeoff})%
\footnote{We chose to display the best \(F_1\)-score for probabilistic models,
  and set \(F_{1}^\text{Top}=F_1^\textit{bbox}\) for TabPedia, for illustrative purposes. These models are labels in \emph{italics}.} (Q2).
To estimate costs, we tracked CPU and NVIDIA A100 GPU time and
applied equivalent cloud computing rates from AWS\footnote{\url{https://aws.amazon.com/ec2/pricing/on-demand/}}.
The full inference cost study (CPU/GPU time and API pricing across methods) is reported in \autoref{app:cost}.
This enables a cost comparison with proprietary general VLM APIs.
Note that XY is faster than TATR because
it processes fewer potential table images during the TSR phase.
The Pareto front depends on the dataset, but it is mainly composed of
Grobid, PyMuPDF, XY, TATR, Docling, MonkeyOCR and Mathpix (Q4).

As expected, results obtained through \emph{bbox} TD (\autoref{tab:te-f1})
tend to be better than those with \emph{txt} TD:
detecting a table by its bounding box is easier than by its content because
alignment between ground truth and model outputs (\emph{e.g.} formula alignment) is a challenge in text-based evaluation.
However, for Qwen2.5-VL and GPT-4.1, the opposite is observed:
they manage to correctly extract tables, but output approximate table coordinates (Q3).

\begin{figure}[h]
  \centering
  \includegraphics[width=0.8\linewidth]{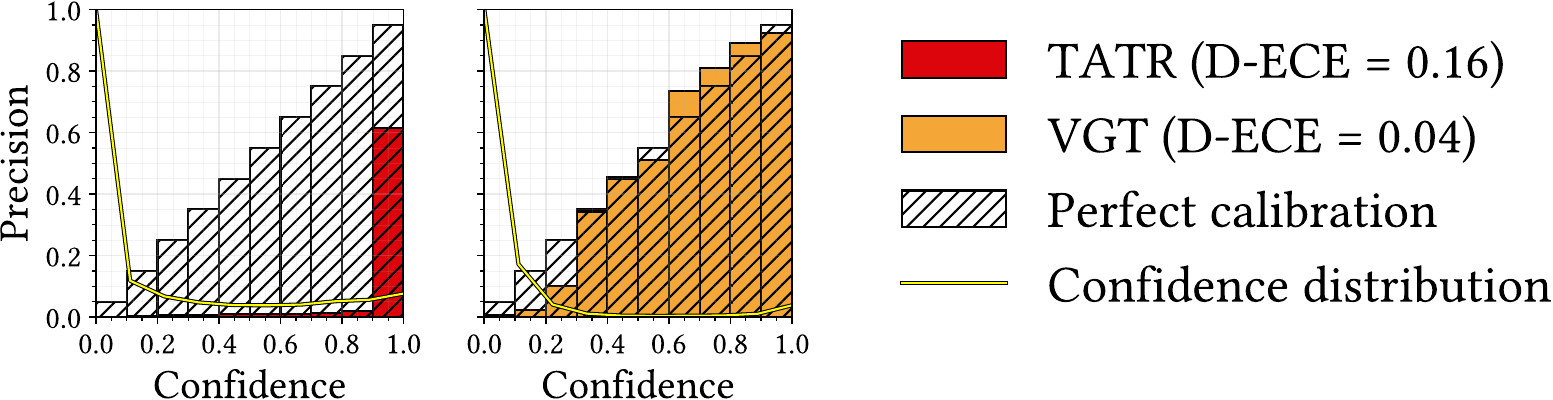}
  \caption{Reliability diagrams on Table-arXiv.}\label{fig:td-conf}
\end{figure}

Figure~\ref{fig:td-conf} illustrates miscalibration for probabilistic
models on Table-arXiv.
Reliability
diagrams~\cite{reliability-diagram} are histograms showing \emph{precision as a function
  of confidence}. %
For perfect calibration,
the histogram should match the identity function (the hatched histogram).
Thus, the gap between the colored and identity histograms
denotes a model's miscalibration (the lower, the better); D-ECE
(Section~\ref{sec:miscalibration}) measures this.
Figure~\ref{fig:td-conf} shows that TATR is a poorly
calibrated model (binary confidence scores make its confidence scores become meaningless) compared to VGT (Q5).
For example, among predictions with confidence scores between 80\% and 90\%, on average, less than 2\% are actual TP.
The poor calibration of XY+TATR (D-ECE=0.45) comes from data drift: the model (TATR-detect) was trained on full-page images but
applied to much smaller cropped page images. Raw TATR itself is already poorly calibrated, and this issue is compounded in XY+TATR.
Post-hoc recalibration is an option but is dataset-dependent and does not generalize reliably.
However, %
we found that TATR is calibrated on PubTables (D-ECE=0.03), which is close to its training set.

From these experiments, we draw the following lessons.

There are \emph{two paradigms for TE probabilistic models usage} in downstream tasks.
First, one can consider only positive predictions \(\mathcal{P}^+\),
which requires setting a confidence threshold.
Second, we can consider the whole set of detected tables \(\mathcal{P}\), with their attached
confidence scores. In the first case, TATR fits
perfectly, and
the \(\theta_c\) choice is easy
according to Figure~\ref{fig:td-conf} (for example \(\mathcal{P}^+=\mathcal{P}_{0.9}\)). Docling and MonkeyOCR also fit this case, without choosing any threshold.
In the second case, VGT ensures interpretable and reliable confidence scores (Q5).

\subsection{Table extraction performance} \label{sec:te-results}
Figure~\ref{fig:td-te} (right) illustrates Precision--Recall weighted by the TSR scores GriTS Topology
(see \autoref{sec:metrics-te}) from \emph{bbox} TD.
We compute \(P^{\rm Top}-R^{\rm Top}\) with \(\theta_J = 0.5\).
In \autoref{tab:te-ap} and \ref{tab:te-f1}  the ``TE metrics'' columns summarize
\(F_1\) and \(AP\) metrics for different TSR scores (denoted \fbox{\(\mathbf{AP}^\text{TSR}\)}, \fbox{\(\mathbf{F_1}^{\!\!\text{TSR}}\)}).
Though scores decrease for metrics with TSR, regardless of the TSR score used, the ranking remains almost the same across methods.

VLMs like GPT suffer from table content hallucination, with performance dropping from \(F_1^\textit{bbox}=0.35\) to \(F_1^\text{Con}=0.27\) on Table-BRGM (Q3).
GriTS Topology scores are closer to vanilla \(P\text{--}R\), which means that when a table is detected (with IoU \(\theta_J=0.5\)),
most of the time, the table topology (rows, columns, spanning cells) is correctly found. However, Precision and Recall
with TEDS are lower, because they take into account partial token match. We obtain similar results using GriTS Content.
TE scores fall for all models,
revising the initial positive assessment of task TD, e.g.~MonkeyOCR and Docling (Q1).

\subsection{Experiment conclusions}

{Method performance varies significantly depending on dataset style}, a dependency captured by our diverse benchmark datasets.
This variability arises from intrinsic design choices: either hard-coded features (for heuristic-based methods) or pretraining data (for probabilistic models).
  {Despite their scale, general VLMs are outperformed by smaller and cheaper, expert VLMs} (MonkeyOCR)
{and computer-vision models} (Docling, TATR) across datasets for TE.
Confidence scores do not always provide interpretable insights,
yet calibration matters in practice for usability:
VGT stands out positively in this regard, while TATR and XY+TATR do not.
For TSR, TCR remains challenging in terms of content accuracy.
Overall, while Docling, MonkeyOCR and Mathpix (the most promising TD models) demonstrate strong performance,
they have yet to achieve perfect TE (\emph{e.g.},~\(F_1^\text{TEDS}<90\%\)).
Our benchmark enables direct, end-to-end comparison of methods across heterogeneous data and from multiple perspectives
(accuracy, cost) at a glance.
\section{Conclusion}
We compared different methods for end-to-end table extraction from scientific PDF documents.
Our main conclusion is that table extraction \emph{remains a
  challenging problem}
(both table detection and table structure recognition).
Our new framework for TE allows better end-to-end evaluation, encompassing TD and TSR, over a wide variety of
model types on new datasets (Table-arXiv and Table-BRGM). This  provides novel resources for advancing TE research and development.

\begin{acks}
  This work was supported by Inria and BRGM via the GéolAug project,
  and was performed using HPC resources from GENCI--IDRIS (Grant 2025-AD010615780R1).
  We are grateful to the CLEPS infrastructure from Inria Paris for providing resources and support.
\end{acks}
\bibliographystyle{ACM-Reference-Format}
\bibliography{sample}
\ifversionExtented
  \appendix
  \onecolumn
  \section{Formal representation of an Extracted Table}\label{app:table-formalism}
We use the following formal representation of the structure of
an extracted table (see
Figure~\ref{fig:table-presentation} for illustration).
Given two natural numbers \(k < n\), we denote by \([k\!:\!n]\)
the set \(\{k, k+1, \ldots, n\}\), and write simply \([n]\)  for \([1\!:\!n]\).
Let \((n, m) \in \mathbb N^*\) be the shape of a table, i.e., its
number of rows and columns.
Each \emph{cell} is defined by the tuple \((i, j, r, c) \in \mathbb N^4\),
where \((i, j)\) is the location of the cell, and \((r+1, c+1)\) are the
number of rows and columns spanned by the cell.
We require \(i\in[n]\), \(i+r \in [n]\), \(j\in[m]\), \(j+c \in [m]\).
Any two distinct cells defined by \((i, j, r, c)\) and \((i', j', r',
c')\) are disjoint, i.e.,
\([i\!:\!i+r] \cap [i'\!:\!i'+r']=\varnothing\) and \([j\!:\!j+c] \cap
[j'\!:\!j'+c']=\varnothing\).
Since \(r, c \geq 0\), the location of a cell that spans
multiple rows or columns is defined by the location \((i, j)\),
of their top-left corner cells.
We call a cell \emph{simple} if \(r=c=0\), i.e., the cell does not
result from merging. A table has at most \(n\times m\) cells,
but there might be fewer, if some cells are merged (non-simple).

\begin{figure}[htbp]
  \begin{tabular}{|l|c|c|c|}
    \hline
    \((1,1)\) & \((1,2)\)                          & \((1,3)\)                            & \((1,4)\) \\ \hline
    \((2,1)\) & \multirow{2}[1]{*}{\((2,2), r=1\)} & \((2,3)\)                            & \((2,4)\) \\ \cline{0-0}\cline{3-4}
    \((3,1)\) &                                    & \multicolumn{2}{|c|}{\((3,3), c=1\)}             \\
    \hline
  \end{tabular}
  \Description[short]{The figure shows a table where each cell contains a tuple indicating its position.}
  \caption{Sample table where \((i,j)\) pairs indicate the location of
    each cell. When omitted, default values for $r$ and~$c$ are~$0$.} \label{fig:table-presentation}
\end{figure}

\section{Prior Datasets}
\label{app:datasets}
\autoref{tab:all-datasets} shows all prior relevant datasets we found in the literature, with their characteristics and statistics.
``E2E-TE'' (end-to-end TE) means that the dataset can be used for end-to-end TE evaluation, i.e., it contains TD and TSR annotations that are related.
``P. w/o Tab.?'' stands for whether the dataset contains
pages that do not contain any tables.
Private datasets are not included in the table.
\begin{table}[htbp]
  \caption{Summary of datasets in the literature vs.\ the ones used in
    our
    evaluation (*: datasets
    without dedicated test set)}
  \label{tab:all-datasets}
  \begin{tabular}{@{}l c cccc c rrr @{}}
    \toprule
    \multirow{2}[1]{*}{\textbf{Dataset}}     &
    \multirow{2}[1]{*}{\textbf{Text-PDF?}}   & \multirow{2}[1]{*}{\textbf{TD}}        & \multirow{2}[1]{*}{\textbf{TSR}} & \multirow{2}[1]{*}{\textbf{TCR}} & \multirow{2}[1]{*}{\textbf{E2E-TE}} &
    \multirow{2}[1]{*}{\textbf{P w/o Tab.?}} & \multicolumn{3}{c}{\# \textbf{Sample}}                                                                                                                                                                      \\
    \cmidrule(lr){8-10}
                                             &                                        &                                  &                                  &                                     &            &            &
    \textbf{Pages}                           & \textbf{Positives}                     & \textbf{Tables}                                                                                                                                                    \\
    \midrule
    \href{https://github.com/Academic-Hammer/SciTSR}{SciTSR}~\cite{chi2019complicated} %
                                             & \checkmark                             &                                  & \checkmark                       & \checkmark                          &            &            & 3\,716  & 3\,716  & 3\,716   \\
    ICDAR 2019 TRACK 2B*~\cite{ICDAR19} %
                                             &                                        & \checkmark                       & \checkmark                       &                                     & \checkmark &            & 250     & 250     & >250     \\
    TableBank-TD~\cite{li-etal-2020-tablebank} %
                                             &                                        & \checkmark                       &                                  &                                     &            &            & 8\,000  & >8\,000 & >8\,000  \\
    TableBank-TSR~\cite{li-etal-2020-tablebank} %
                                             &                                        &                                  & \checkmark                       &                                     &            &            & 5\,000  & 5\,000  & 5\,000   \\
    PubTabNet~\cite{zhong2020imagebasedtablerecognitiondata} %
                                             &                                        &                                  & \checkmark                       & \checkmark                          &            &            & 9\,138  & 9\,138  & 9\,138   \\
    TabLeX*~\cite{10.1007/978-3-030-86331-9_36} %
                                             &                                        &                                  & \checkmark                       & \checkmark                          &            &            & 1M      & 1M      & 1M       \\
    PubTables-Test
                                             &                                        & \checkmark                       & \checkmark                       &                                     &            &            & 57\,125 & 57\,125 & >57\,125 \\
    FinTabNet~\cite{9423317, smock2023aligning} %
                                             & \checkmark                             & \checkmark                       & \checkmark                       & \checkmark                          & \checkmark &            & 9\,289  & 9\,289  & 9\,289   \\
    CTE* (Pub)~\cite{gemelli2023ctedatasetcontextualizedtable}
                                             &                                        & \checkmark                       & \checkmark                       & \checkmark                          & \checkmark & \checkmark & 75\,000 & ?       & 35\,000  \\
    WTW~\cite{Long_2021_ICCV} %
                                             &                                        &                                  & \checkmark                       & \checkmark                          &            &            & 3\,611  & 3\,611  & 3\,611   \\
    SynthTabNet*~\cite{TableFormer2022} %
                                             &                                        &                                  & \checkmark                       & \checkmark                          &            &            & 14\,984 & 14\,984 & 14\,984  \\
    TNCR*~\cite{ABDALLAH2021} %
                                             &                                        & \checkmark                       &                                  &                                     &            &            & 6\,621  & 6\,621  & 9\,428   \\
    TabRecSet*~\cite{yang2023large} %
                                             &                                        & \checkmark                       & \checkmark                       & \checkmark                          & \checkmark &            & 32\,072 & 32\,072 & 38\,177  \\
    iFLYTAB*~\cite{zhang2023semv2} %
                                             &                                        &                                  & \checkmark                       &                                     &            &            & 17\,291 & 17\,291 & 17\,291  \\
    OmniDocBench*~\cite{11091844}
                                             &                                        & \checkmark                       & \checkmark                       & \checkmark                          & \checkmark & \checkmark & 1\,355  & 351     & 512      \\
    \midrule
    \textbf{PubTables-PDF}
                                             & \checkmark                             & \checkmark                       & \checkmark                       & \checkmark                          & \checkmark &            & 46\,942 & 46\,942 & 55\,990  \\
    \textbf{Table-arXiv}*
                                             & \checkmark                             & \checkmark                       & \checkmark                       & \checkmark                          & \checkmark & \checkmark & 36\,869 & 5\,214  & 6\,308   \\
    \textbf{Table-BRGM}*
                                             & \checkmark                             & \checkmark                       & \checkmark                       & \checkmark                          & \checkmark & \checkmark & 2\,003  & 256     & 361      \\
    \textbf{ICDAR 2013}*
                                             & \checkmark                             & \checkmark                       & \checkmark                       & \checkmark                          & \checkmark & \checkmark & 238     & 135     & 163      \\
    \bottomrule
  \end{tabular}
\end{table}

Note that we are looking for datasets suitable for end-to-end table
extraction (i.e., with ground truth for both TD and either TSR or TCR,
ideally both), that contain text-PDFs and not just images. Datasets that
include
pages without tables are preferable, as they better reflect performance in
practice.

FinTabNet~\cite{9423317} is a dataset containing complex tables from the annual reports of the S\&P 500 companies.
While FinTabNet meets our criteria (that is: text-based PDFs, with end-to-end TE annotations),
its first version suffered from table oversegmentation.
FinTabNet.c~\cite{smock2023aligning} fixed this problem, but the dataset is still noisy (\autoref{fig:FinTabNet}).
First, it considers tables of contents as tables, which does not fit
our setting (\autoref{fig:fin-e}).
Second, for pages with multiple tables, not all tables have been annotated, which may lead to FPs during evaluation
(Figures~\ref{fig:fin-a}, \ref{fig:fin-b}, \ref{fig:fin-c}, \ref{fig:fin-d}).
Finally, some bounding boxes were inaccurate, suffering from a shift in position
(as we illustrate in Figures~\ref{fig:fin-b}, \ref{fig:fin-c} and \ref{fig:fin-d}).

For the datasets used in the evaluation, \autoref{fig:count-dataset} shows the distribution of tables in the datasets according to their number of rows and columns.

\begin{figure}[htbp]
  \centering
  \begin{subfigure}[b]{0.18\textwidth}
    \centering
    \includegraphics[width=\linewidth]{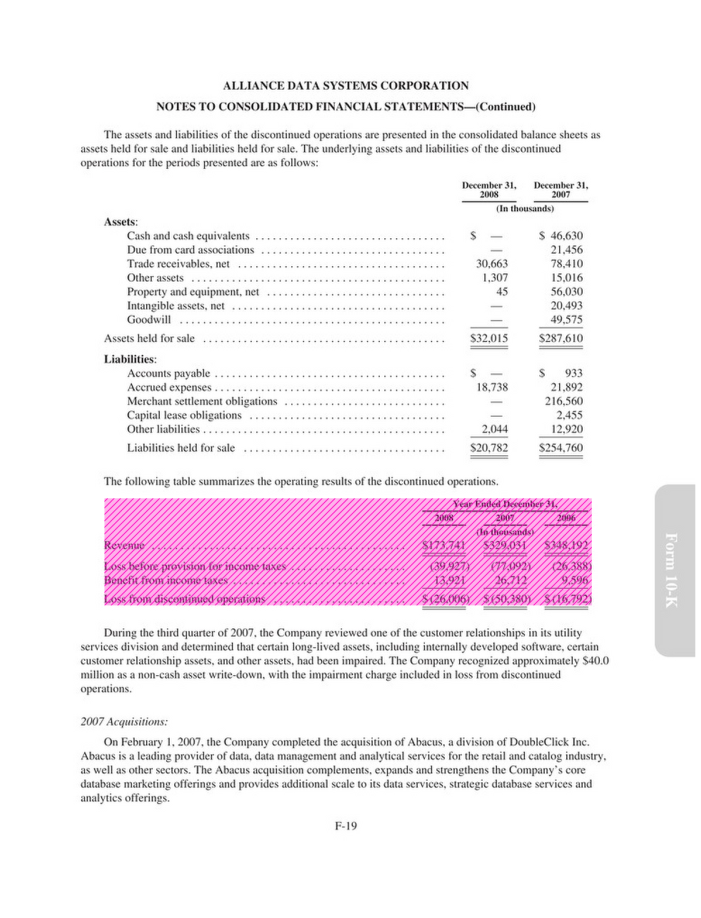}
    \caption{\texttt{ADS\_2008\_19.png}}\label{fig:fin-a}
  \end{subfigure}
  \hfill
  \begin{subfigure}[b]{0.18\textwidth}
    \centering
    \includegraphics[width=\linewidth]{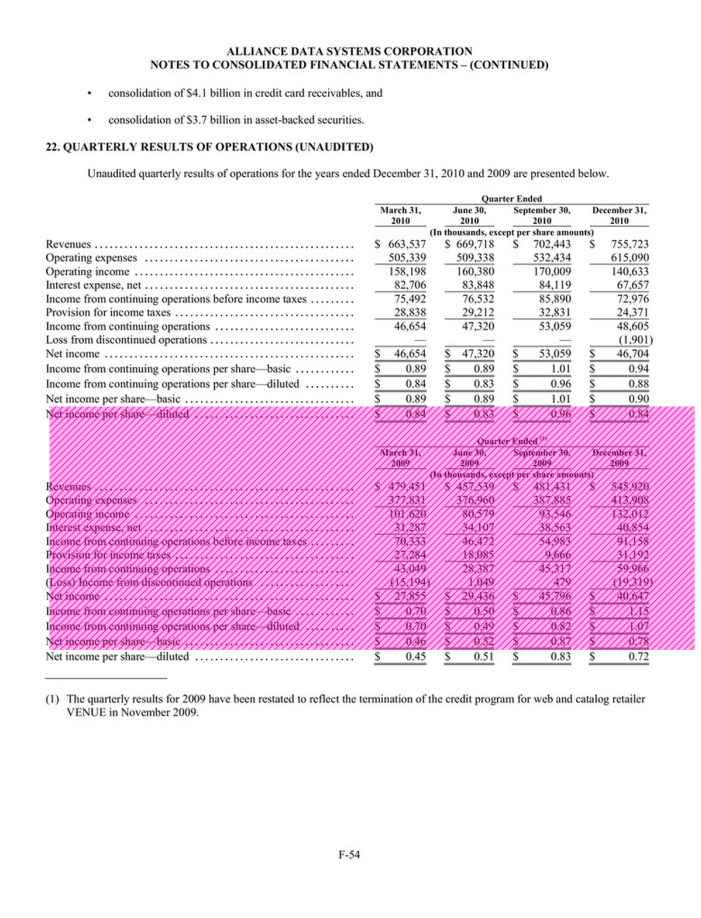}
    \caption{\texttt{ADS\_2010\_8.png}}\label{fig:fin-b}
  \end{subfigure}
  \hfill
  \begin{subfigure}[b]{0.18\textwidth}
    \centering
    \includegraphics[width=\linewidth]{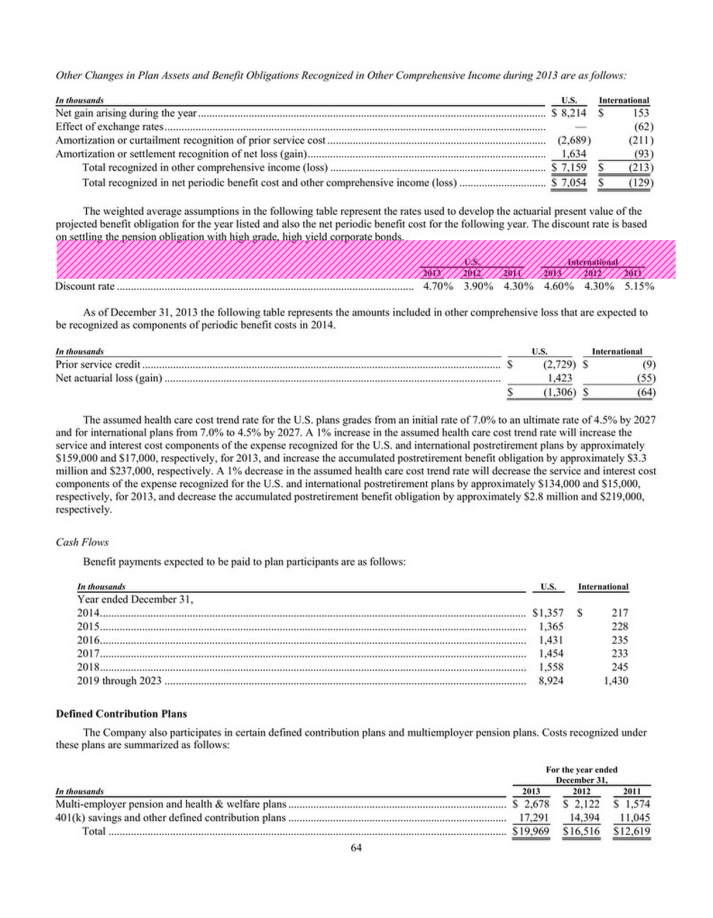}
    \caption{\texttt{AKAM\_2014\_7.png}}\label{fig:fin-c}
  \end{subfigure}
  \hfill
  \begin{subfigure}[b]{0.18\textwidth}
    \centering
    \includegraphics[width=\linewidth]{figure/fintabnet/ADS_2010_8.png}
    \caption{\texttt{ADS\_2010\_8.png}}\label{fig:fin-d}
  \end{subfigure}
  \hfill
  \begin{subfigure}[b]{0.18\textwidth}
    \centering
    \includegraphics[width=\linewidth]{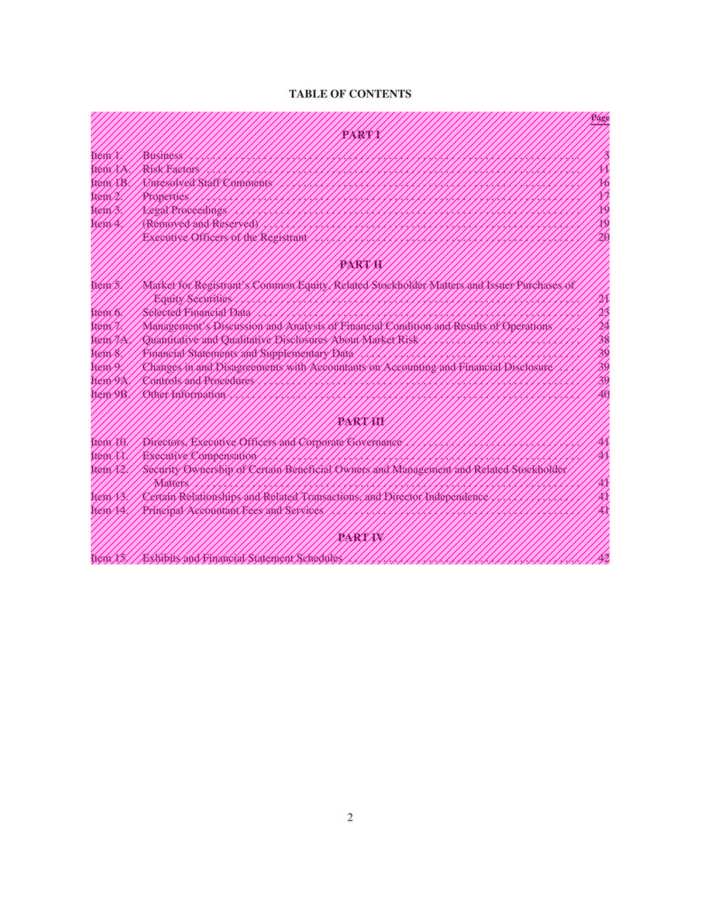}
    \caption{\texttt{WAB\_2010\_0.png}}\label{fig:fin-e}
  \end{subfigure}
  \caption{Sample misannotated tables from FinTabNet.c}\label{fig:FinTabNet}
\end{figure}

\begin{figure}[htbp]
  \centering
  \includegraphics[width=\linewidth]{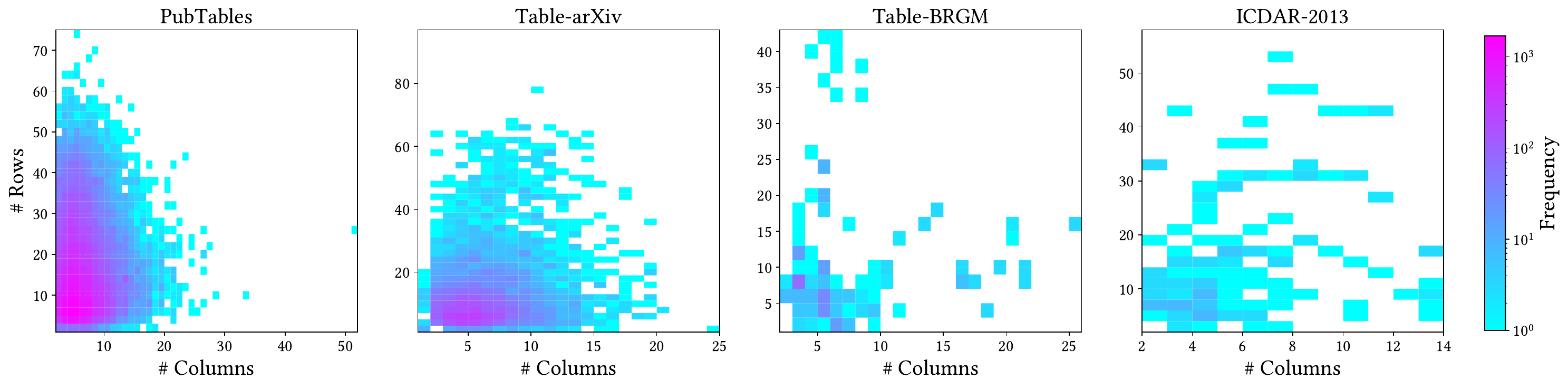}
  \caption{Distribution of the numbers of rows and columns in our datasets}\label{fig:count-dataset}
\end{figure}

\section{Model Details and Settings}\label{sec:setting}
Some specialized VLMs were excluded from our study because they do not
support end-to-end table extraction (TE). These include:
LayoutLLM~\cite{luo2024layoutllm} TD only;
Table-LLaVA~\cite{zheng2024multimodal} which requires table images as input, not full pages.
We selected MonkeyOCR for its top performance on OmniDocBench, including TSR.
According to the original paper, TabPedia achieved performance comparable to TATR in both TD and TSR.

We describe the settings used for the models under test.
The datasets used are described in \autoref{sec:datasets}.
Our code is available
\href{https://gitlab.inria.fr/msoric/table-extraction-benchmark}{on our
  GitLab repository}; we implemented our pipelines using~\cite{smock2021tabletransformer} as a starting point.
Our benchmark measures method performance using the previously described metrics in \autoref{sec:metrics}.

\subsection{Baseline Models}
We used \PdfPlumber{} and \Pymupdf{} with the default configuration. We used \Camelot{}
with the \verb|Lattice| method, which relies on demarcated lines between cells and outperformed other methods.
However, this method is limited, as not all tables include visible lines.

We used \Grobid's default configuration (\autoref{sec:grobid}), which is a feature-engineered ML approach with a cascading of
linear chains of Conditional Random Fields (CRF). The command \verb|processFullText| extracts and structures in \href{https://tei-c.org/}{TEI} (an XML dialect) the full text
of PDF files. From the TEI output, we retrieve table's locations (page numbers) and coordinates (in points) and convert them
into pixels with respect to GT image sizes for comparison.
We encountered problems in table locations during \emph{bbox}
TD evaluation: some extracted tables were positioned on the wrong page. Since page location is used for matching GT tables during
evaluation, this led to extra (undue) FNs. To avoid that, we tried to split multi-page PDF files into single-page PDF files and input them one by one.
However, this degraded the results, as some previously detected tables were now missed.  This can be explained by the Grobid training: the model expects the first page to be a front cover with titles, authors, etc.; feeding it separate pages degrades performance.
Thus, we kept the whole PDF as input despite the above problem.

\subsection{Computer Vision based methods}
We provide in this section some details about methods and pipelines used.

For \Docling, we use the default \verb|convert| method on PDF files,
to retrieve extracted tables (location and HTML).

For object detection-based methods, we use TATR-e, XY+TATR-e and VGT+TATR-s, respectively denoted as \TATRs{}, \XYs{} and \VGTs{} in
figures. Their pipelines are respectively described in
Figures~\ref{fig:pip-tatr}, \ref{fig:pip-xy} and
\ref{fig:pip-vgt}. All rely on TATR-s for TSR; we use PDFAlto to retrieve text content (token) from the initial PDF.
Regarding the parameters, we set \(p_\mathrm{px,s}=100\) pixels, which determines how much padding in pixels will be added around a detected
table before outputting a cropped image of the table. For XY+TATR-e, cropped images, which are XY-cut outputs, are padded with
\(p_\mathrm{px,d}=10\) pixels before being fed to TATR-d. These hyperparameters have been adjusted based on TATR-s.

In \autoref{fig:pip-xy},
we pre-process a page as an image with
the XY-cut algorithm that creates sub-images. TATR-detect takes these image chunks one by one in
input and predicts table bounding boxes with confidence scores. Only the top 2 predictions (with the
highest confidence scores) are kept (not shown in the figure).
Then, the initial image is cropped with the bounding boxes predicted (adding the XY-cut offset) to get cropped tables images that are individually sent
to TATR-structure, with tokens content and location from the new input.
Finally, we add the output of XY+TATR-detect to obtain the final output
of \XY{}, the end-to-end model.

In \autoref{fig:pip-vgt}, VGT takes a page as an image and a grid (obtained through PDF parser, tokenizer and embedding model) as input and predicts table
bounding boxes with confidence scores. Post-processing is not shown.
Then, the initial image is cropped with the bounding boxes predicted to get cropped tables images that are individually sent
into TATR-structure with tokens content and location from the new input.
Finally, we add the output of VGT to obtain the final output of \VGT, the end-to-end model.
\begin{figure}[htbp]
  \centering
  \begin{subfigure}[b]{.3\linewidth}
    \includegraphics[width=\linewidth]{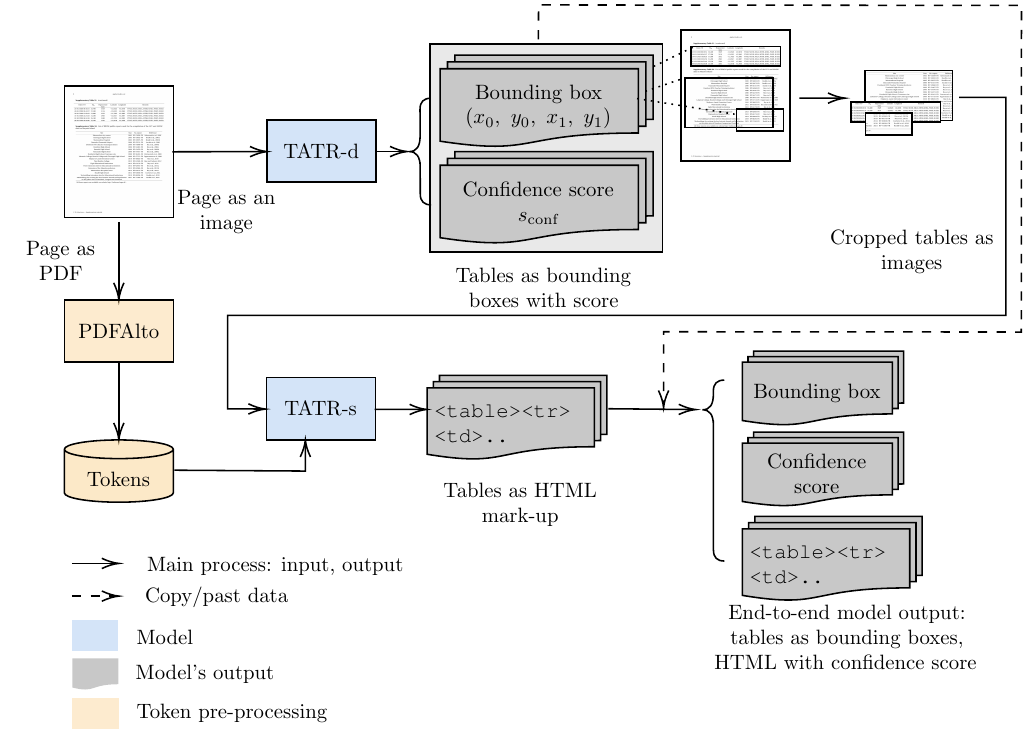}
    \caption{TATR-extract pipeline} \label{fig:pip-tatr}
  \end{subfigure}
  \hfill
  \begin{subfigure}[b]{.3\linewidth}
    \includegraphics[width=\linewidth]{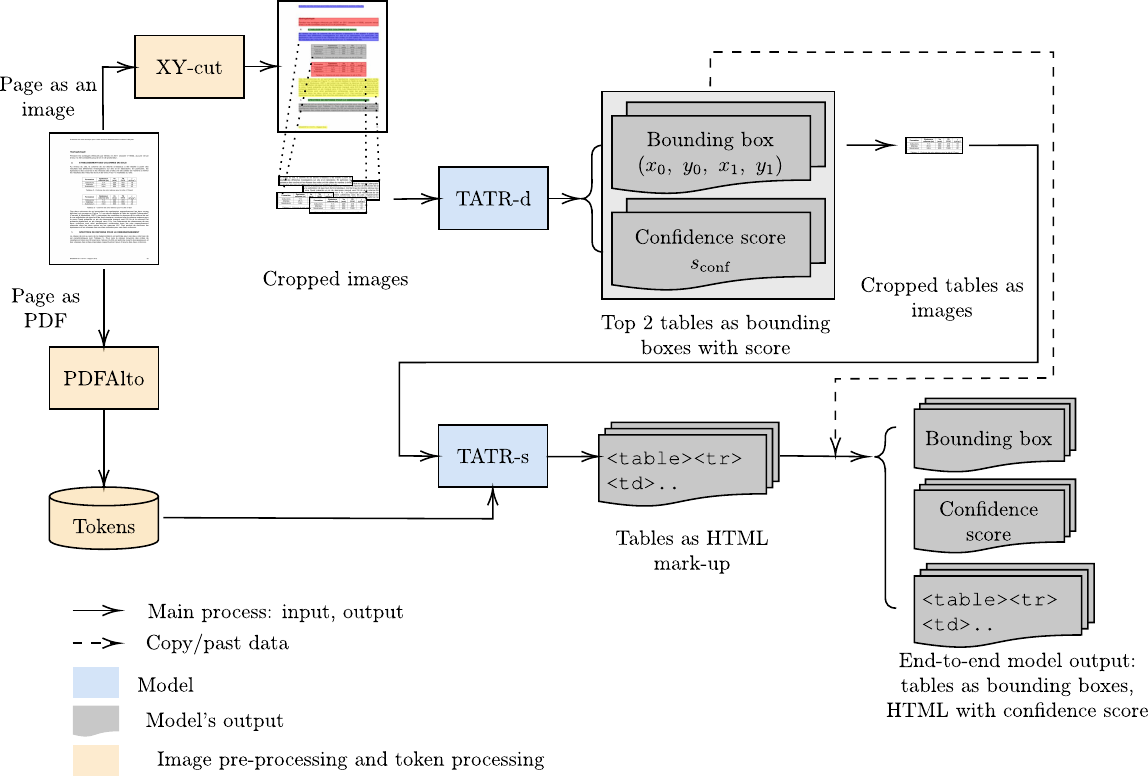}
    \caption{XY+TATR-extract pipeline} \label{fig:pip-xy}
  \end{subfigure}
  \hfill
  \begin{subfigure}[b]{.3\linewidth}
    \includegraphics[width=\linewidth]{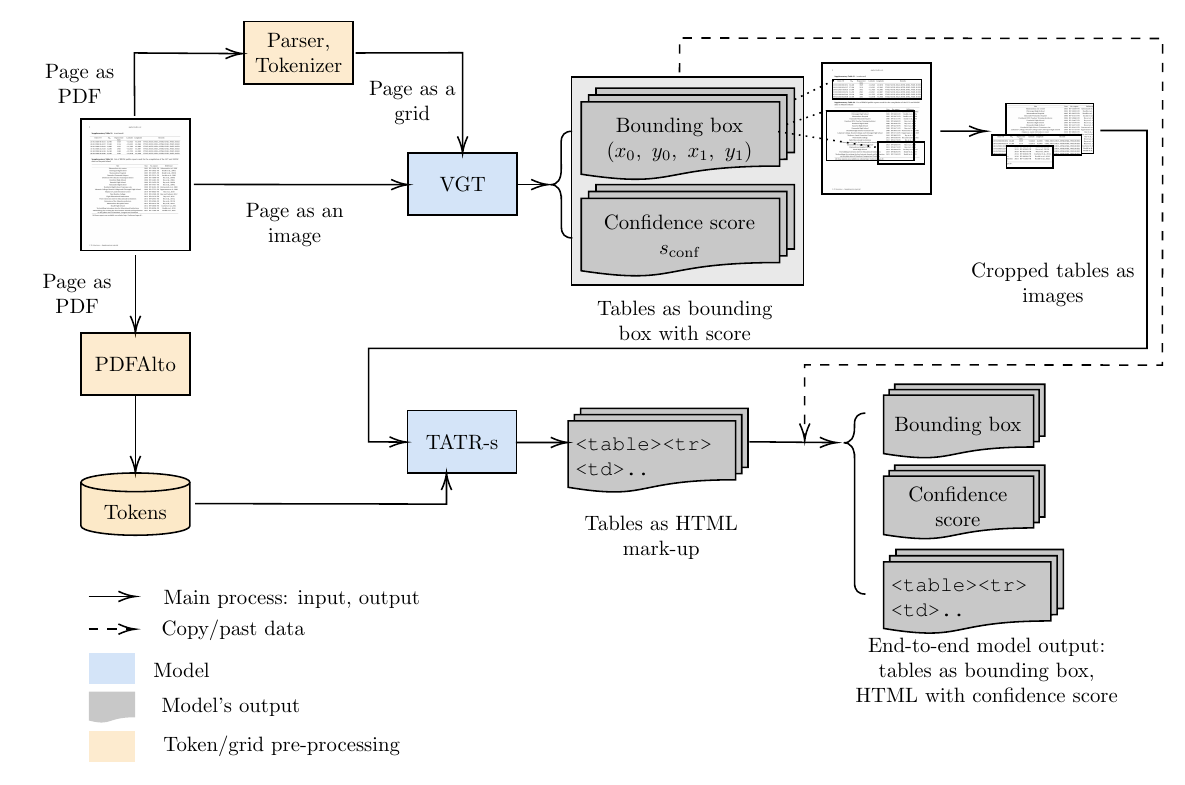}
    \caption{VGT+TATR-structure pipeline} \label{fig:pip-vgt}
  \end{subfigure}
  \caption{Pipeline for table extraction}
\end{figure}

We use VGT fine-tuned on the PubLayNet benchmark, DiT-base weights for ViT, BERT tokenizer with
LayoutLM~\cite{xu2019layoutlm} for word embeddings grid model to obtain the same pretrained model from the VGT repository~\cite{da2023visiongridtransformerdocument}.
We used the pipeline of~\cite{da2023visiongridtransformerdocument}, including the PDF parser, to obtain a grid input (tokens' position and content on pages).
For \emph{bbox} TD evaluation, we use the pixel coordinates of images.

\subsubsection{\XY{} motivation}
Let us analyze \emph{how} TATR-detect ``sees'', and \emph{where} it looks for tables.
The bounding boxes used for prediction come from the transformer decoder output.
Object queries
(decoder input) are positional embeddings learned during training, which contain abstract information about spatial
locations, or ``where should the model look?''
Thus, the
decoder performs cross attention between object queries, and the
encoder outputs 15 embeddings.
These are fed into prediction heads that output bounding boxes with associated classes. Thus, each output, which is directly associated with an object
query, is roughly specialized in an area (Figures~\ref{fig:xy-black} and~\ref{fig:xy-white}).

As queries pay attention to different regions, we provide image chunks
(sub images) produced by XY-cut instead of whole pages
as input for TATR-detect (see Figures \ref{fig:xy-cut-whole}, \ref{fig:xy-cut-top} and \ref{fig:xy-cut-bot} for example). We cut a page into two chunks and give each of them to the model.
We also experimented with keeping the initial page shape (by adding padding to the image chunk), but results
without padding were more promising.

\begin{figure}[htbp]
  \centering
  \begin{tabular}{llcc}
     &
    \begin{subfigure}[t]{0.22\textwidth}
      \frame{\includegraphics[width=\linewidth]{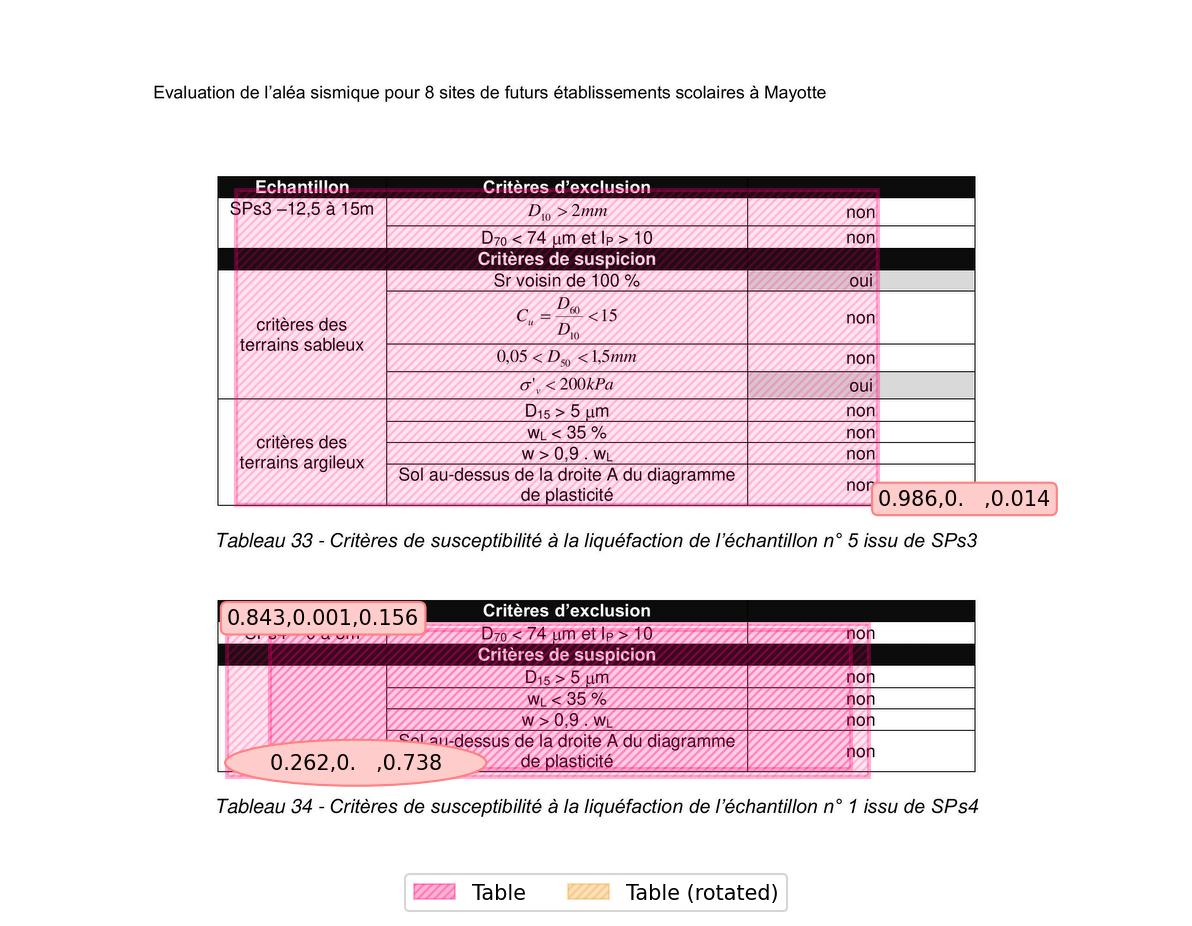}}
      \setcounter{subfigure}{1}
      \caption{Top chunk as input} \label{fig:xy-cut-top}
    \end{subfigure}
     &
    \multicolumn{2}{c}{
      \begin{subfigure}[t]{0.5\textwidth}
        \includegraphics[width=\linewidth]{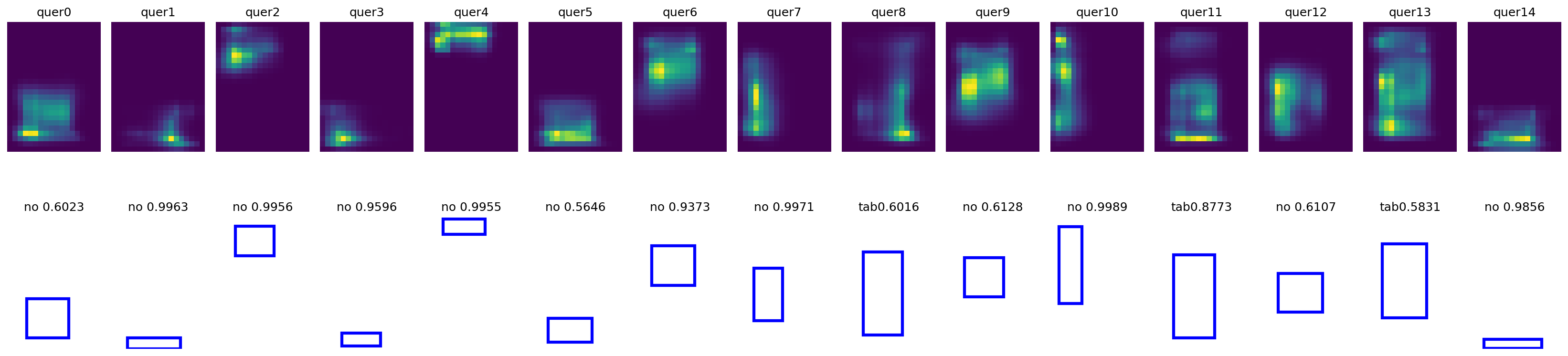}
        \setcounter{subfigure}{3}
        \caption{White image as input}\label{fig:xy-white}
      \end{subfigure}
    }
    \\[0.5em]
    \raisebox{0pt}[0pt][0pt]{
      \begin{subfigure}[t]{0.22\textwidth}
        \frame{\includegraphics[width=\linewidth]{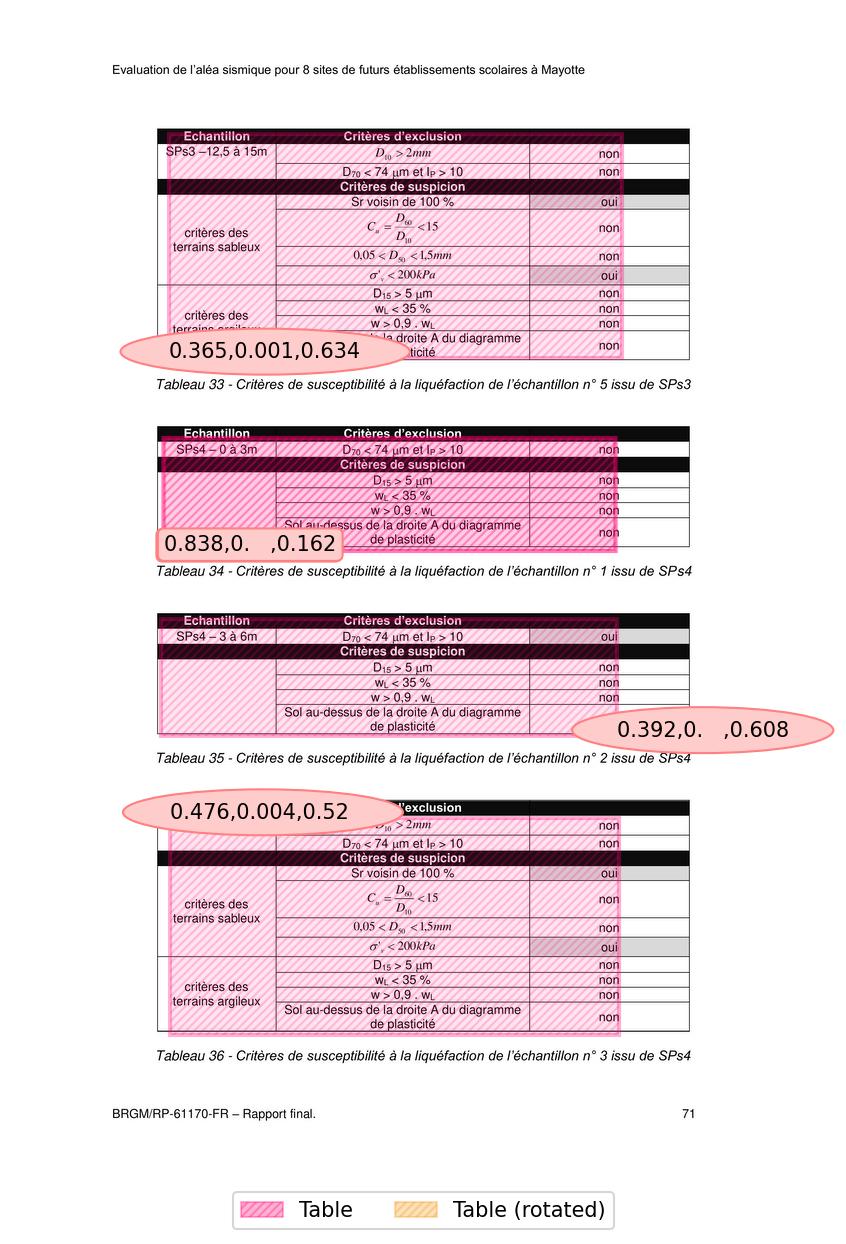}}
        \setcounter{subfigure}{0}
        \caption{Whole page as input} \label{fig:xy-cut-whole}
      \end{subfigure}}
     &
    \begin{subfigure}[t]{0.22\textwidth}
      \frame{\includegraphics[width=\linewidth]{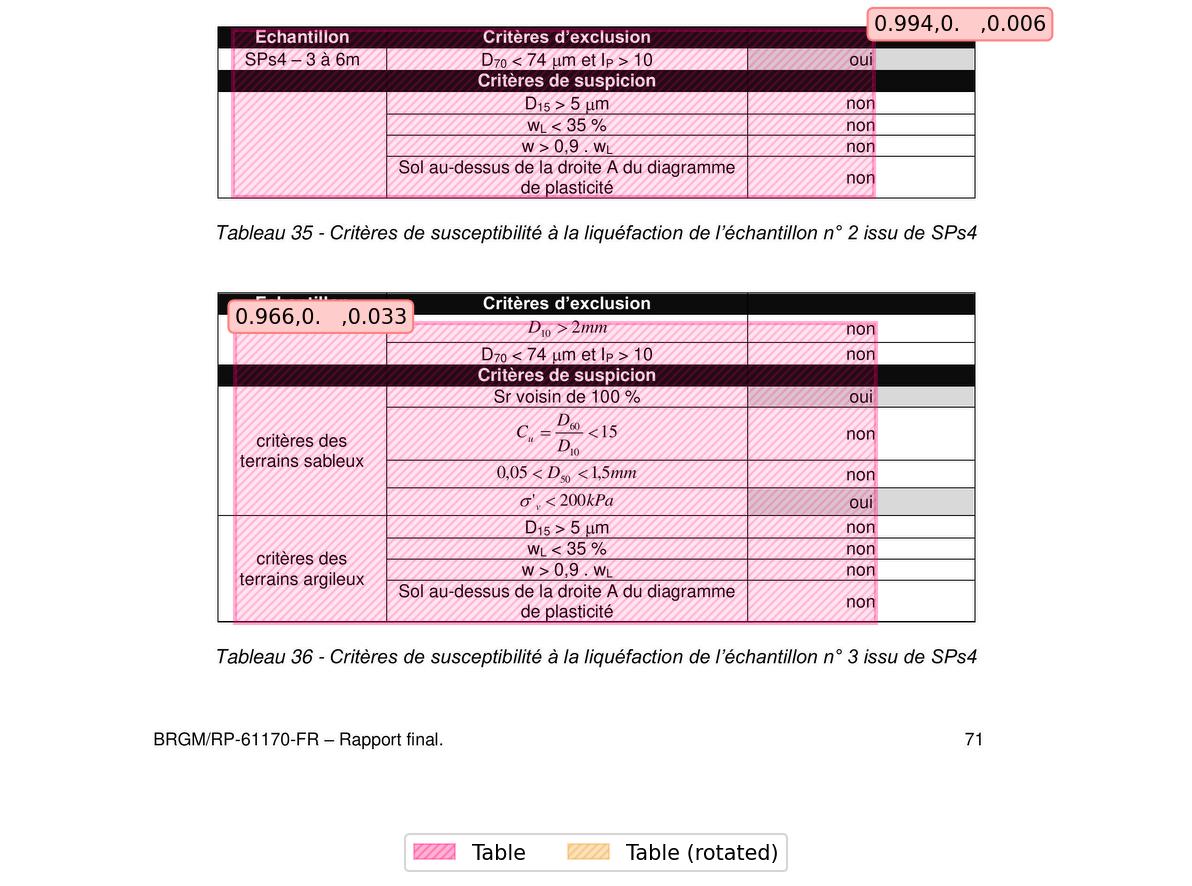}}
      \setcounter{subfigure}{2}
      \caption{Bottom chunk as input} \label{fig:xy-cut-bot}
    \end{subfigure}
     &
    \multicolumn{2}{c}{
      \begin{subfigure}[t]{0.5\textwidth}
        \includegraphics[width=\linewidth]{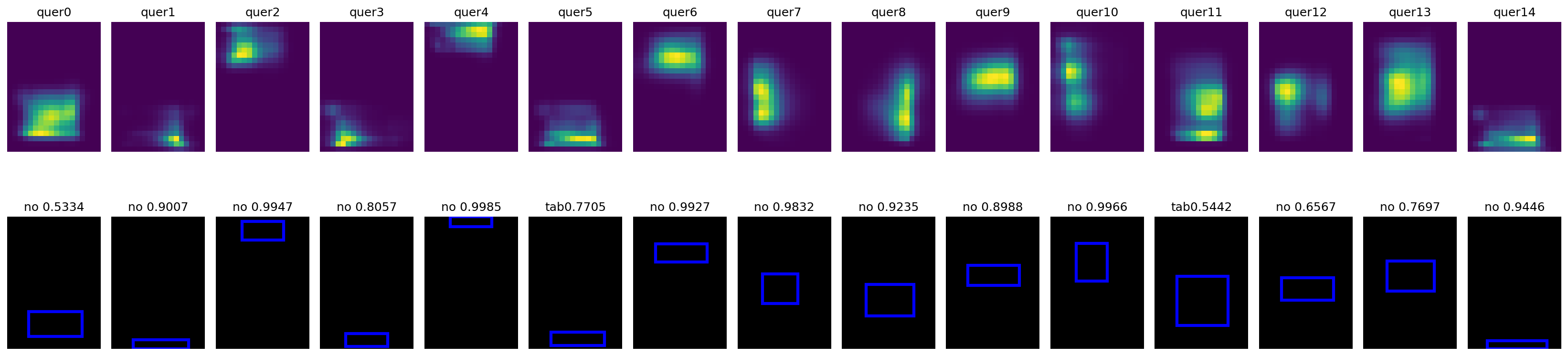}
        \setcounter{subfigure}{4}
        \caption{Black image as input} \label{fig:xy-black}
      \end{subfigure}
    }
  \end{tabular}

  \caption{Left part (\ref{fig:xy-cut-whole}, \ref{fig:xy-cut-top}, \ref{fig:xy-cut-bot}) shows how TATR's confidence scores change.
    Confidence scores are ordered following the class ``table'', ``rotated table'' and ``no object''.
    If the \fbox{dominant class is ``tables''}, the distribution is surrounded by a frame, otherwise (``no object'' dominant) it is an ellipse.
    Right part (\ref{fig:xy-white}, \ref{fig:xy-black}) shows query location.}
  \label{fig:xy}
\end{figure}

Notice that for the first table ``Tableau 33'',
TATR-detect confidence scores are: $c_\mathrm{table}=0.37,c_\mathrm{no  \, object}=0.63$.
With the regular \TATR{} pipeline (\autoref{fig:pip-tatr}), this table would not have been detected.

In Figures \ref{fig:xy-black} and \ref{fig:xy-white}, above each ``query \(i\)'', we plot the activation map of attention weights from the last transformer decoder layer.
This corresponds to visualizing, for each detected object,
which part of the image the model was looking at to predict this specific bounding box and class.
Under each heatmap, we visualize the output: bounding box with dominant class (``tab'' or ``no'')
and its confidence score. We can roughly count 5 queries focused on the page bottom, 2 on the top, and 8 in the middle.

\subsection{General VLMs}\label{app:vlm-prompt}
\begin{tcolorbox}[float, colback=blue!5!white,colframe=blue!75!black,title=Sample prompt used for general VLMs.]
  \begin{verbatim}
You are an expert table extraction and HTML formatting engine.
Task: Analyze the provided image to identify all tables.
For each table found, extract the data as a single, valid, and complete HTML table.
If no tables are present, output empty lists for both keys.
Output Format must be a single JSON object, with keys:
 - `html`: list[str] (one string per extracted table),
 - `bbox`: list[list[int, int, int, int]] (one bounding box list per extracted table)
The lists for `html` and `bbox` MUST have the same length.
'html' elements: Each string must be the complete, valid HTML table code, starting with `<table>` and ending
with `</table>`.
'bbox' elements: Each inner list must be the bounding box
`[x_top_left, y_top_left, x_bottom_right, y_bottom_right]` for the corresponding table in the 'html' list,
with coordinates normalized to the 0-1000 range.
If no tables are found, output the JSON: `{"html": [], "bbox": []}`.
EACH string value inside the `html` list MUST NOT contain any newline characters (`\n` or similar).
It must be a single continuous line of text.
The output HTML must be perfectly rectangular: every row must have the same column count.
Use `rowspan` and `colspan` only when necessary to achieve correct structure and alignment.
All HTML special characters within table cells (e.g., in `<td>` or `<th>` content)
**MUST** be properly escaped: `<` must be `&lt;`, `>` must be `&gt;`, `&` must be `&amp;`.
\end{verbatim}
\end{tcolorbox}

For \gemini, \gpt{} and \qwen{} we used zero-shot learning, with the
prompt shown above.
Note that all images are standardized to have a height of 1\,000 pixels.

\subsection{Expert VLMs}
For \pedia{}, \got{} and \monkey{}, we used the default configuration and pipelines provided by the authors.

\subsection{Commercial tool}
For \mathpix{}, we used the official \texttt{/v3/pdf} API endpoint with its
default configuration, additionally enabling HTML conversion and table
extraction (\texttt{include\_table\_html} and \texttt{enable\_tables\_fallback}).
Each document is submitted directly as a PDF; unlike the VLMs, Mathpix ingests
the PDF natively. We read the predicted
table bounding boxes and their content from the returned
output, converting the \LaTeX{} \texttt{tabular} strings to HTML for evaluation.
Mathpix returns two confidence scores, \texttt{confidence} and
\texttt{confidence\_rate}, but both mix structure and content, as they are
computed over all tokens of the output \LaTeX{} table (markup and cell text
alike). Since this does not reflect table-detection confidence, as for
MonkeyOCR, we treat Mathpix as a non-probabilistic model.

\section{Metrics}\label{app:metrics}
In this section, we present new metrics that can be used for TD, which generalize the commonly used metrics based on a fixed IoU threshold~\(\theta_J\).

\paragraph{Probabilistic metrics}
The value \(\theta_J=0.5\) is a natural choice, because if we consider non-overlapping predictions, %
we can get only one prediction with IoU greater than 50\% (TP) per ground truth.
However,  different \(\theta_J\) values are also sometimes used.  For instance, the ICDAR 2019 competition cTDaR~\cite{ICDAR19} computes \(F_1\)-scores
with \(\theta_J \in \{0.6, 0.7, 0.8, 0.9\}\); then, they compute
\({\rm WAvg}(F_1)\), which is a weighted average of \(F_1\)-scores for these four $\theta_J$ values, with weights proportional to their thresholds.
Thus, the \(F_1\)-scores based on higher thresholds are given more importance. %

Let us discuss the mathematical basis for choosing such a set of \(\theta_J\) values.
A general, principled way to compute a threshold-dependent metric \(X_\theta\), such as Precision, Recall or \(F_1\)-score,
is an \emph{integral} over the interval comprising all possible $\theta_J$ values.
\({\rm WAvg}(X)\)
can be seen  as a \emph{discrete approximation} of such an integral. %
Further, the weighted average can be interpreted as the \emph{expected value of the random variable (RV) \(X_\theta\)}
according to a \emph{piecewise linear probability density function} (\emph{pdf}, in short) \(f_s\),
where \(s\) is a parameter between \(0\) and \(1\) and \(f_s: x \mapsto \alpha_s \cdot x \cdot \mathbf{1}_{[s,1]}(x)\).
Here, \(\mathbf{1}_{[s,1]}\) is the \emph{indicator} function whose value is \(1\) on \([s,1]\) and  \(0\) everywhere else,
and \(\alpha_s > 0\) is a normalization factor to ensure \(f_s\) is a \emph{pdf}.

Now, let us view  \(\theta_J\) as a real RV with values in \([0,1]\).
As \emph{pdf}, we chose two different functions: \(f_{0}\) the simplest and most naive version, and \(f_{0.5}\) that generalizes \({\rm WAvg}\).
After some normalization, we get \(f_0(\theta)=2\cdot \theta \cdot \mathbf 1_{[0,1]}(\theta)\) and
\(f_{0.5}(\theta)=\frac{8}{3} \cdot \theta \cdot \mathbf 1_{[0.5,1]}(\theta)\).
For a random variable \(\theta_J\) distributed according to \emph{pdf} \(f\), we write \(\theta_J \sim f\).

\begin{definition}[Expected Precision and Recall] \label{def:excepted-metrics}
  Let \fbox{\(P_{\theta_J},R_{\theta_J},F_{1,\theta_J}\)} be the %
  Precision, Recall and \(F_1\)-score obtained with the random variable threshold \(\theta_J\). Under the assumption that \(\theta_J\) is sampled according to the probability distribution function \(f\), we have:
  \[
    \mathbb E_{\theta_J\sim f} [P_{\theta_J}]= \frac{1}{\lvert \mathcal{P}^+ \rvert}\sum_{i\in \mathcal{P}^+}\mathbb E_{\theta_J\sim f} [ \mathbf{1}_{[J_{i}>\theta_J]}]
  \]\vspace{-.8em}\[
    \mathbb E_{\theta_J\sim f} [R_{\theta_J}] = \frac{1}{\lvert \mathcal{G} \rvert}\sum_{i\in \mathcal{P}^+}\mathbb E_{\theta_J\sim f} [ \mathbf{1}_{[J_{i}>\theta_J]}]
  \]
  \noindent where \(\lvert \mathcal{G} \rvert\) is the number of ground truth tables and
  \(J_i\) the IoU for the positive sample \(i\).
\end{definition}

From now on, we use \fbox{\(\mathbb E_s [X]\)} as shorthand for \(\mathbb E_{\theta_J \sim f_s} [X_{\theta_J}]\), for some \(s \in [0,1]\) and metric \(X_{\theta_J}\in \{P_{\theta_J}, R_{\theta_J}, F_{1,\theta_J}\}\).
Note that \(\mathbb E_s [X]\) becomes stricter as \(s\) increases.

\paragraph{Model confidence and (mis)calibration}
To compute D-ECE efficiently,
\eqref{eq:ece-th} approximates (discretizes) \eqref{eq:ece} by partitioning predictions into \(M\) equal %
bins, and taking a weighted average of the bins' (Precision -- Confidence) gap.  \(\lvert \mathcal P \rvert\) is the number of predictions, and \(B_m\) is
the set of indices of predictions whose confidence falls between \(\frac{m-1}{M}\) and \(\frac{m}{M}\).
The \({\rm prec}\) and the \({\rm conf}\) functions compute the mean precision and confidence, respectively:
\begin{equation} \label{eq:ece}
  \operatorname{D-ECE} = \mathbb E_{\widehat{P}} \left[
    \left\lvert \mathbb P\left[\widehat Y=Y=1 \,\middle|\, \widehat P = p\right]-p \right\rvert \right].
\end{equation}
\begin{equation}
  \label{eq:ece-th}
  \operatorname{D-ECE} \approx \sum_{m=1}^M \frac{\lvert B_m \rvert}{\lvert \mathcal P \rvert} \left\lvert {\rm prec}(B_m)- {\rm conf}(B_m) \right\rvert.
\end{equation}
We compute this score over \(\mathcal P\) and take \(M=10\) to obtain an overall calibration error.

\section{Methodology}\label{app:methodology}
\begin{figure}[htbp]
  \begin{tikzpicture}[
      edge from parent/.style={draw, -latex},
      edge label/.style={pos=0.1, sloped, above, font=\small},
      level 1/.style={sibling distance=4cm, level distance=1cm},
      level 2/.style={sibling distance=3cm, level distance=1.5cm}
    ]

    \node {\(\mathcal P\)}
    child {node {\(\mathcal P \supseteq \mathcal P^+ \in \left\{\mathcal P_{\theta_c}\right\}_{\theta_c}\)}
    child {node {\(\mathcal P_\textit{bbox}^{++}\)}
        edge from parent
        node [edge label, above left] {Bbox}
      }
    child {node {\(\mathcal P_\textit{txt}^{++}\)}
        edge from parent
        node [edge label, above right] {Text}
      }
    edge from parent
    node [edge label, above left] {Conf. score}
    }
    child {node {\(\mathcal P^+=\mathcal P\)}
        child {node {\(\mathcal P_\textit{bbox}^{++}\)}
            edge from parent
            node [edge label, above left] {Bbox}
          }
        child {node {\(\mathcal P_\textit{txt}^{++}\)}
            edge from parent
            node [edge label, above right] {Text}
          }
        edge from parent
        node [edge label, above right] {No conf. score}
      };
  \end{tikzpicture}
  \Description[short]{The table compiles the following information for each dataset (PubTables, Table-arXiv, Table-BRGM):
    topic, language, type, and counts (number of PDFs, pages, positive samples, and tables).}
  \caption{TSR inputs depending on the TD model type (with or without a confidence score), and
    evaluation type (\emph{bbox}-based or \emph{txt}-based).}
  \label{fig:prediction-set}
\end{figure}

\paragraph{Text Table Detection}%
For \textbf{methods that output table locations} (bounding boxes), we can use the metrics
presented in Section~\ref{sec:metrics-td}, based on IoU.
We denote the output of such a method by \fbox{\(\mathcal
  P^{++}_\textit{bbox}\)}, whose elements are tuples that include bounding boxes,
HTML table representation and, if available,  confidence scores.
For models which \textbf{do not output bounding boxes} (recall \autoref{tab:all-model}),
based on the \emph{text}, we obtain  \fbox{\(\mathcal P^{++}_\textit{txt}\)}
whose elements are tuples that include HTML table representation
and, if available, confidence scores.
Accordingly, we call ``\emph{bbox} TD''
and ``\emph{txt} TD'' the TD evaluations based on
bounding boxes, respectively,  table content. After the TD evaluation,
TSR evaluation  inputs elements from either  \(\mathcal
P^{++}_\textit{bbox}\) or  \(\mathcal P^{++}_\textit{txt}\).
\autoref{fig:prediction-set} depicts an overview of possible paths for evaluation.
Probabilistic models offer different ways to evaluate their performance.

To evaluate TD results from general VLMs and GOT, we characterize tables by their \emph{text} (ignoring the table
structure tags), by building from each table a \emph{multiset of 2-grams} from the content of present cells.
We ignore table structure tags here because they simply reflect HTML syntax, thus their similarity is not informative; we want to measure how much cell
\emph{content} is shared by the prediction and GT.
Given the multisets obtained from the prediction  $(S_P)$, respectively, from the GT $(S_{GT})$,
we define:
\(
\text{Jaccard}(S_P, S_{GT})= \frac{|S_P \Cap S_{GT} |}{|S_P \Cup S_{GT}|}
\)
where \(| \cdot |\) denotes the cardinality of a multiset,
\(\Cup\) is the multiset union (retaining each multiset element with its maximum multiplicity), and \(\Cap\)
the multiset intersection (based on minimum multiplicity).
We use \emph{multisets} rather than regular sets to better characterize cell content, which may contain some repetitions. %
We use \emph{character-level} $n$-grams instead of the more common word-level
$n$-grams since the former better reflect character-level errors, which
happen frequently with VLMs, due to hallucination or token extraction issues.
Because of frequent alignment issues for token extraction (through PDFAlto),
especially within mathematical
formulas, we do not treat a whitespace as a character.

From now on, we call \textbf{text-Jaccard}  the \emph{Jaccard index  over 2-gram-2-character multisets}.
Thus, for all models, we can define TP and FP using \emph{text-Jaccard}, which leads to \(\mathcal P^{++}_\textit{txt}\).
Here is an example: for a given HTML table, the associated \emph{2-gram-2-character multisets}:
\begin{verbatim}
<table>
  <tr>
    <td>Location</td>
    <td>Time</td>
    <td>Times</td>
    ...
</table>
\end{verbatim}
We define \(S_P\) as:
\[
  S_P =\left\{\!\!\left\{ (\verb|"Lo"|,\verb|"ca"|), (\verb|"ca"|,\verb|"ti"|), (\verb|"ti"|,\verb|"on"|), (\verb|"on"|,\verb|"Ti"|), \right.\right.  \left.\left.  (\verb|"Ti"|,\verb|"me"|), (\verb|"me"|,\verb|"Ti"|),\dots \right\}\!\!\right\}
\]

\subsection{Evaluating TSR Impacted by TD}
Traditionally, TD and TSR are evaluated independently:\\
\[
  \mathcal{X}_{\mathrm{TD}} \xmapsto{\text{TD Model}}
  \widehat{\mathcal{Y}}_{\mathrm{TD}} \qquad \qquad
  \mathcal{X}_{\mathrm{TSR}} \xmapsto{\text{TSR Model}}
  \widehat{\mathcal{Y}}_{\mathrm{TSR}}
\]
The TD test dataset consists of a set of  page images
\(\mathcal{X}_{\mathrm{TD}}\) and GT table location
annotations
$\mathcal{Y}_{\mathrm{TD}}$.
The TSR test dataset consists of a set of cropped table images
\(\mathcal{X}_{\mathrm{TSR}}\) and GT table structures
$\mathcal{Y}_{\mathrm{TSR}}$.

However, in downstream tasks, models are used sequentially to perform
end-to-end TE, %
thus TSR evaluation depends on TD:
\[
  {}\hfill
  \mathcal{X}_{\mathrm{TD}} \overbrace{\xmapsto{\text{TD Model}} \left.
  \mathcal{P}\right|_{\mathrm{TD}} \xmapsto{\text{TSR
      Model}}}^{\text{TE Model}} \mathcal{P}\hfill
\]
where \(\mathcal{P}\) contains the HTML tables, and optionally, the bounding box coordinates and the confidence score,
as shown in \autoref{tab:all-model}. Note that for one-step TE methods, we may not be able to separate (discern) the two sub-models.  %

Aiming for an end-to-end evaluation, we choose not to use a predefined dataset \(\mathcal{X}_{\mathrm{TSR}}\) as input for TSR.
Instead, we use the set of true positives \(\left.\mathcal{P}^{++}\right|_{\mathrm{TSR}} \subseteq \mathcal{P}^{+}\) obtained from
the TD inference part; we use \(\theta_J=0.5\) to distinguish TP from FP. In this way, the TSR score better reflects what can be observed in the use of the end-to-end model.
In other words, the evaluation pairs are: \( \left( \left.\mathcal{P}^+\right|_{\mathrm{TD}}, \mathcal{Y}_{\mathrm{TD}} \right) \) and
\( \left( \left.\mathcal{P}^{++}\right|_{\mathrm{TSR}}, \mathcal{Y}_{\mathrm{TSR}} \right) \). This has two consequences:
($i$)~\emph{TSR evaluation is not independent of TD}; we denote those scores \fbox{\(\left.\text{TSR}\right|_{\mathrm{TD}}\)};  %
($ii$)~We can \emph{no longer directly compare \(\left.\mathrm{TSR}\right|_{\mathrm{TD}}\) results between models},
since the input dataset \( \left.\mathcal{P}^{++}\right|_{\mathrm{TSR}}\) is different for each model (determined by the model's performance on
TD).
Consequently, if TD fails to correctly\footnote{For instance, a TP with IoU at 51\% may miss part of the table's content.}
detect tables, the structure model will be penalized. %
While this may seem ``unfair'' to TSR models, it reflects the reality of end-to-end TE result quality, which is measured on the final results.
Moreover, when using one-step (black-box) models, we cannot separate the detection part from the structure part.

Given a \({\rm TSR}\) metric,
for each TP \(i\), we compute a
score \(s_i^{\rm TSR}\). The final \(\left.\text{TSR}\right|_{\mathrm{TD}}\) score for each model is then obtained
by averaging the scores over the set of TP \(\mathcal{P}^{++}\):
\(
\left.\text{TSR}\right|_{\mathrm{TD}}=\frac{1}{\lvert \mathcal{P}^{++}
  \rvert}\sum_{i \in \mathcal{P}^{++}}s_i^{\rm TSR}.
\)

\subsection{\(\mathcal P_\textit{bbox}\) and \(\mathcal P_\textit{txt}\)} \label{sec:jaccard-bbox-content}
We use IoU as our main box-based metric. But some methods cannot be trusted
on bounding boxes alone: \got{} does not output them at all
(\autoref{tab:all-model}), while \gpt{} and \qwen{} output coordinates but
tend to hallucinate them (see \autoref{tab:te-f1} and \autoref{fig:gpt-hallu}).
For these cases, we use \emph{text}-Jaccard instead, so all models can still
be compared fairly.
\autoref{fig:jaccard-bbox-content-fig} checks how well this works: for each
sample, we plot IoU against \emph{text}-Jaccard, treat IoU (at
\(\theta_J=0.5\)) as ground truth, and treat \emph{text}-Jaccard as a
classifier trying to predict it. We report Precision and Recall for each
dataset.

Recall stays high everywhere (0.89-0.93): if a table is well
localized (high IoU), it almost always also gets a good \emph{text}-Jaccard
score. Precision drops more, from 0.92 on PubTables to 0.65 on Table-arXiv,
where PDFAlto struggles with dense math notation, lowering the text score
even when the box is correct. Overall, \emph{text}-Jaccard is close enough
to IoU for a fair comparison, even for methods whose coordinates cannot be
trusted.

\begin{figure}[htbp]
  \centering
  \includegraphics[width=\linewidth]{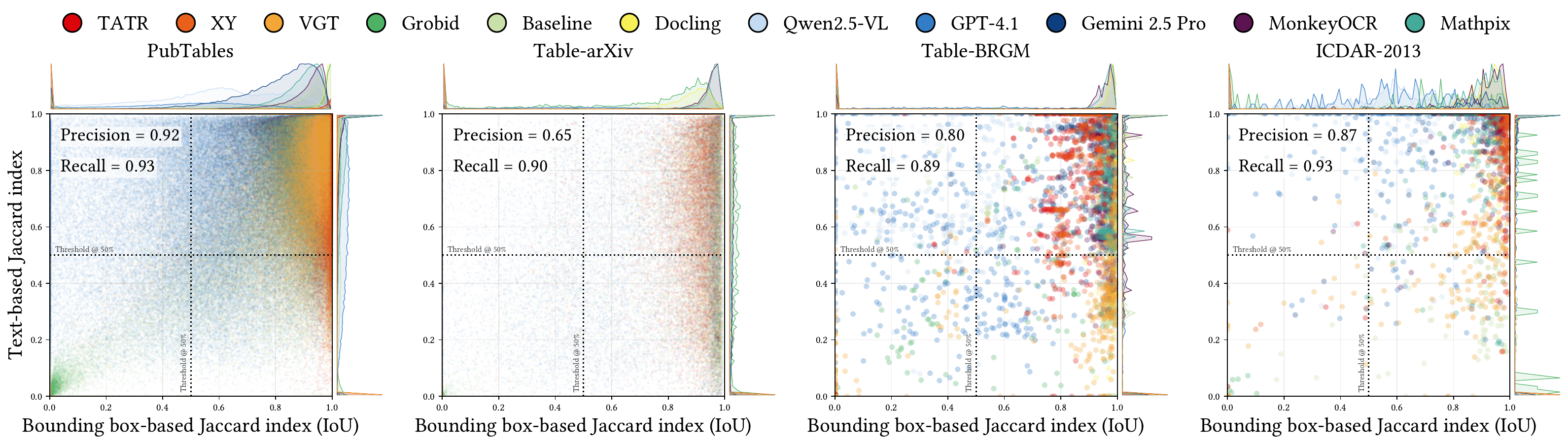}
  \caption{\emph{text}-Jaccard distribution over bounding \emph{bbox} Jaccard index (IoU) distribution
    for (left to right): PubTables, Table-arXiv dataset, Table-BRGM and ICDAR-2013.} \label{fig:jaccard-bbox-content-fig}
\end{figure}

\section{Cost Estimation (Speed Study)}\label{app:cost}
CPU rates are based on the AWS \verb|t3.large| instance pricing\footnote{\url{https://aws.amazon.com/ec2/pricing/on-demand/}}.
GPU rates are derived from the Hugging Face Inference Endpoints documentation\footnote{\url{https://huggingface.co/docs/inference-endpoints/en/pricing}}.
TabPedia-TD$^\dag$ required two Nvidia A100 SXM4 80 GB GPUs, doubling the GPU Time.
For TabPedia-TSR$^\ddag$, we used the same hardware as TabPedia-TD
but encountered timeout issues, as the TSR inference took significantly longer than the TD phase.
While we did not complete the full inference for TabPedia-TSR, we included the projected total cost in \autoref{fig:tradeoff}.

\begin{table}[htbp]
  \centering
  \caption{CPU and GPU (Nvidia A100 SXM4 80 GB) time consumed during inference on 1\,000 samples}
  \resizebox{0.35\columnwidth}{!}{\begin{tabular}{l ccc}
      \toprule
      \multirow{2}[1]{*}{\textbf{Method}}     & \textbf{CPU Time}   & \textbf{GPU Time}   & \textbf{API Cost} \\
                                              & \textbf{(hh:mm:ss)} & \textbf{(hh:mm:ss)} & \textbf{(USD)}    \\
      \midrule
      \Camelot                                & 00:13:19            & --                  & --                \\
      \Pymupdf                                & 00:02:57            & --                  & --                \\
      \PdfPlumber                             & 00:03:19            & --                  & --                \\
      \Grobid                                 & 00:01:00            & --                  & --                \\
      \TATR                                   & 01:32:20            & --                  & --                \\
      \XY                                     & 01:00:54            & --                  & --                \\
      \VGT                                    & 05:13:08            & --                  & --                \\
      \Docling                                & 02:10:38            & --                  & --                \\
      \qwen                                   & --                  & 02:47:40            & --                \\
      \gpt                                    & --                  & --                  & 03.62             \\
      \gemini                                 & --                  & --                  & 15.17             \\
      \textcolor{pedia}{TabPedia-TD}$^\dag$   & --                  & 04:20:26            & --                \\
      \textcolor{pedia}{TabPedia-TSR}$^\ddag$ & --                  & 40:00:00            & --                \\
      \got                                    & --                  & 02:23:10            & --                \\
      \monkey                                 & --                  & 00:41:38            & --                \\
      \mathpix                                & --                  & --                  & 03.85             \\
      \midrule
      \textbf{Rate (USD / hours)}             & 0.0832              & 2.5000              & --                \\
      \bottomrule                                                                                             \\
    \end{tabular}}
\end{table}

\section{Additional Experimental Results}\label{app:more-results}

\autoref{fig:conf-all} plots reliability diagrams for all probabilistic models across datasets,
and \autoref{tab:conf-all} gathers D-ECE TE scores (lower is better) for these models.
Note that MonkeyOCR achieves a low D-ECE score because it outputs very high confidence scores for all tables and also has very high precision.
TATR's D-ECE score is nearly 0.00 on PubTables because, according to its confidence distribution, it outputs binary scores (0 or 1), and on average,
achieves perfect calibration on this dataset.

\begin{figure}[htbp]
  \centering
  \includegraphics[width=\linewidth]{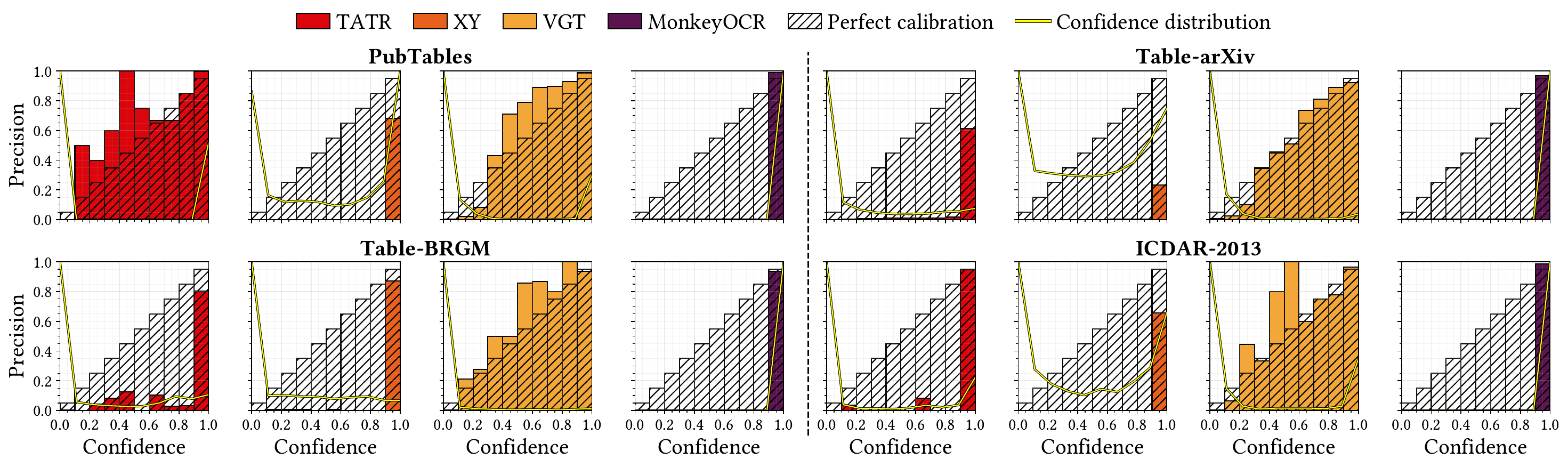}
  \caption{Reliability diagrams of probabilistic models across datasets}\label{fig:conf-all}
\end{figure}
\begin{table*}[htbp]
  \caption{D-ECE TE across datasets for probabilistic models}\label{tab:conf-all}
  \begin{tabular}{l cccc} \toprule
    \textbf{Method} & \textbf{PubTables} & \textbf{Table-arXiv} & \textbf{Table-BRGM} & \textbf{ICDAR-2013} \\
    \midrule
    \TATR           & \textbf{0.00}      & 0.16                 & 0.16                & 0.08                \\
    \XY             & 0.31               & 0.45                 & 0.21                & 0.31                \\
    \VGT            & 0.03               & \underline{0.04}     & \textbf{0.01}       & \underline{0.04}    \\
    \monkey         & \underline{0.01}   & \textbf{0.03}        & \underline{0.06}    & \textbf{0.01}       \\
    \bottomrule
  \end{tabular}
\end{table*}

\autoref{tab:te-f1txt} gathers results for expected TD and TE metrics across datasets for models, using \emph{txt} TD.

\autoref{tab:expected-metrics} gathers results for expected TD metrics (Precision, Recall and F1) across datasets for models, using \emph{bbox} and \emph{txt} TD.
We denote \fbox{\(\mathbb E_0[\mathbf{F_1}]\)} the expected score obtained with the \emph{pdf} \(f_0\) (see \autoref{def:excepted-metrics}).
With the expected metrics, we can better distinguish between models' IoU distributions than with a single threshold or a range of values.
For instance, for \emph{bbox} TD, on Table-arXiv, Docling and MonkeyOCR have similar \(F_1^\textit{bbox}\)-scores with \(\theta_J=0.5\) (0.90 and 0.88 respectively),
but their expected \(F_1\)-scores are different (0.62 and 0.75 respectively), which reflects the fact that MonkeyOCR has higher IoU distribution than Docling.
Regarding Qwen2.5-VL and GPT-4.1, the expected metrics on \emph{bbox} TD reveal the poor spatial understanding.

\begin{table*}[htbp]
  \caption{\(\mathbf{F_1}\)-scores for \emph{txt} TD and \(\mathbf{F_1}^{\!\!\text{TSR}}\)-scores for \emph{txt} TE across datasets for models} \label{tab:te-f1txt}
  \resizebox{\columnwidth}{!}{
    \begin{tabular}{@{} l c ccc  c ccc  c ccc  c ccc @{}} \toprule
      \multirow{3}[4]{*}{\textbf{Method}} & \multicolumn{4}{c}{\textbf{PubTables}} & \multicolumn{4}{c}{\textbf{Table-arXiv}}               & \multicolumn{4}{c}{\textbf{Table-BRGM}} & \multicolumn{4}{c}{\textbf{ICDAR-2013}}                                                                                                                                                                                                                                                                                                                                \\
      \cmidrule(lr){2-5} \cmidrule(lr){6-9} \cmidrule(lr){10-13} \cmidrule(lr){14-17}
                                          & \textbf{TD}                            & \multicolumn{3}{c}{\textbf{TE metrics \textit{(txt)}}} & \textbf{TD}                             & \multicolumn{3}{c}{\textbf{TE metrics \textit{(txt)}}} & \textbf{TD}      & \multicolumn{3}{c}{\textbf{TE metrics \textit{(txt)}}} & \textbf{TD}      & \multicolumn{3}{c}{\textbf{TE metrics \textit{(txt)}}}                                                                                                                                                         \\
      \cmidrule(lr){2-2} \cmidrule(lr){3-5} \cmidrule(lr){6-6} \cmidrule(lr){7-9} \cmidrule(lr){10-10} \cmidrule(lr){11-13} \cmidrule(lr){14-14} \cmidrule(lr){15-17}
                                          & \Ftxt                                  & \FTop                                                  & \FCon                                   & \FTEDS                                                 & \Ftxt            & \FTop                                                  & \FCon            & \FTEDS                                                 & \Ftxt            & \FTop            & \FCon            & \FTEDS           & \Ftxt            & \FTop            & \FCon            & \FTEDS           \\
      \midrule
      \Camelot                            & 0.23                                   & 0.22                                                   & 0.20                                    & 0.20                                                   & 0.11             & 0.09                                                   & 0.08             & 0.08                                                   & 0.75             & 0.74             & 0.68             & 0.67             & 0.54             & 0.53             & 0.50             & 0.50             \\
      \Pymupdf                            & 0.30                                   & 0.27                                                   & 0.24                                    & 0.23                                                   & 0.13             & 0.12                                                   & 0.11             & 0.10                                                   & 0.63             & 0.61             & 0.58             & 0.56             & 0.55             & 0.51             & 0.49             & 0.47             \\
      \PdfPlumber                         & 0.30                                   & 0.24                                                   & 0.21                                    & 0.19                                                   & 0.09             & 0.08                                                   & 0.08             & 0.07                                                   & 0.52             & 0.51             & 0.49             & 0.48             & 0.46             & 0.41             & 0.40             & 0.38             \\
      \Grobid                             & 0.61                                   & 0.56                                                   & 0.49                                    & 0.46                                                   & 0.33             & 0.31                                                   & 0.27             & 0.24                                                   & 0.06             & 0.05             & 0.04             & 0.04             & 0.12             & 0.11             & 0.10             & 0.09             \\
      \Docling                            & \textbf{0.96}                          & \textbf{0.93}                                          & \textbf{0.86}                           & \underline{0.85}                                       & \textbf{0.70}    & \textbf{0.68}                                          & \textbf{0.62}    & \underline{0.60}                                       & 0.78             & 0.77             & 0.74             & \underline{0.74} & \textbf{0.99}    & \textbf{0.99}    & \textbf{0.94}    & \textbf{0.94}    \\
      \qwen                               & 0.90                                   & 0.86                                                   & 0.74                                    & 0.77                                                   & 0.50             & 0.49                                                   & 0.44             & 0.44                                                   & 0.68             & 0.67             & 0.61             & 0.62             & 0.92             & 0.89             & 0.85             & 0.85             \\
      \gpt                                & 0.68*                                  & 0.63*                                                  & 0.55*                                   & 0.56*                                                  & 0.29             & 0.28                                                   & 0.25             & 0.25                                                   & 0.43             & 0.42             & 0.39             & 0.39             & 0.86             & 0.84             & 0.79             & 0.79             \\
      \gemini                             & 0.85*                                  & 0.78*                                                  & 0.73*                                   & 0.75*                                                  & 0.27             & 0.26                                                   & 0.24             & 0.24                                                   & 0.45             & 0.41             & 0.38             & 0.39             & 0.80             & 0.75             & 0.73             & 0.74             \\
      \got                                & 0.39                                   & 0.35                                                   & 0.30                                    & 0.31                                                   & 0.14             & 0.13                                                   & 0.12             & 0.12                                                   & 0.07             & 0.06             & 0.05             & 0.05             & 0.28             & 0.26             & 0.25             & 0.25             \\
      \monkey                             & \textbf{0.96}                          & \underline{0.89}                                       & \underline{0.84}                        & \textbf{0.86}                                          & 0.67             & \underline{0.65}                                       & \underline{0.61} & 0.59                                                   & \underline{0.82} & \underline{0.78} & \underline{0.76} & 0.73             & \underline{0.98} & \underline{0.95} & \underline{0.92} & \underline{0.91} \\
      \mathpix                            & \underline{0.93}                       & 0.88                                                   & 0.81                                    & \underline{0.85}                                       & \underline{0.68} & \underline{0.65}                                       & 0.59             & \textbf{0.61}                                          & \textbf{0.91}    & \textbf{0.86}    & \textbf{0.77}    & \textbf{0.81}    & 0.95             & 0.90             & 0.86             & 0.90             \\
      \bottomrule
    \end{tabular}}
\end{table*}

\begin{table*}[htbp]
  \caption{\(\mathbf{F_1}\) and \(\mathbb E_0[\mathbf{F_1}]\)-scores for \emph{bbox} and \emph{txt} TD across datasets}\label{tab:expected-metrics}

  \resizebox{\columnwidth}{!}{
    \begin{tabular}{@{} l cccc cccc cccc cccc @{}} \toprule
      \multirow{3}[4]{*}{\textbf{Method}} & \multicolumn{4}{c}{\textbf{PubTables}}          & \multicolumn{4}{c}{\textbf{Table-arXiv}}          & \multicolumn{4}{c}{\textbf{Table-BRGM}} & \multicolumn{4}{c}{\textbf{ICDAR-2013}}                                                                                                                                                                                                                                              \\
      \cmidrule(lr){2-5} \cmidrule(lr){6-9} \cmidrule(lr){10-13} \cmidrule(lr){14-17}
                                          & \multicolumn{2}{c}{\textbf{TD \textit{(bbox)}}} & \multicolumn{2}{c}{\textbf{TD \textit{(txt)}}}
                                          & \multicolumn{2}{c}{\textbf{TD \textit{(bbox)}}} & \multicolumn{2}{c}{\textbf{TD \textit{(txt)}}}
                                          & \multicolumn{2}{c}{\textbf{TD \textit{(bbox)}}} & \multicolumn{2}{c}{\textbf{TD \textit{(txt)}}}
                                          & \multicolumn{2}{c}{\textbf{TD \textit{(bbox)}}} & \multicolumn{2}{c}{\textbf{TD \textit{(txt)}}}                                                                                                                                                                                                                                                                                                                                     \\
      \cmidrule(lr){2-3} \cmidrule(lr){4-5} \cmidrule(lr){6-7} \cmidrule(lr){8-9} \cmidrule(lr){10-11} \cmidrule(lr){12-13} \cmidrule(lr){14-15} \cmidrule(lr){16-17}
                                          & \(\mathbf{F_1}^{\!\!\textit{bbox}}\)            & \(\mathbb E_0[\mathbf{F_1}^{\!\!\textit{bbox}}]\) & \(\mathbf{F_1}^{\!\!\textit{txt}}\)     & \(\mathbb E_0[\mathbf{F_1}^{\!\!\textit{txt}}]\)
                                          & \(\mathbf{F_1}^{\!\!\textit{bbox}}\)            & \(\mathbb E_0[\mathbf{F_1}^{\!\!\textit{bbox}}]\) & \(\mathbf{F_1}^{\!\!\textit{txt}}\)     & \(\mathbb E_0[\mathbf{F_1}^{\!\!\textit{txt}}]\)
                                          & \(\mathbf{F_1}^{\!\!\textit{bbox}}\)            & \(\mathbb E_0[\mathbf{F_1}^{\!\!\textit{bbox}}]\) & \(\mathbf{F_1}^{\!\!\textit{txt}}\)     & \(\mathbb E_0[\mathbf{F_1}^{\!\!\textit{txt}}]\)
                                          & \(\mathbf{F_1}^{\!\!\textit{bbox}}\)            & \(\mathbb E_0[\mathbf{F_1}^{\!\!\textit{bbox}}]\) & \(\mathbf{F_1}^{\!\!\textit{txt}}\)     & \(\mathbb E_0[\mathbf{F_1}^{\!\!\textit{txt}}]\)                                                                                                                                                                                                                                     \\
      \midrule
      \Camelot                            & 0.25                                            & 0.21                                              & 0.23                                    & 0.21                                             & 0.33             & 0.29             & 0.11             & 0.09             & 0.83             & 0.78             & 0.75             & \underline{0.59} & 0.66             & 0.53             & 0.54             & 0.51             \\
      \Pymupdf                            & 0.42                                            & 0.31                                              & 0.30                                    & 0.23                                             & 0.38             & 0.35             & 0.13             & 0.10             & 0.85             & 0.80             & 0.63             & 0.53             & 0.76             & 0.61             & 0.55             & 0.50             \\
      \PdfPlumber                         & 0.41                                            & 0.30                                              & 0.30                                    & 0.23                                             & 0.24             & 0.21             & 0.09             & 0.07             & 0.74             & 0.69             & 0.52             & 0.44             & 0.63             & 0.51             & 0.46             & 0.40             \\
      \Grobid                             & 0.67                                            & 0.59                                              & 0.61                                    & 0.45                                             & 0.43             & 0.31             & 0.33             & 0.22             & 0.15             & 0.11             & 0.06             & 0.05             & 0.16             & 0.10             & 0.12             & 0.08             \\
      \Docling                            & \textbf{0.99}                                   & \textbf{0.95}                                     & \textbf{0.96}                           & \textbf{0.83}                                    & 0.90             & \underline{0.70} & \textbf{0.70}    & \textbf{0.52}    & 0.86             & 0.81             & 0.78             & \textbf{0.63}    & \textbf{0.99}    & \textbf{0.83}    & \textbf{0.99}    & \textbf{0.88}    \\
      \qwen                               & 0.65                                            & 0.37                                              & 0.90                                    & 0.69                                             & 0.43             & 0.23             & 0.50             & 0.35             & 0.29             & 0.18             & 0.68             & 0.47             & 0.73             & 0.36             & 0.92             & 0.82             \\
      \gpt                                & 0.42*                                           & 0.23*                                             & 0.68*                                   & 0.44*                                            & 0.28             & 0.17             & 0.29             & 0.20             & 0.35             & 0.19             & 0.43             & 0.28             & 0.65             & 0.34             & 0.86             & 0.69             \\
      \gemini                             & \underline{0.77}*                               & 0.54*                                             & 0.85*                                   & 0.69*                                            & 0.42             & 0.29             & 0.27             & 0.20             & 0.45             & 0.35             & 0.45             & 0.36             & 0.79             & 0.60             & 0.80             & 0.74             \\
      \pedia                              & \textbf{0.99}                                   & \underline{0.87}                                  & TO                                      & TO                                               & 0.26             & 0.18             & TO               & TO               & 0.26             & 0.17             & TO               & TO               & 0.69             & 0.59             & TO               & TO               \\
      \got                                & --                                              & --                                                & 0.39                                    & 0.25                                             & --               & --               & 0.14             & 0.10             & --               & --               & 0.07             & 0.06             & --               & --               & 0.28             & 0.20             \\
      \monkey                             & \textbf{0.99}                                   & 0.82                                              & \textbf{0.96}                           & \underline{0.80}                                 & \textbf{0.97}    & \textbf{0.87}    & 0.67             & \underline{0.48} & \textbf{0.97}    & \textbf{0.88}    & \underline{0.82} & 0.51             & \underline{0.98} & \underline{0.82} & \underline{0.98} & \underline{0.86} \\
      \mathpix                            & \textbf{0.99}                                   & 0.78                                              & \underline{0.93}                        & 0.67                                             & \underline{0.95} & \textbf{0.87}    & \underline{0.68} & \underline{0.48} & \underline{0.95} & \underline{0.87} & \textbf{0.91}    & 0.58             & \underline{0.98} & 0.77             & 0.95             & 0.83             \\
      \bottomrule
    \end{tabular}}
\end{table*}

\autoref{fig:te-bbox2} plots TE results for \emph{bbox} TD with GriTS Content and TEDS.
\autoref{fig:te-txt} plots TE results for \emph{txt} TD with GriTS Topology, Content and TEDS.

\begin{figure}[htbp]
  \centering
  \begin{subfigure}[b]{\textwidth}
    \centering
    \includegraphics[width=0.7\linewidth]{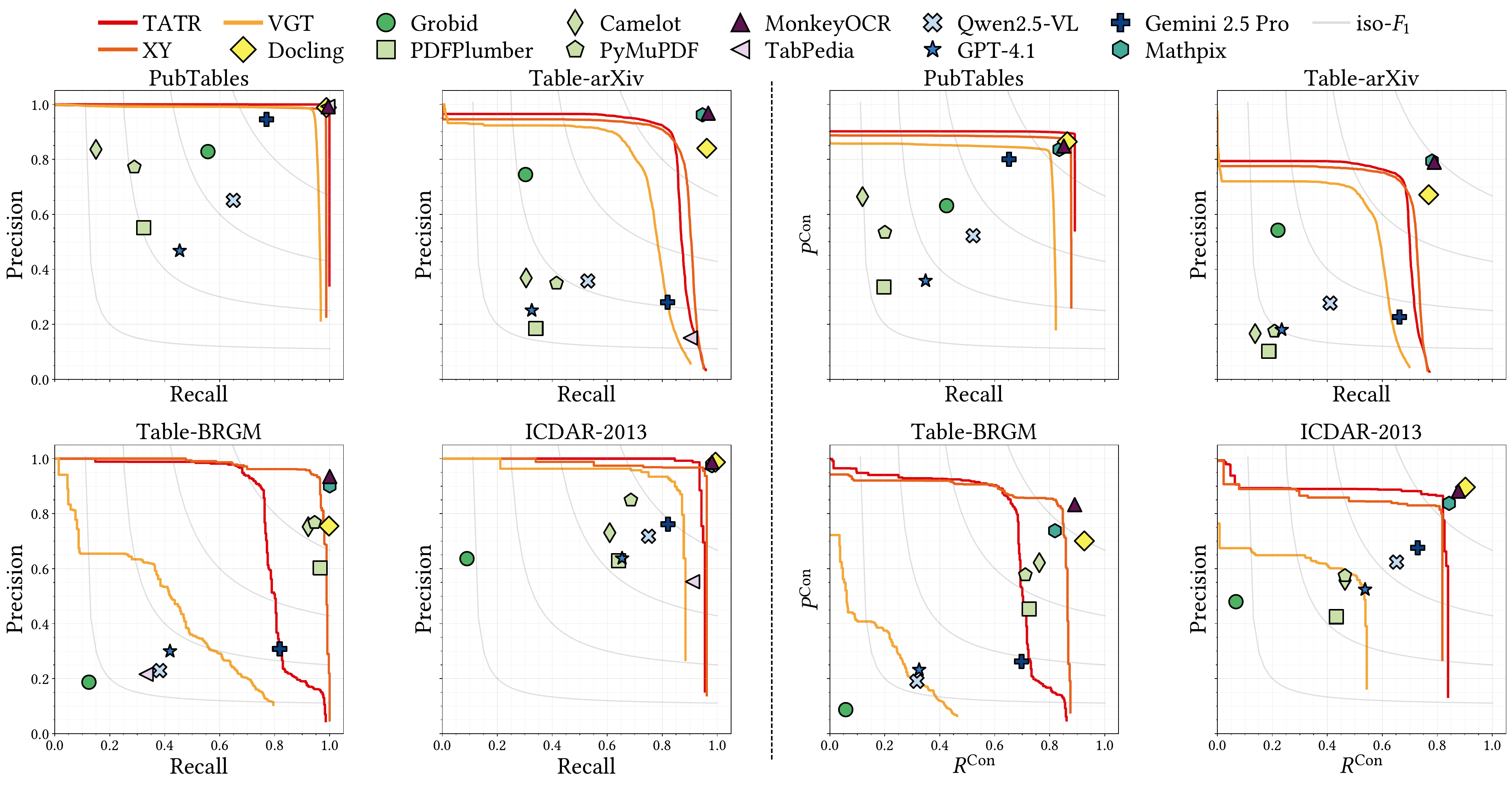}
    \caption{(right) \(P^{\text{Con}}-R^{\text{Con}}\) for \emph{bbox} TE}
  \end{subfigure}
  \begin{subfigure}[b]{\textwidth}
    \centering
    \includegraphics[width=0.7\linewidth]{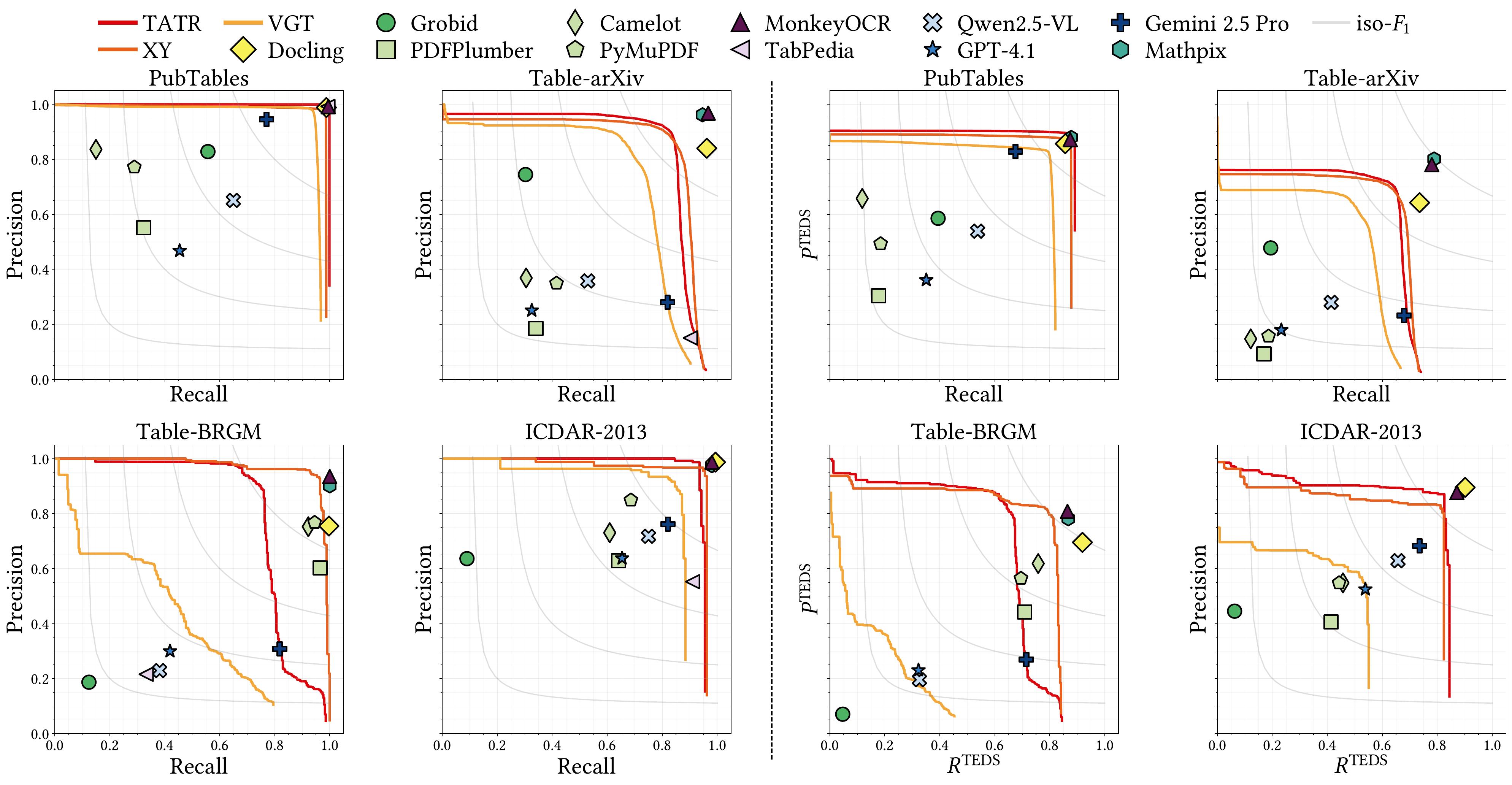}
    \caption{(right) \(P^{\text{TEDS}}-R^{\text{TEDS}}\) for \emph{bbox} TE}
  \end{subfigure}
  \caption{(left) Precision--Recall for \emph{bbox} TD; (right) \(P^{\text{TSR}}-R^{\text{TSR}}\) for \emph{bbox} TE}\label{fig:te-bbox2}
\end{figure}

\begin{figure}[htbp]
  \centering
  \begin{subfigure}[b]{\textwidth}
    \centering
    \includegraphics[width=0.7\linewidth]{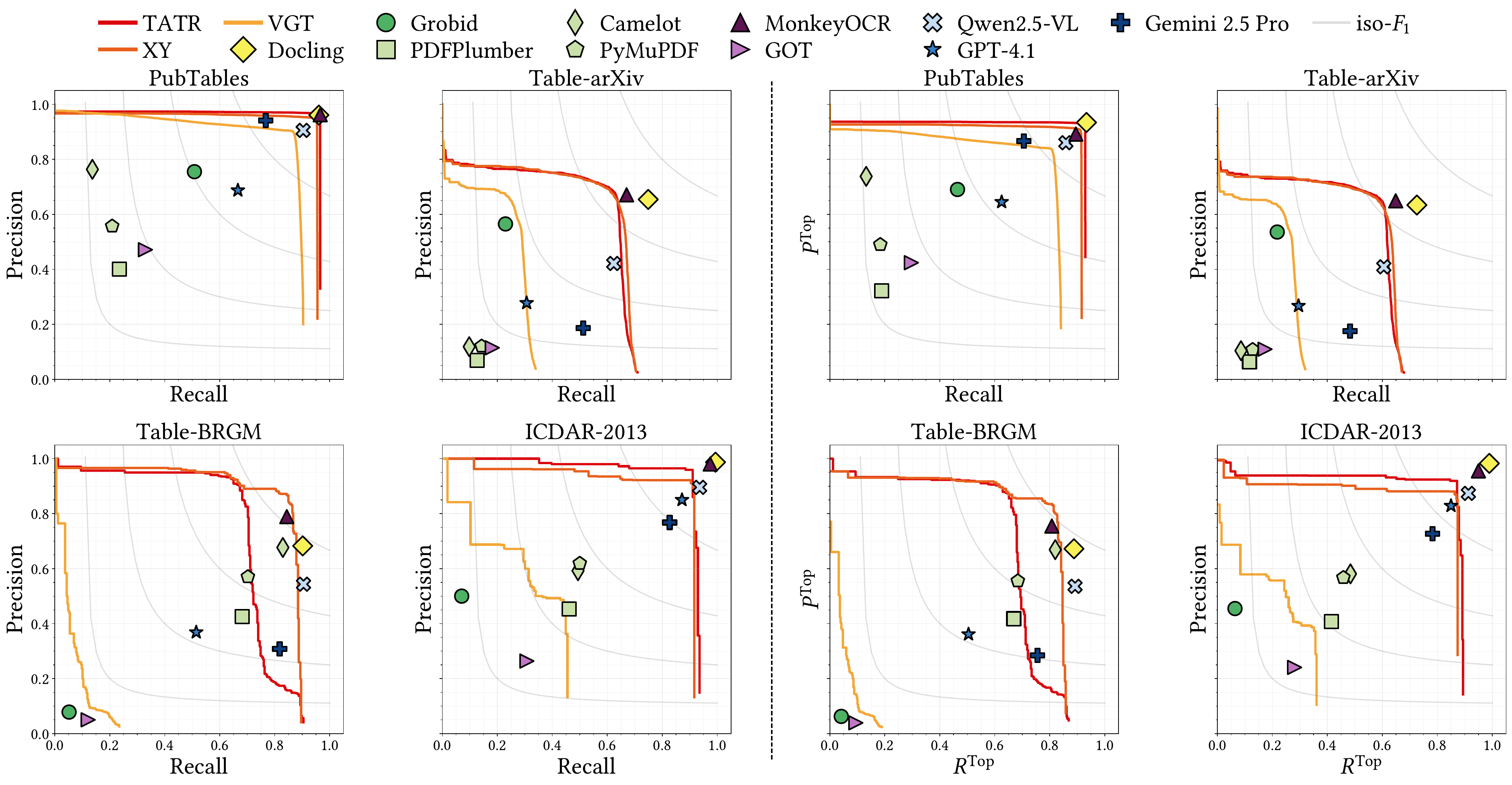}
    \caption{(right) \(P^{\text{Top}}-R^{\text{Top}}\) for \emph{txt} TE}
  \end{subfigure}
  \begin{subfigure}[b]{\textwidth}
    \centering
    \includegraphics[width=0.7\linewidth]{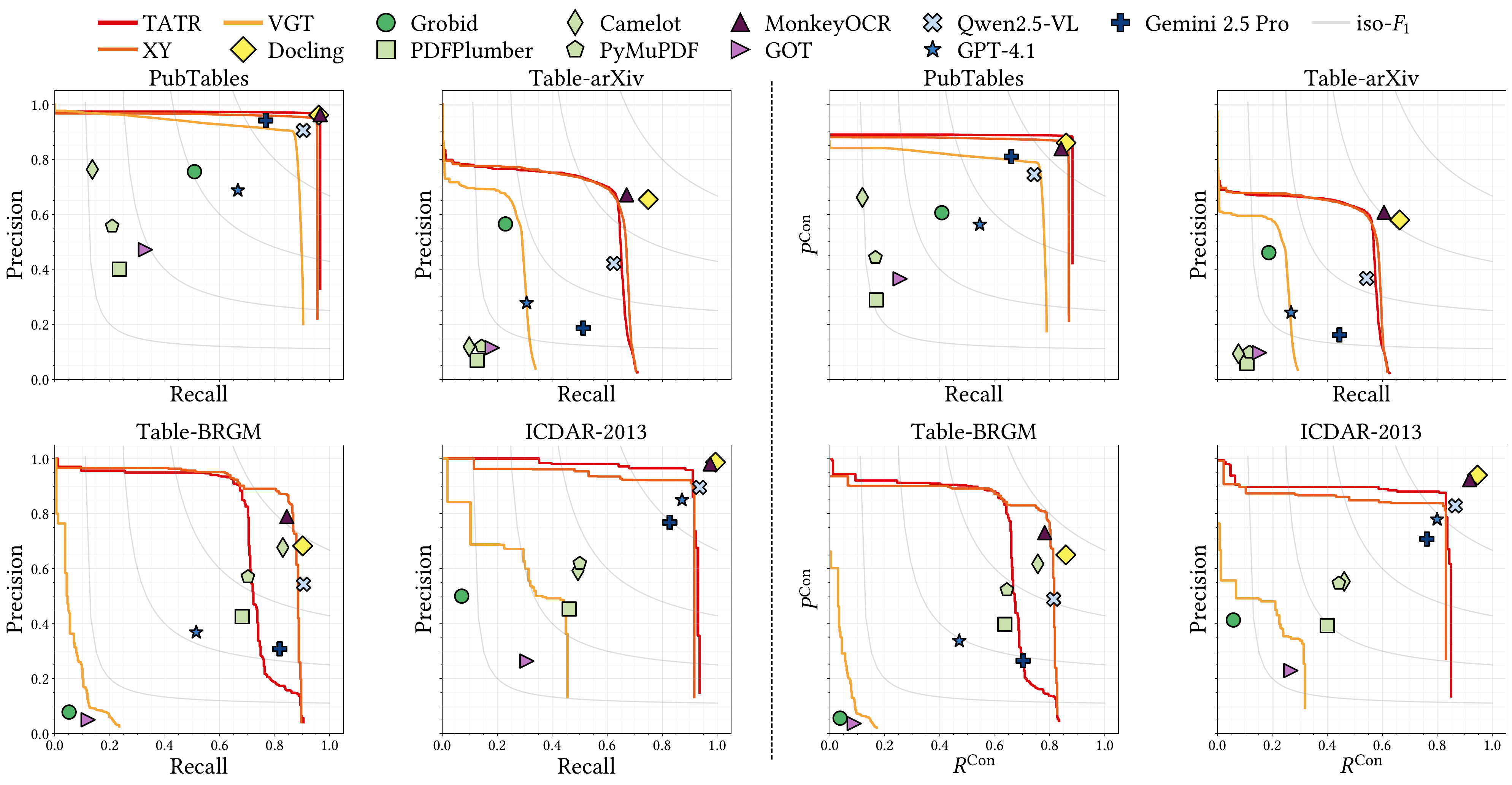}
    \caption{(right) \(P^{\text{Con}}-R^{\text{Con}}\) for \emph{txt} TE}
  \end{subfigure}
  \begin{subfigure}[b]{\textwidth}
    \centering
    \includegraphics[width=0.7\linewidth]{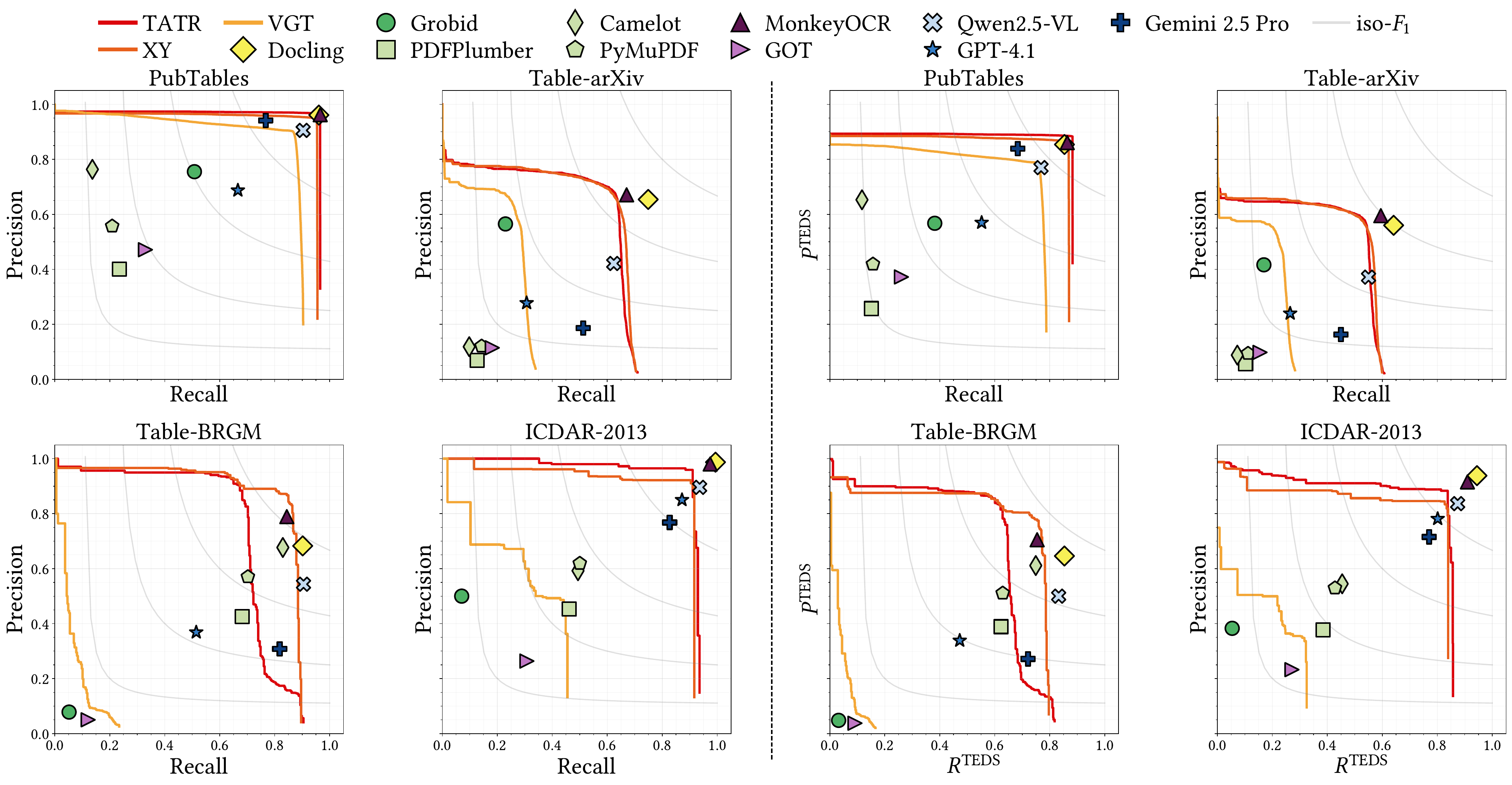}
    \caption{(right) \(P^{\text{TEDS}}-R^{\text{TEDS}}\) for \emph{txt} TE}
  \end{subfigure}
  \caption{(left) Precision--Recall for \emph{txt} TD; (right) \(P^{\text{TSR}}-R^{\text{TSR}}\) for \emph{txt} TE}\label{fig:te-txt}
\end{figure}

\autoref{fig:tradeoff-all} gathers tradeoff plots (performance vs. computational cost) for each TE metrics and datasets, using \emph{bbox} TD.
\autoref{fig:tradeoff-all-txt} similarly gathers tradeoff plots for \emph{txt} TD.
\autoref{fig:gpt-hallu} shows an example of hallucination (reporting tokens that
were not present in the actual input) and visual capacities (approximate bounding boxes) by \gpt on the Table-BRGM dataset.

\begin{figure}[htbp]
  \centering
  \begin{subfigure}[b]{0.3\textwidth}
    \centering
    \includegraphics[width=\linewidth]{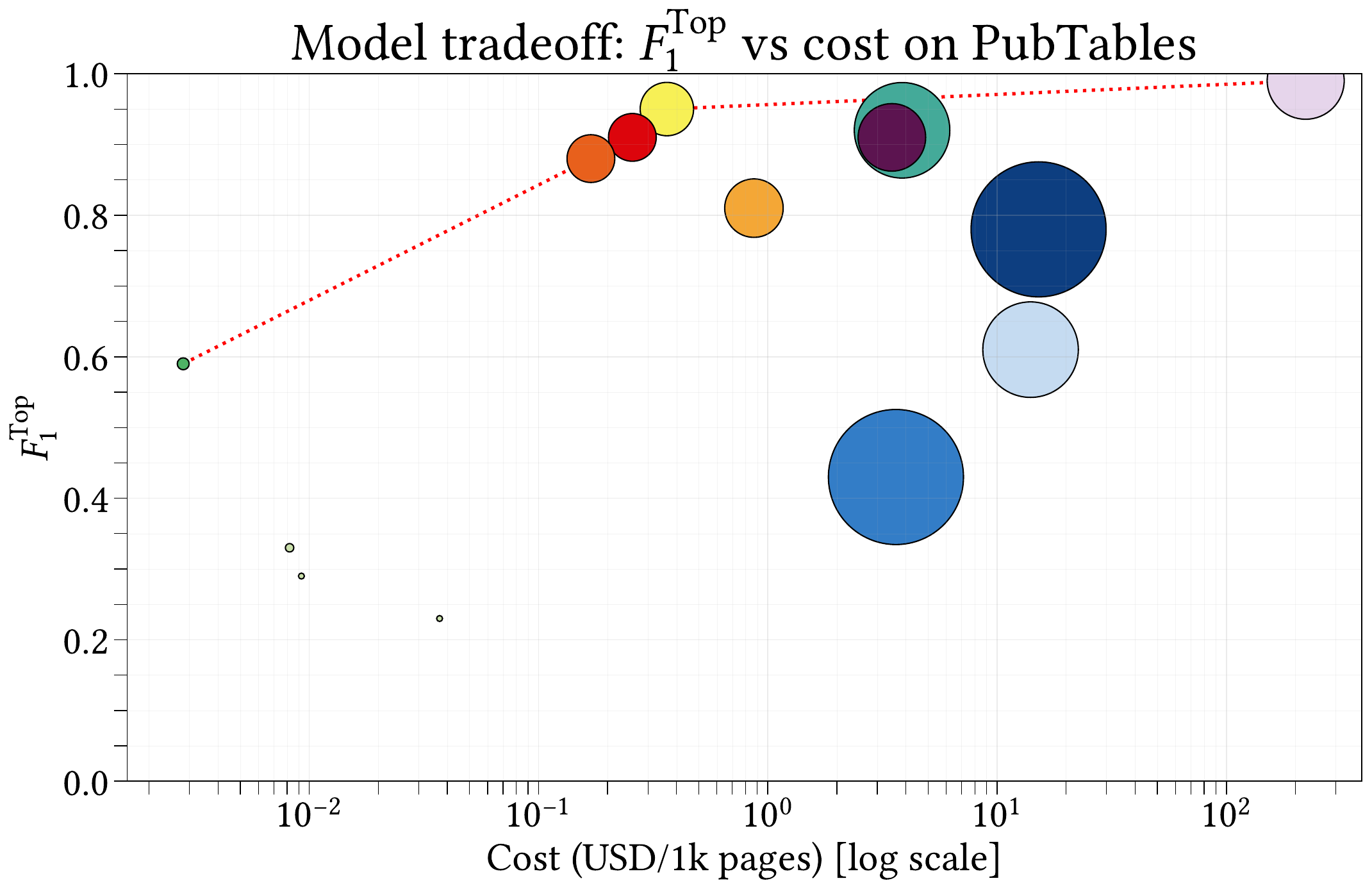}
    \caption{\FTop{} on PubTables}
  \end{subfigure}
  \hfill
  \begin{subfigure}[b]{0.3\textwidth}
    \centering
    \includegraphics[width=\linewidth]{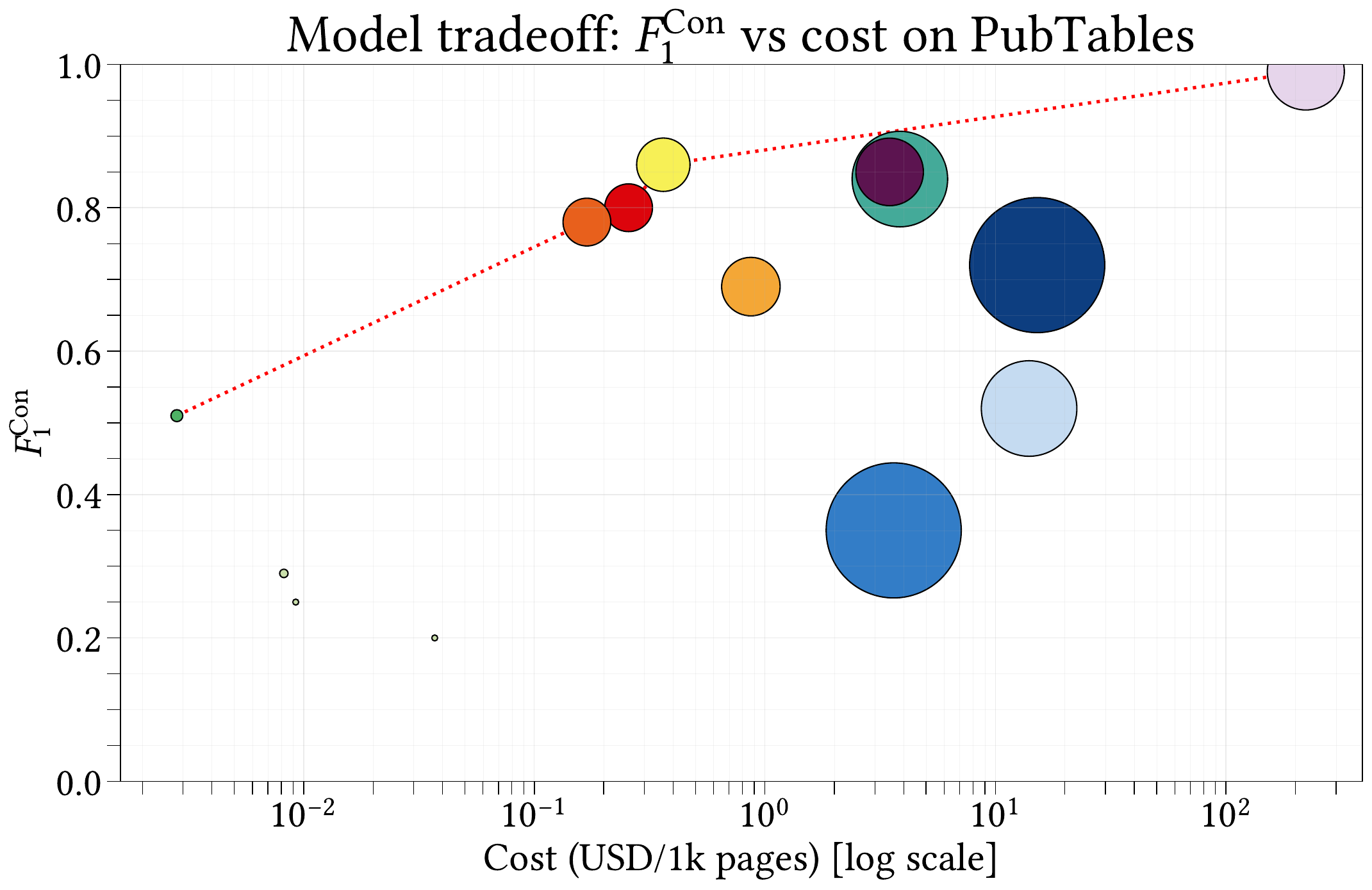}
    \caption{\FCon{} on PubTables}
  \end{subfigure}
  \hfill
  \begin{subfigure}[b]{0.3\textwidth}
    \centering
    \includegraphics[width=\linewidth]{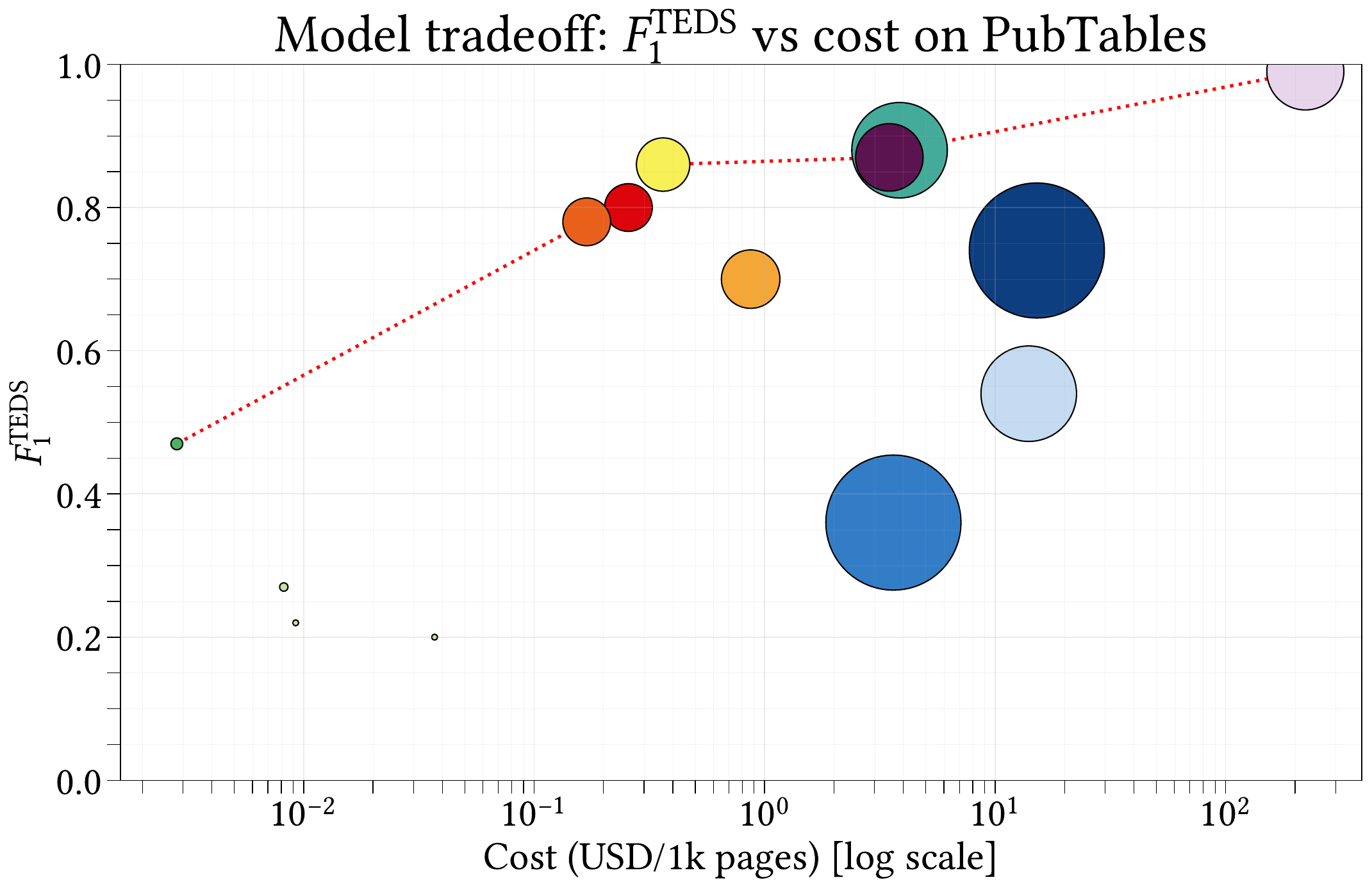}
    \caption{\FTEDS{} on PubTables}
  \end{subfigure}
  \vspace{1em}
  \begin{subfigure}[b]{0.3\textwidth}
    \centering
    \includegraphics[width=\linewidth]{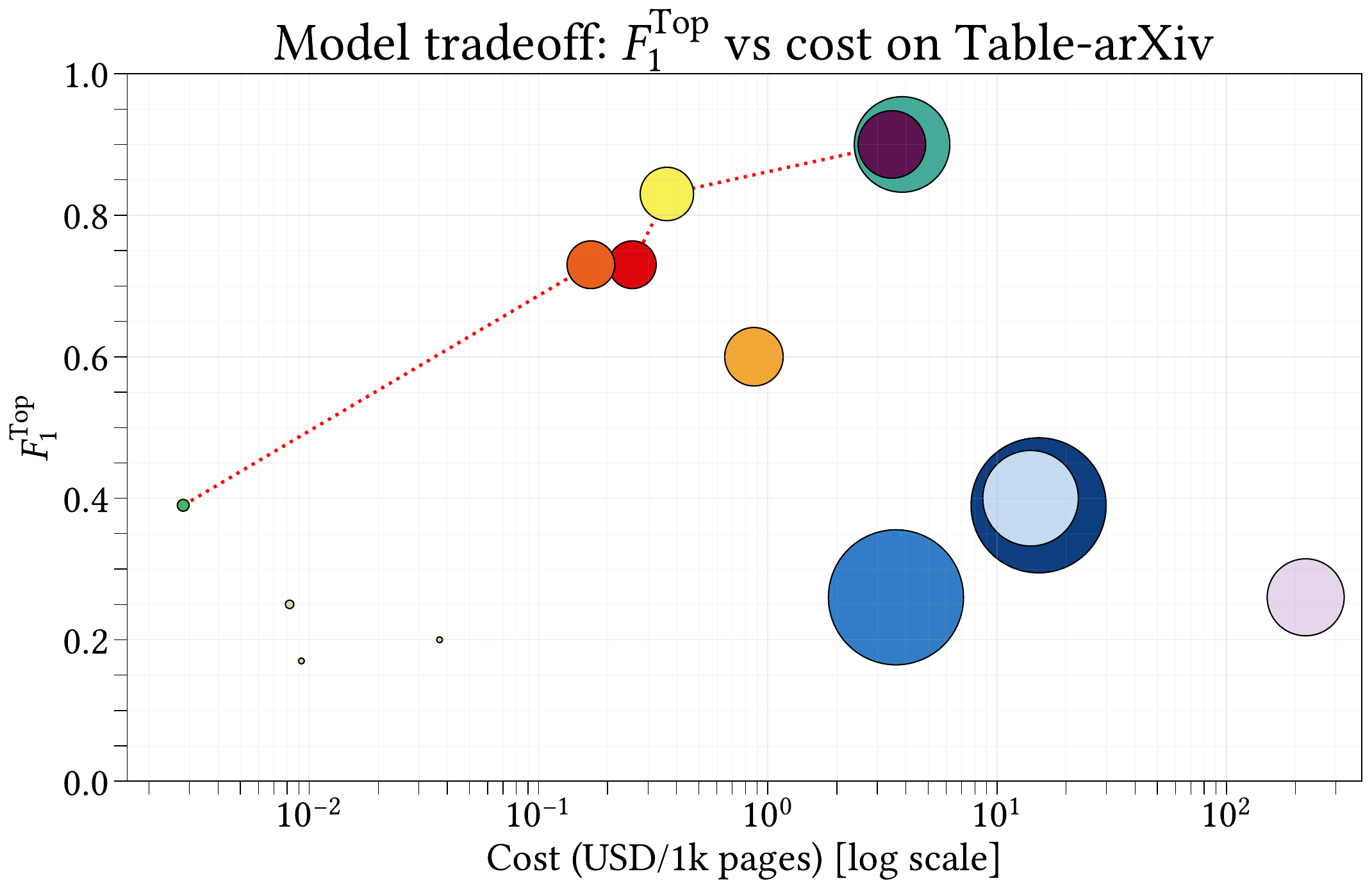}
    \caption{\FTop{} on Table-arXiv}
  \end{subfigure}
  \hfill
  \begin{subfigure}[b]{0.3\textwidth}
    \centering
    \includegraphics[width=\linewidth]{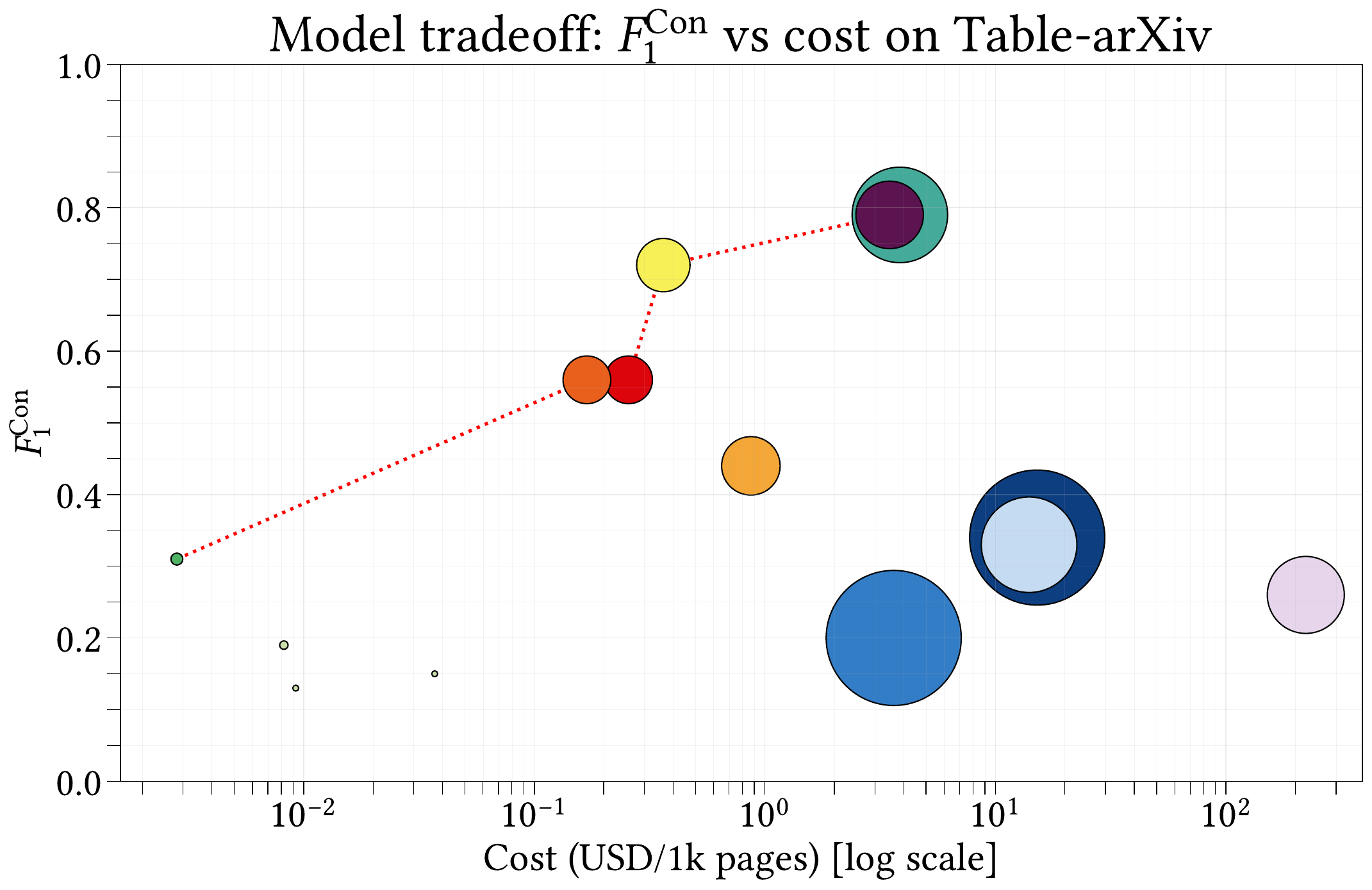}
    \caption{\FCon{} on Table-arXiv}
  \end{subfigure}
  \hfill
  \begin{subfigure}[b]{0.3\textwidth}
    \centering
    \includegraphics[width=\linewidth]{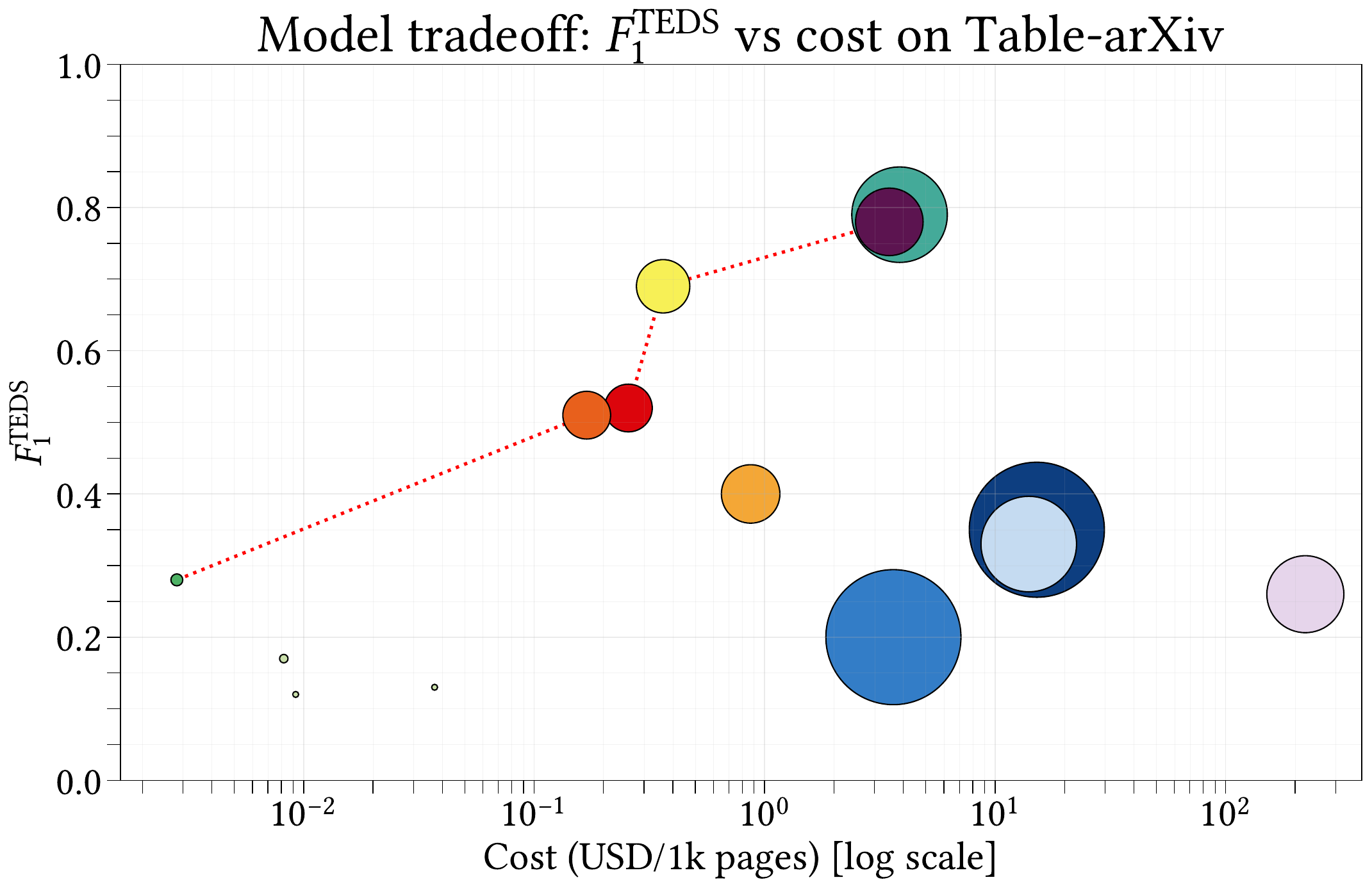}
    \caption{\FTEDS{} on Table-arXiv}
  \end{subfigure}
  \vspace{1em}
  \begin{subfigure}[b]{0.3\textwidth}
    \centering
    \includegraphics[width=\linewidth]{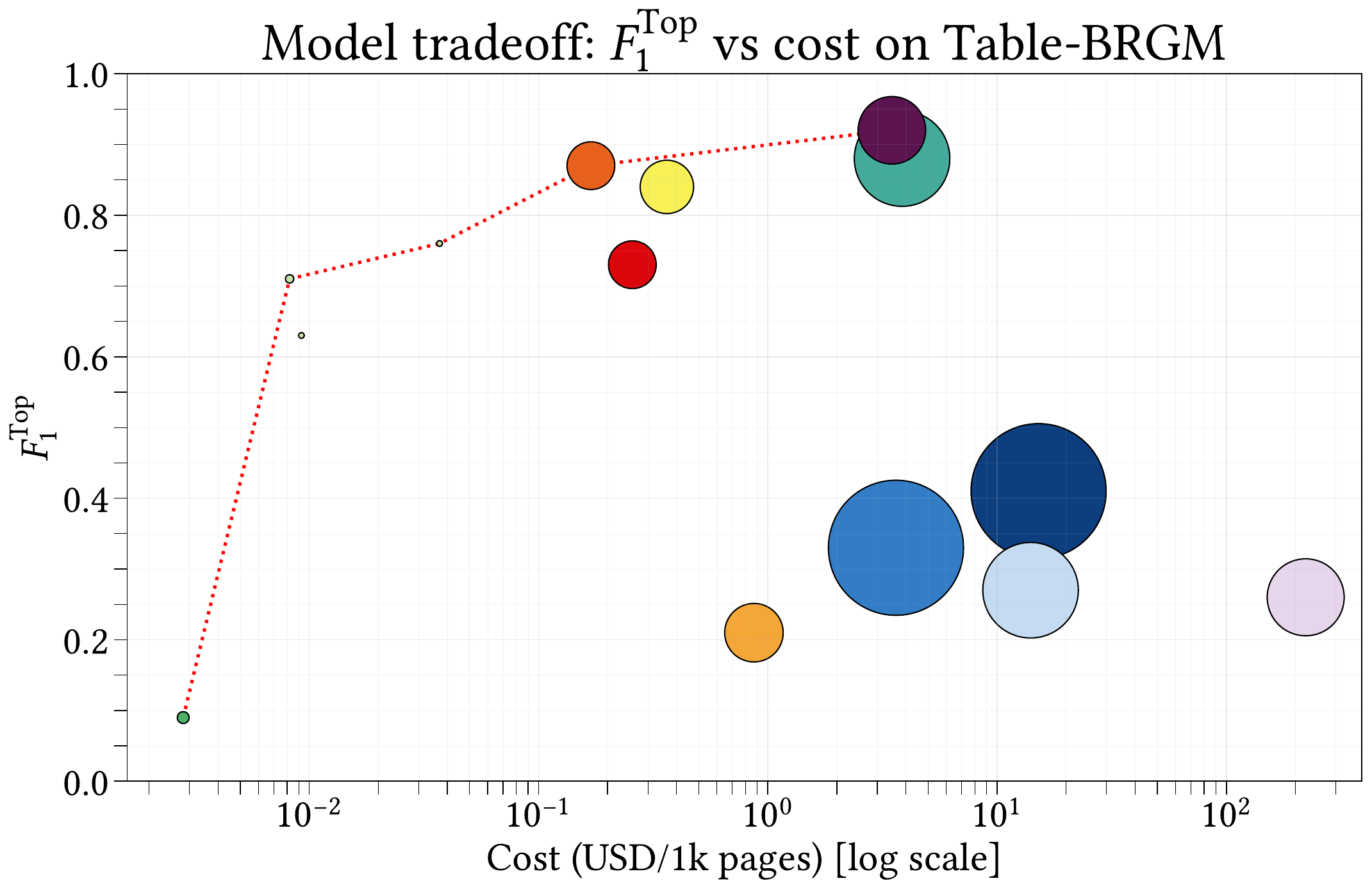}
    \caption{\FTop{} on Table-BRGM}
  \end{subfigure}
  \hfill
  \begin{subfigure}[b]{0.3\textwidth}
    \centering
    \includegraphics[width=\linewidth]{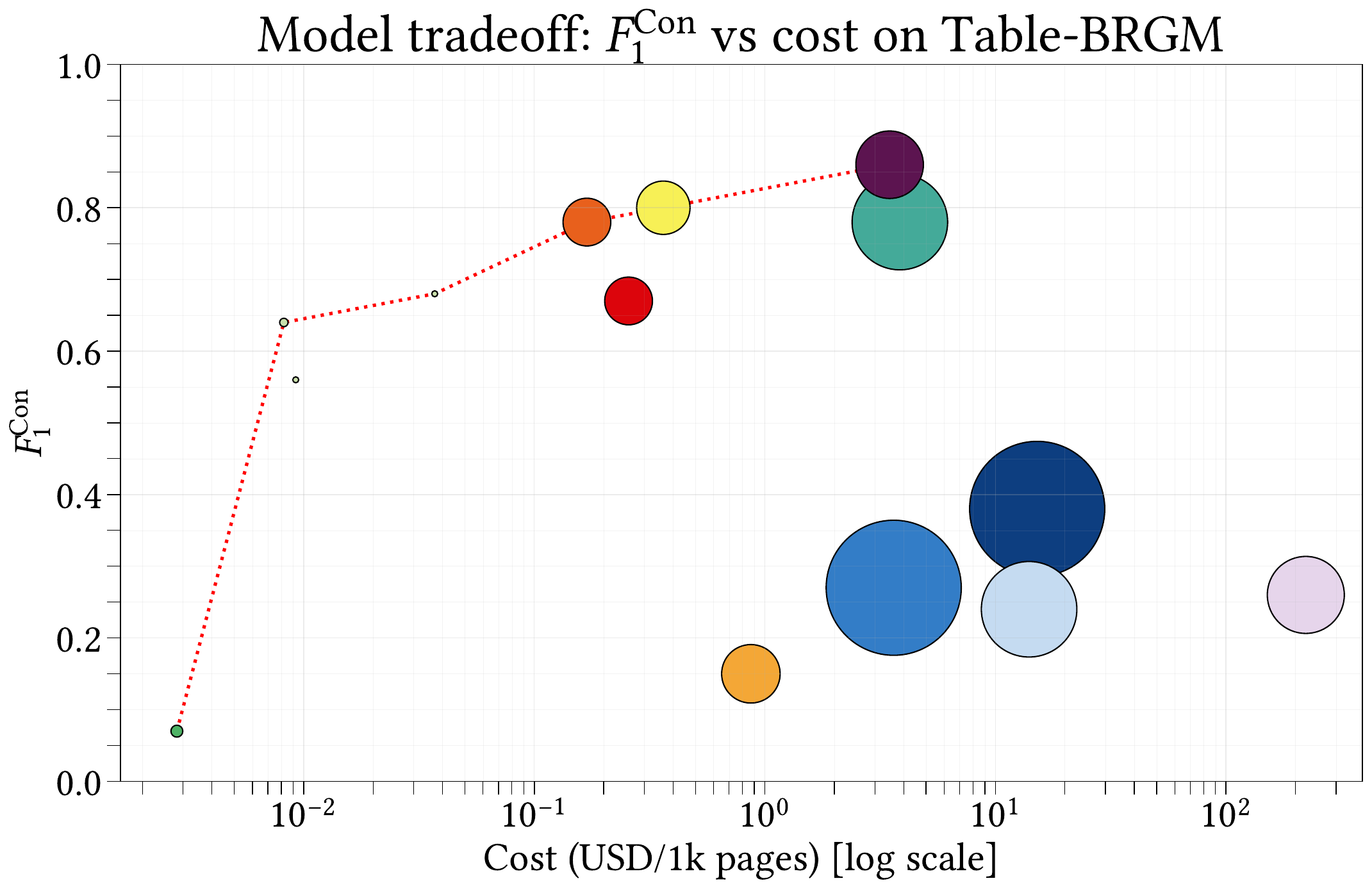}
    \caption{\FCon{} on Table-BRGM}
  \end{subfigure}
  \hfill
  \begin{subfigure}[b]{0.3\textwidth}
    \centering
    \includegraphics[width=\linewidth]{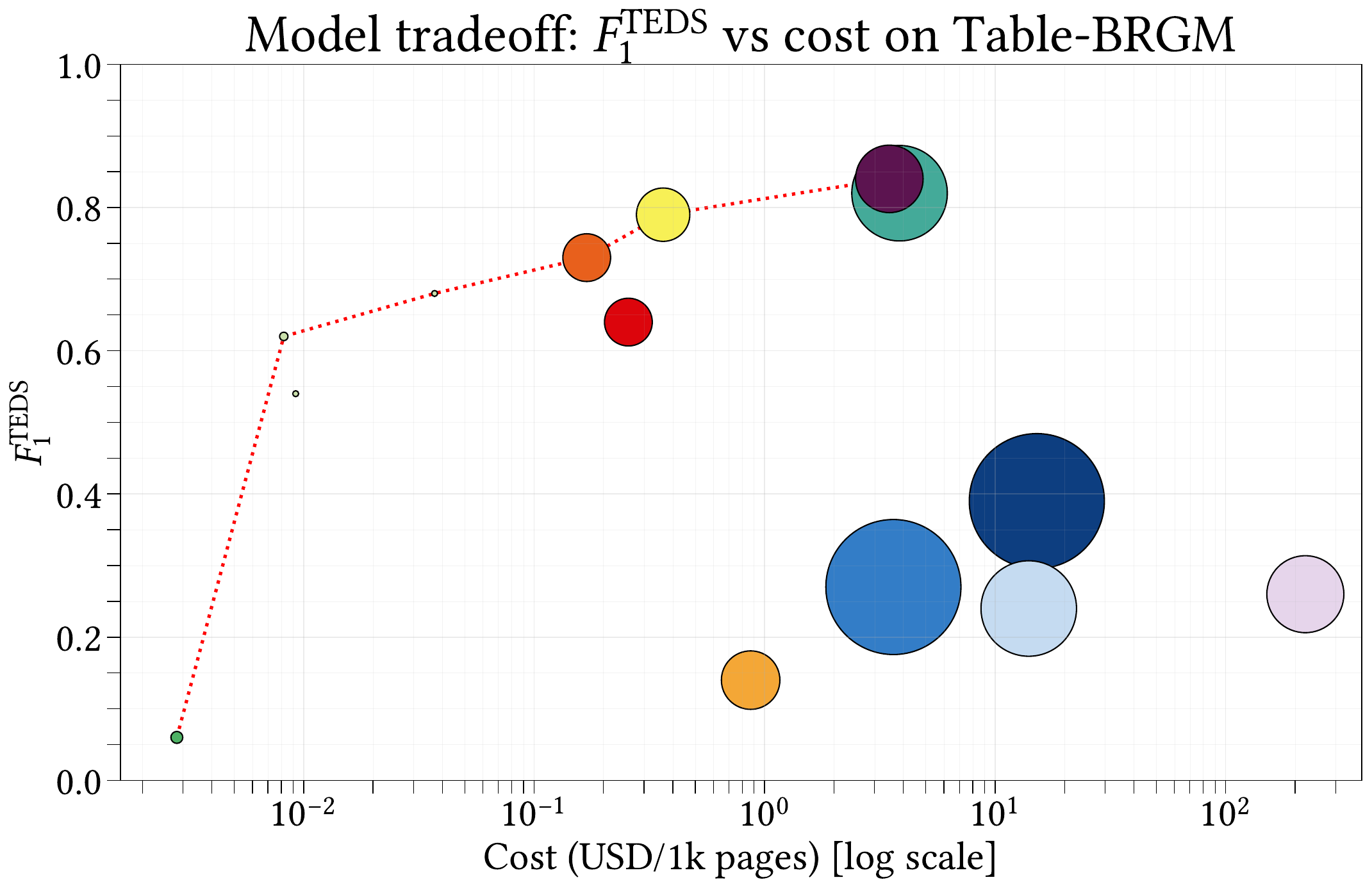}
    \caption{\FTEDS{} on Table-BRGM}
  \end{subfigure}
  \vspace{1em}
  \begin{subfigure}[b]{0.3\textwidth}
    \centering
    \includegraphics[width=\linewidth]{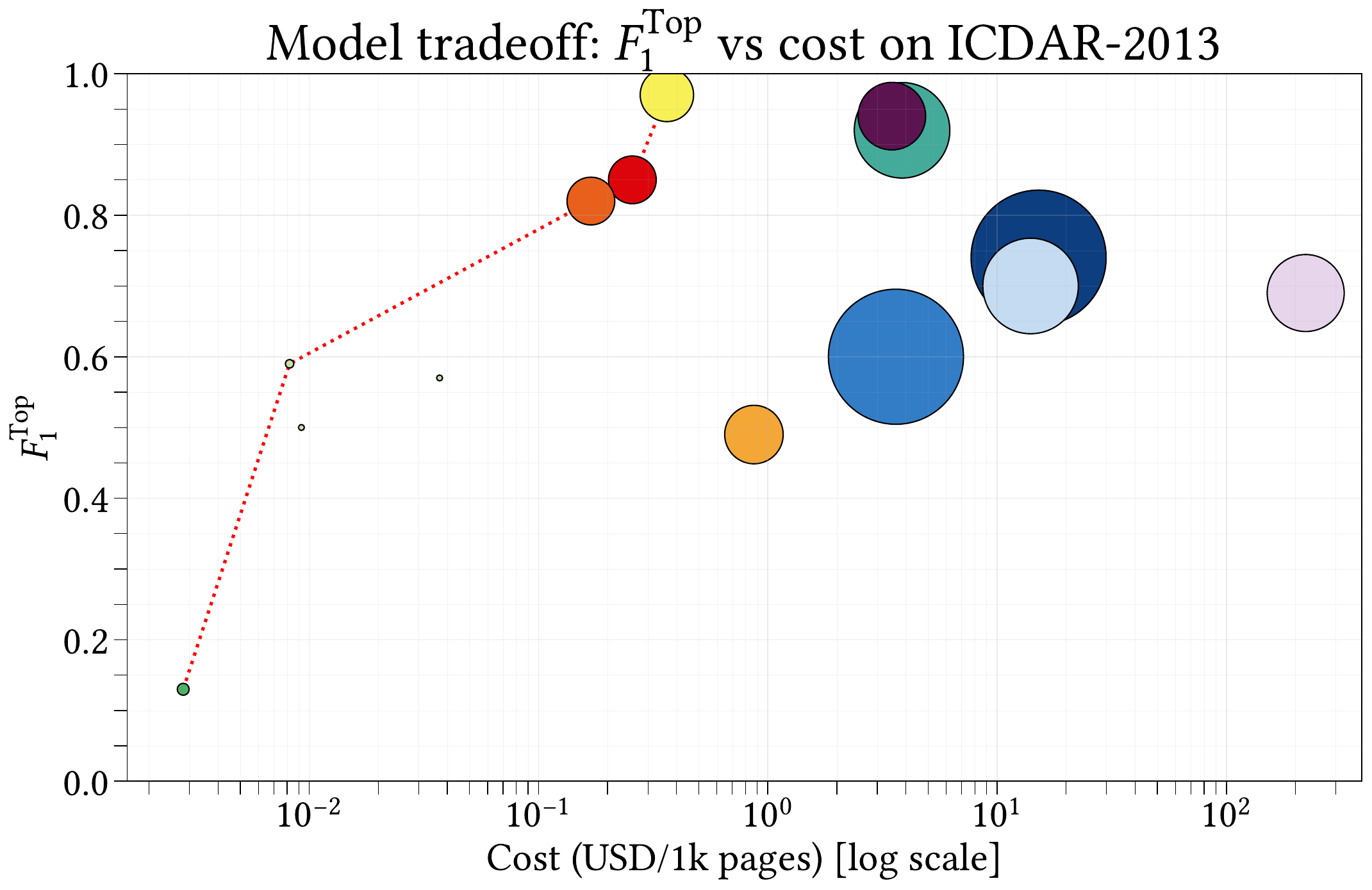}
    \caption{\FTop{} on ICDAR-2013}
  \end{subfigure}
  \hfill
  \begin{subfigure}[b]{0.3\textwidth}
    \centering
    \includegraphics[width=\linewidth]{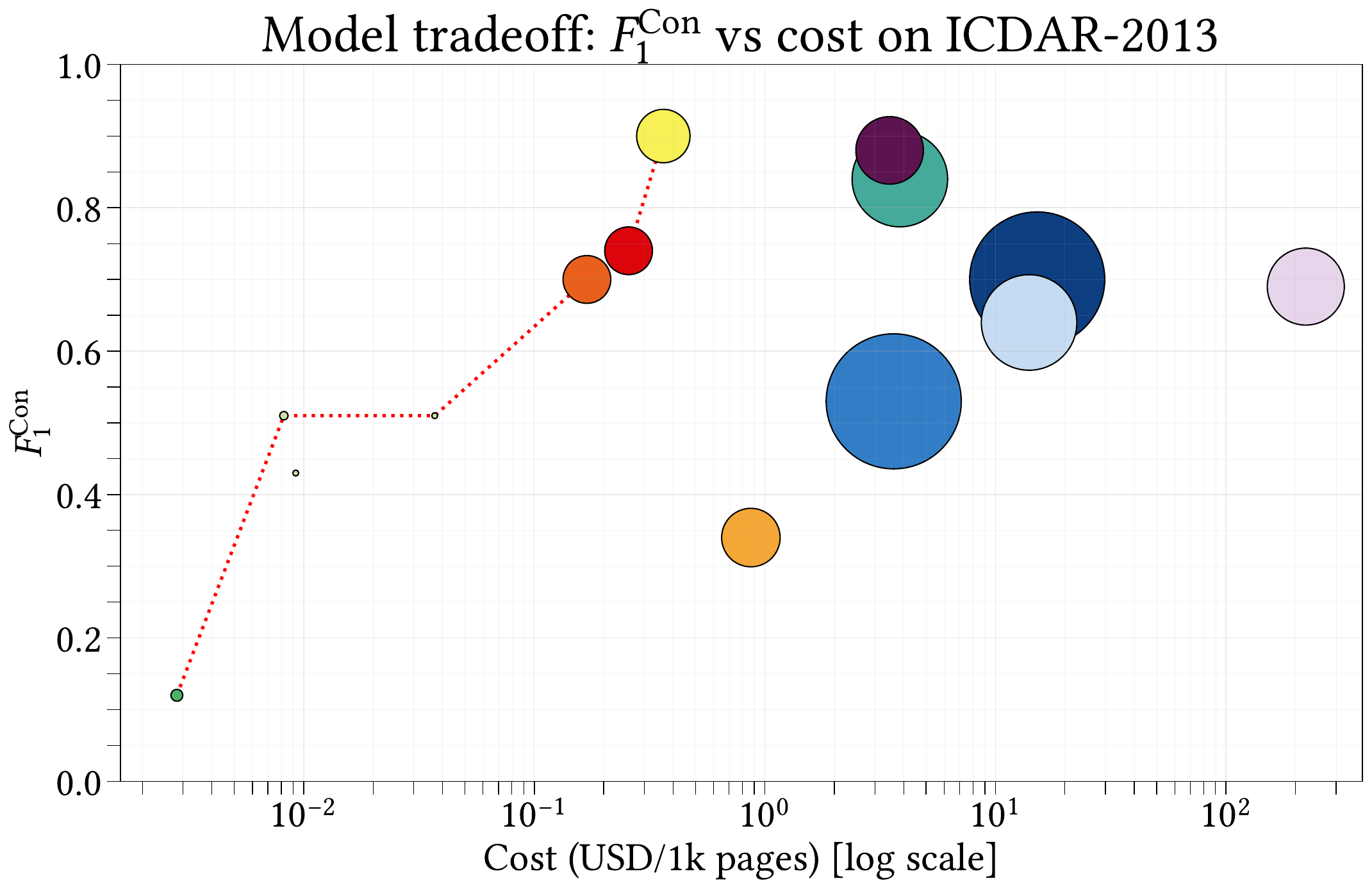}
    \caption{\FCon{} on ICDAR-2013}
  \end{subfigure}
  \hfill
  \begin{subfigure}[b]{0.3\textwidth}
    \centering
    \includegraphics[width=\linewidth]{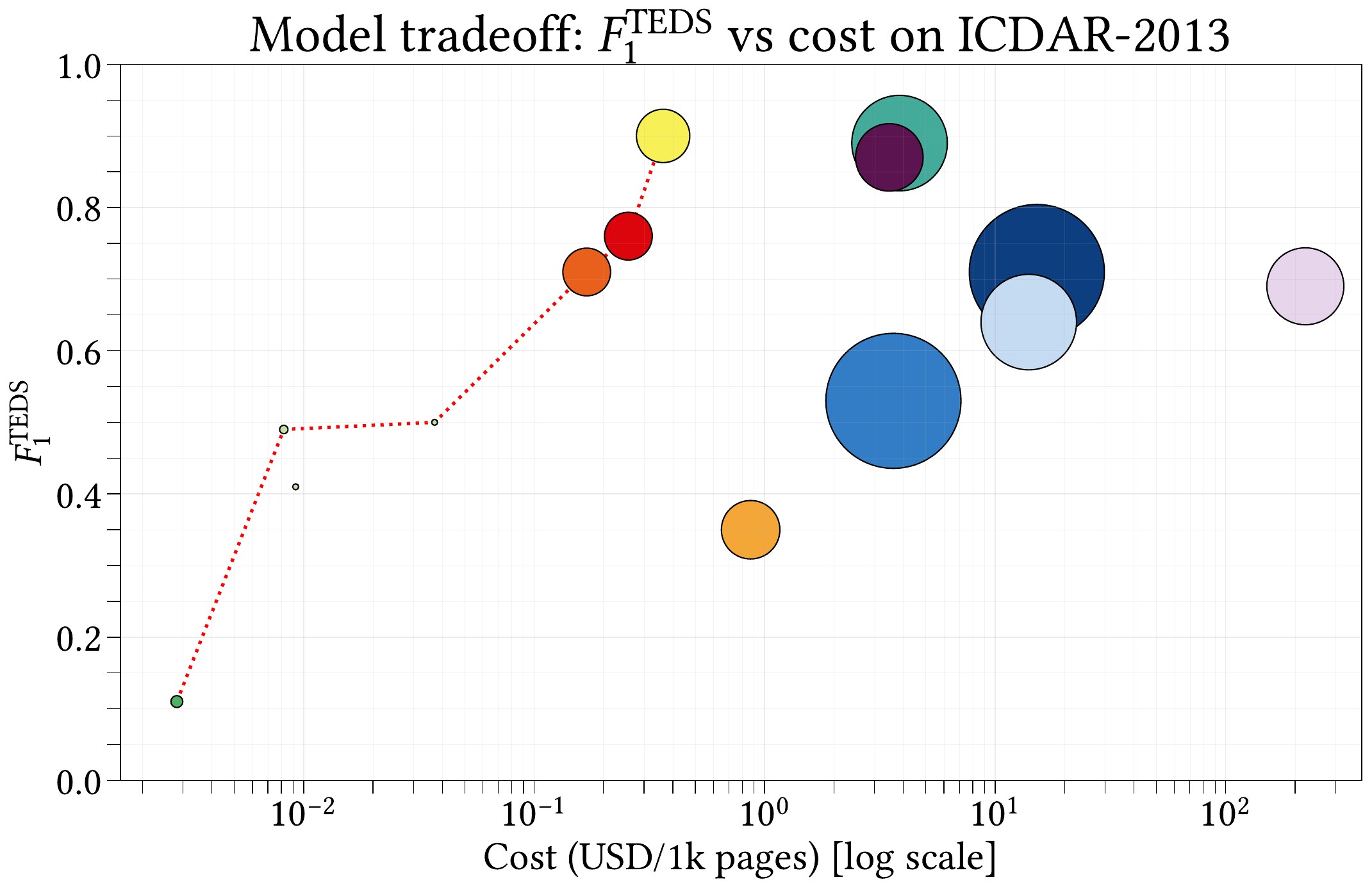}
    \caption{\FTEDS{} on ICDAR-2013}
  \end{subfigure}
  \caption{Tradeoff: TE metrics vs. computational cost for each \emph{bbox} TE metric across datasets.
    Bubble size is roughly proportional to the logarithm number of parameters (see \autoref{fig:tradeoff}).
    Colors are the same as in \autoref{fig:tradeoff} and \autoref{tab:all-model}.}
  \label{fig:tradeoff-all}
\end{figure}

\begin{figure}[htbp]
  \centering
  \begin{subfigure}[b]{0.3\textwidth}
    \centering
    \includegraphics[width=\linewidth]{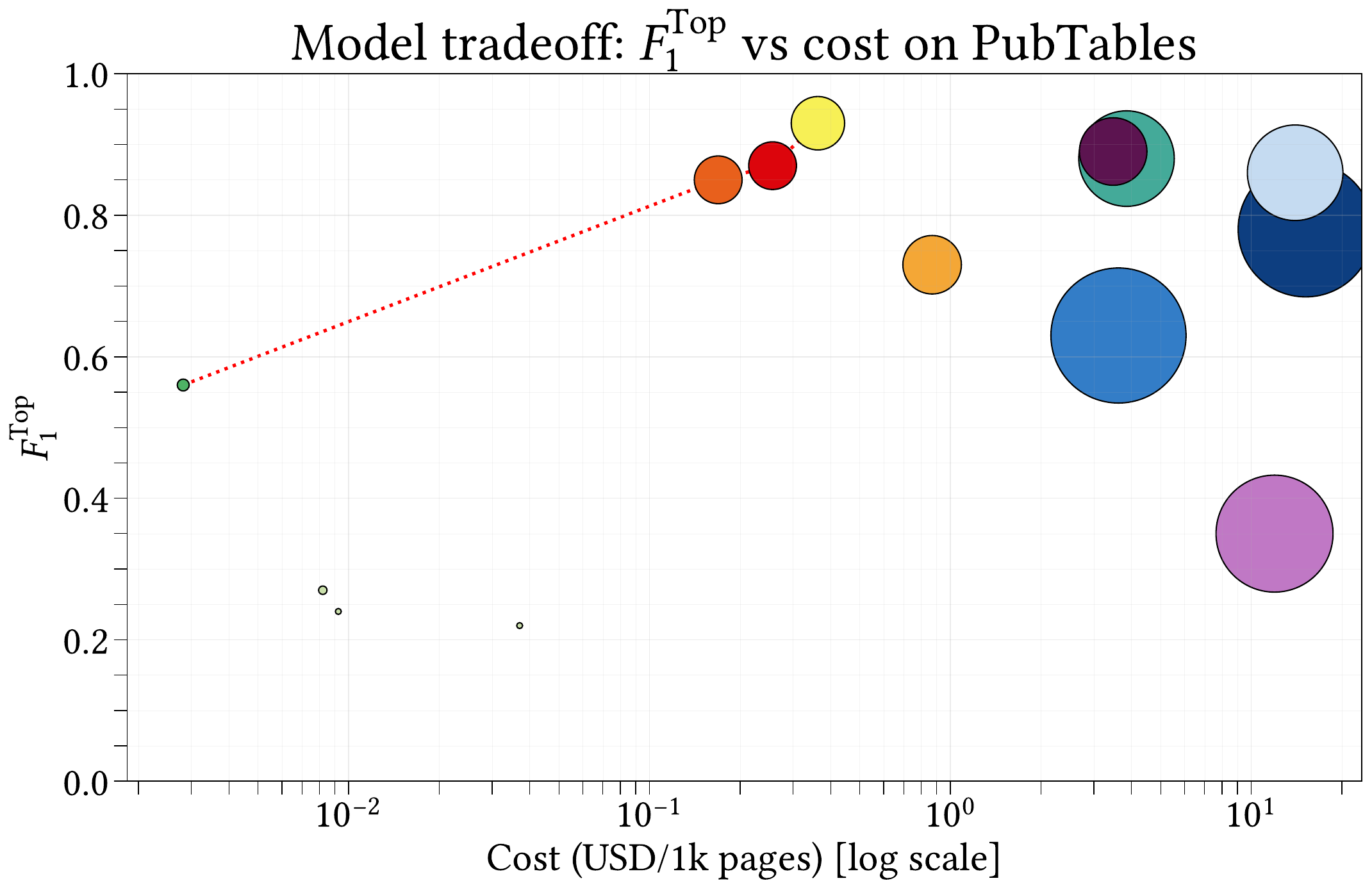}
    \caption{\FTop{} on PubTables}
  \end{subfigure}
  \hfill
  \begin{subfigure}[b]{0.3\textwidth}
    \centering
    \includegraphics[width=\linewidth]{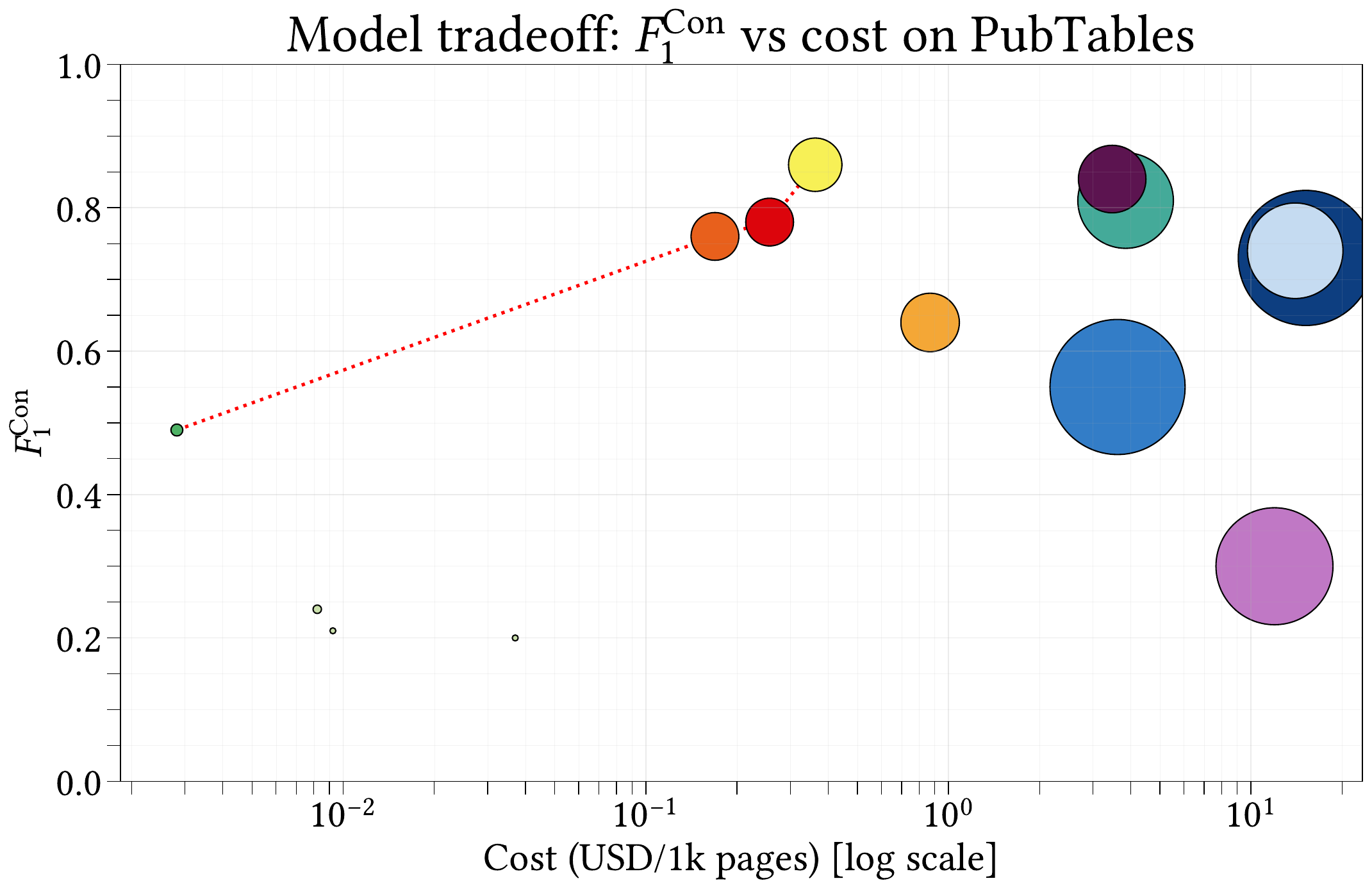}
    \caption{\FCon{} on PubTables}
  \end{subfigure}
  \hfill
  \begin{subfigure}[b]{0.3\textwidth}
    \centering
    \includegraphics[width=\linewidth]{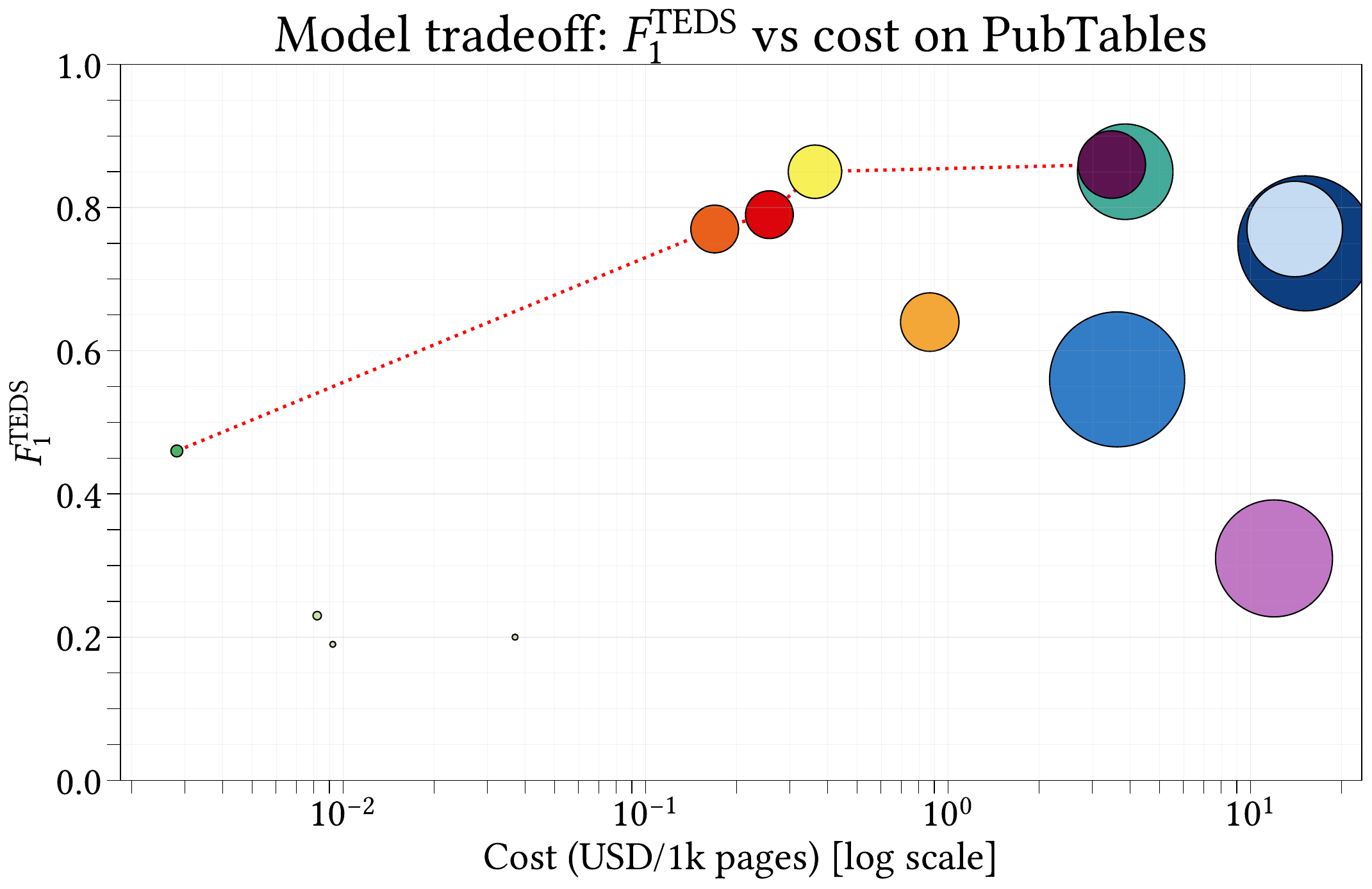}
    \caption{\FTEDS{} on PubTables}
  \end{subfigure}
  \vspace{1em}
  \begin{subfigure}[b]{0.3\textwidth}
    \centering
    \includegraphics[width=\linewidth]{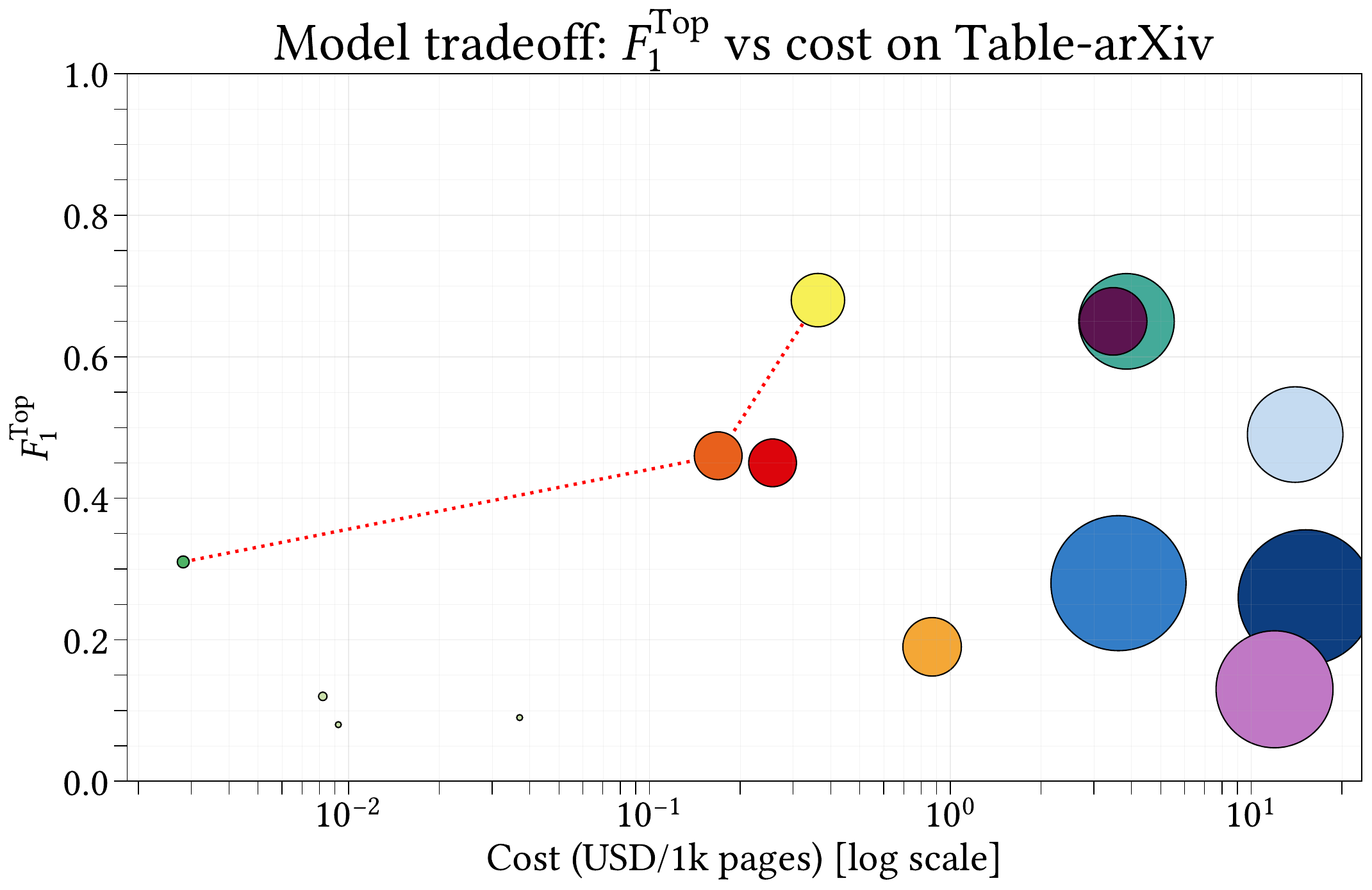}
    \caption{\FTop{} on Table-arXiv}
  \end{subfigure}
  \hfill
  \begin{subfigure}[b]{0.3\textwidth}
    \centering
    \includegraphics[width=\linewidth]{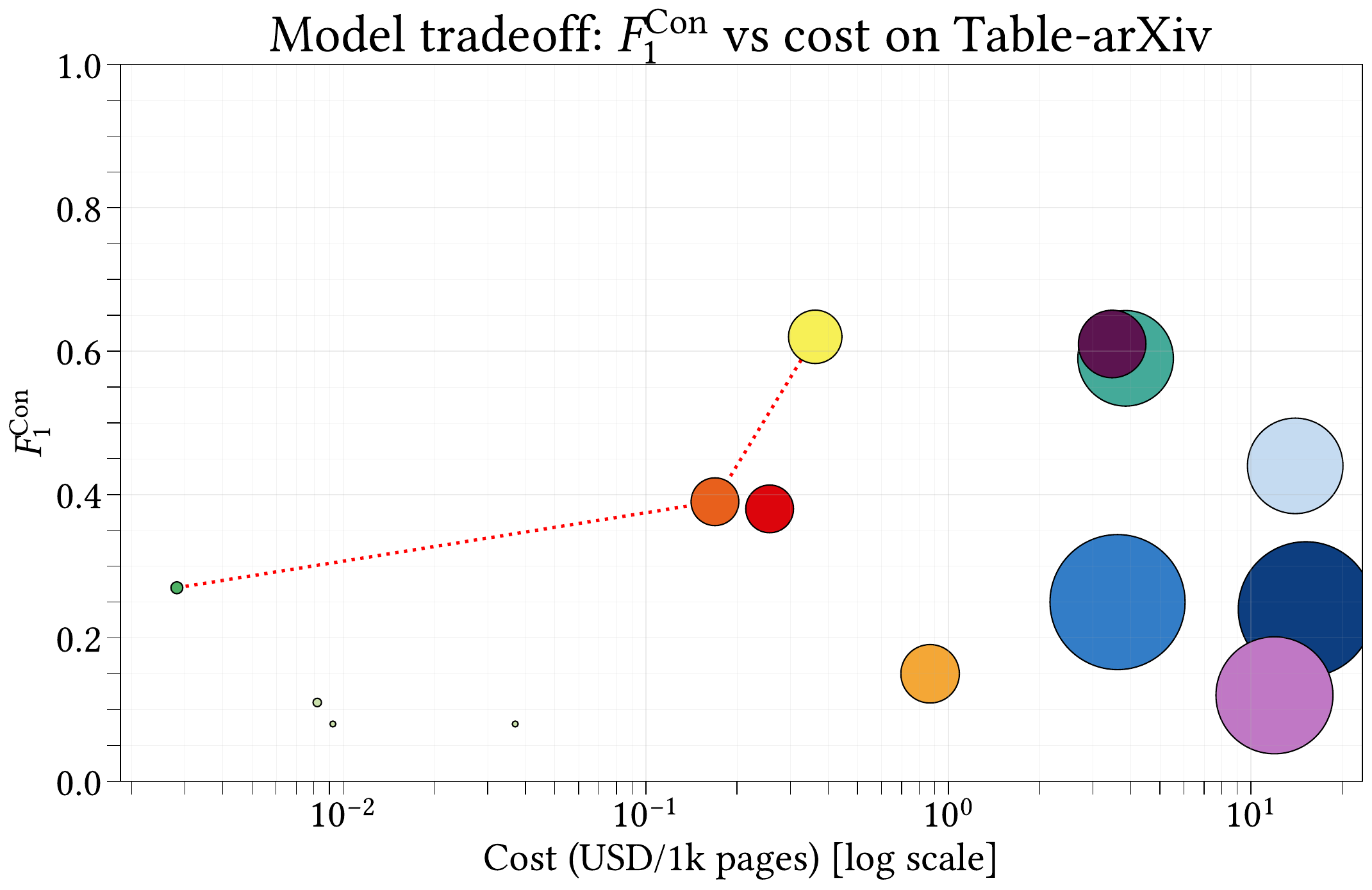}
    \caption{\FCon{} on Table-arXiv}
  \end{subfigure}
  \hfill
  \begin{subfigure}[b]{0.3\textwidth}
    \centering
    \includegraphics[width=\linewidth]{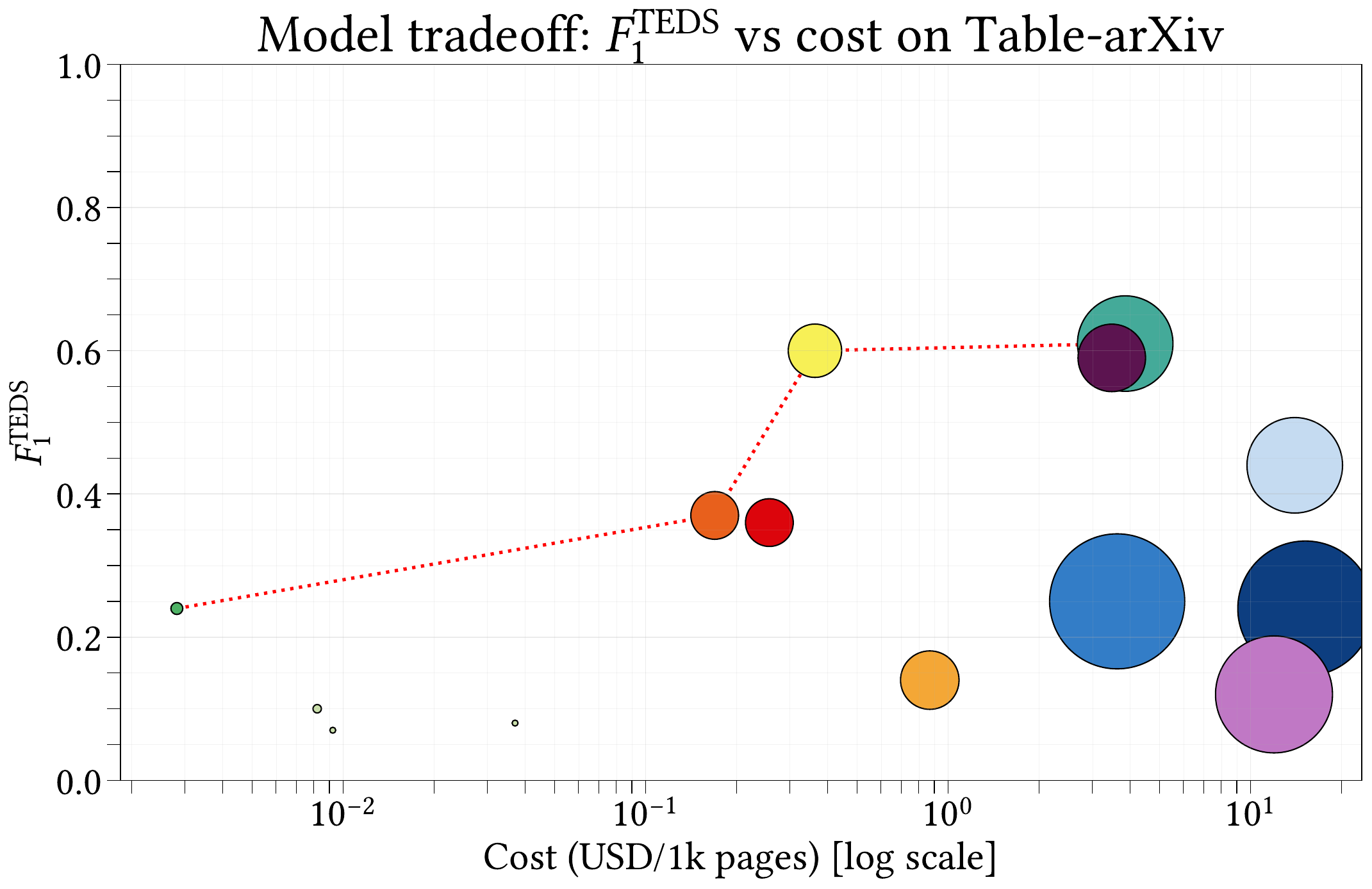}
    \caption{\FTEDS{} on Table-arXiv}
  \end{subfigure}
  \vspace{1em}
  \begin{subfigure}[b]{0.3\textwidth}
    \centering
    \includegraphics[width=\linewidth]{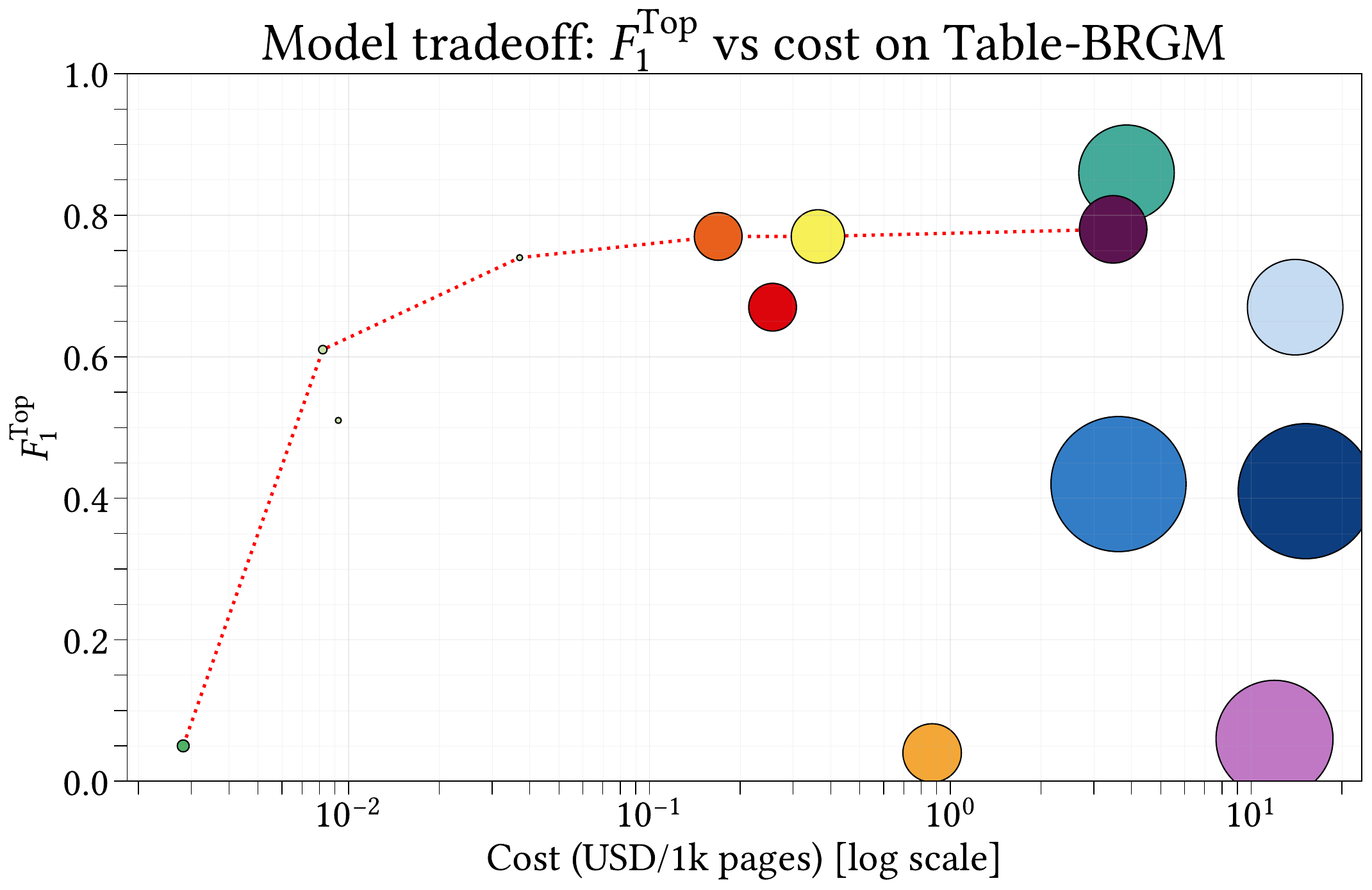}
    \caption{\FTop{} on Table-BRGM}
  \end{subfigure}
  \hfill
  \begin{subfigure}[b]{0.3\textwidth}
    \centering
    \includegraphics[width=\linewidth]{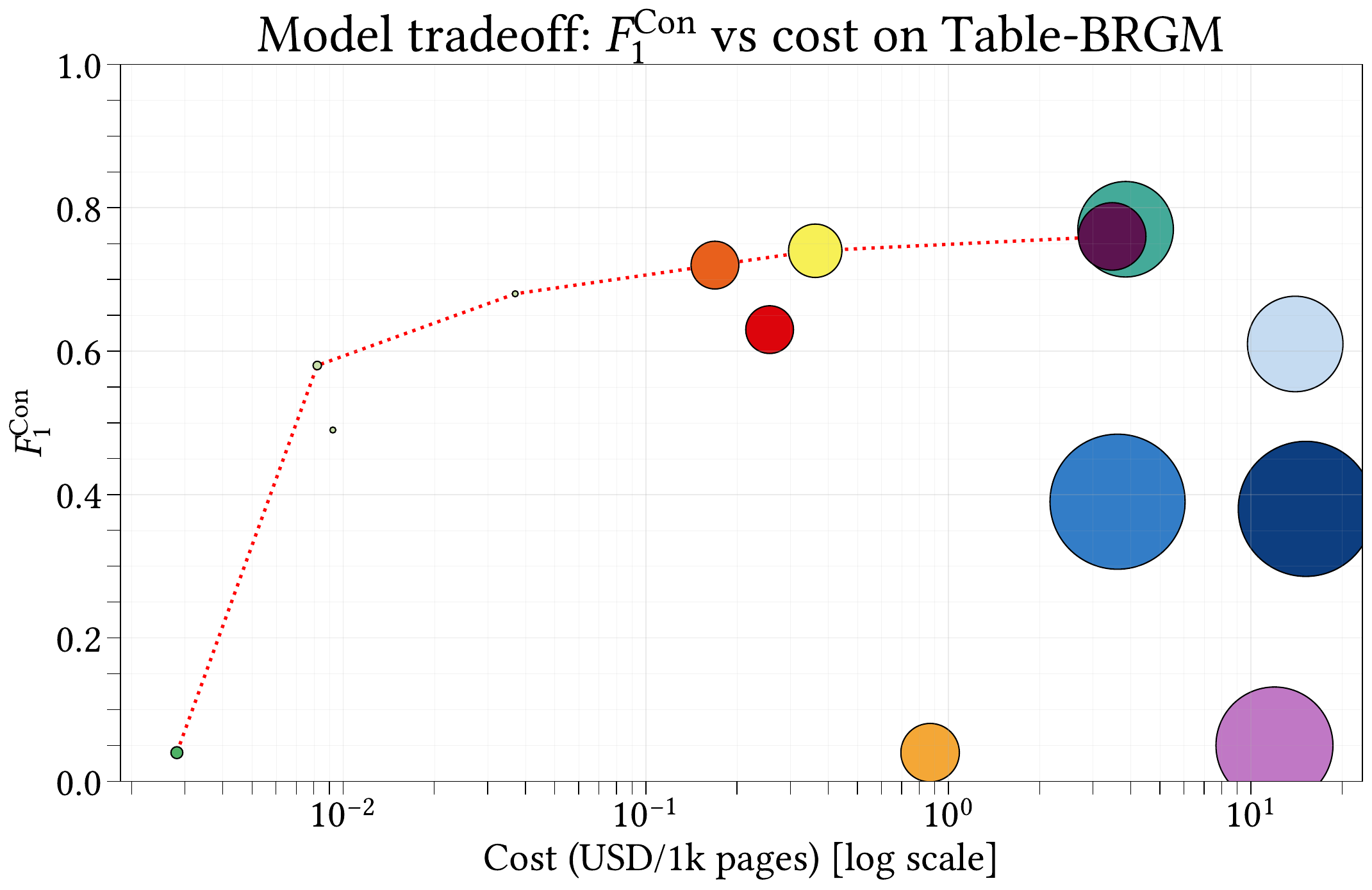}
    \caption{\FCon{} on Table-BRGM}
  \end{subfigure}
  \hfill
  \begin{subfigure}[b]{0.3\textwidth}
    \centering
    \includegraphics[width=\linewidth]{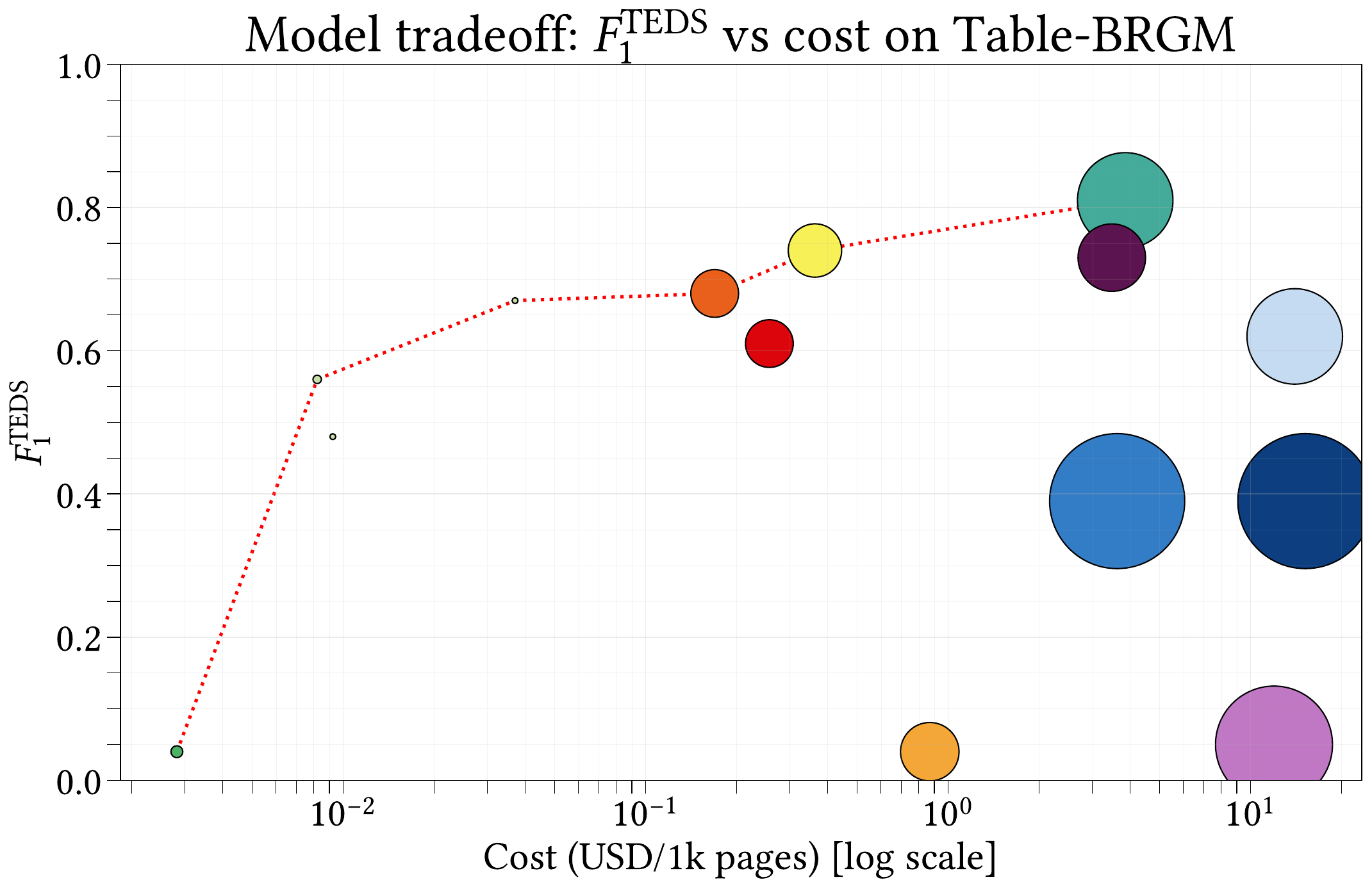}
    \caption{\FTEDS{} on Table-BRGM}
  \end{subfigure}
  \vspace{1em}
  \begin{subfigure}[b]{0.3\textwidth}
    \centering
    \includegraphics[width=\linewidth]{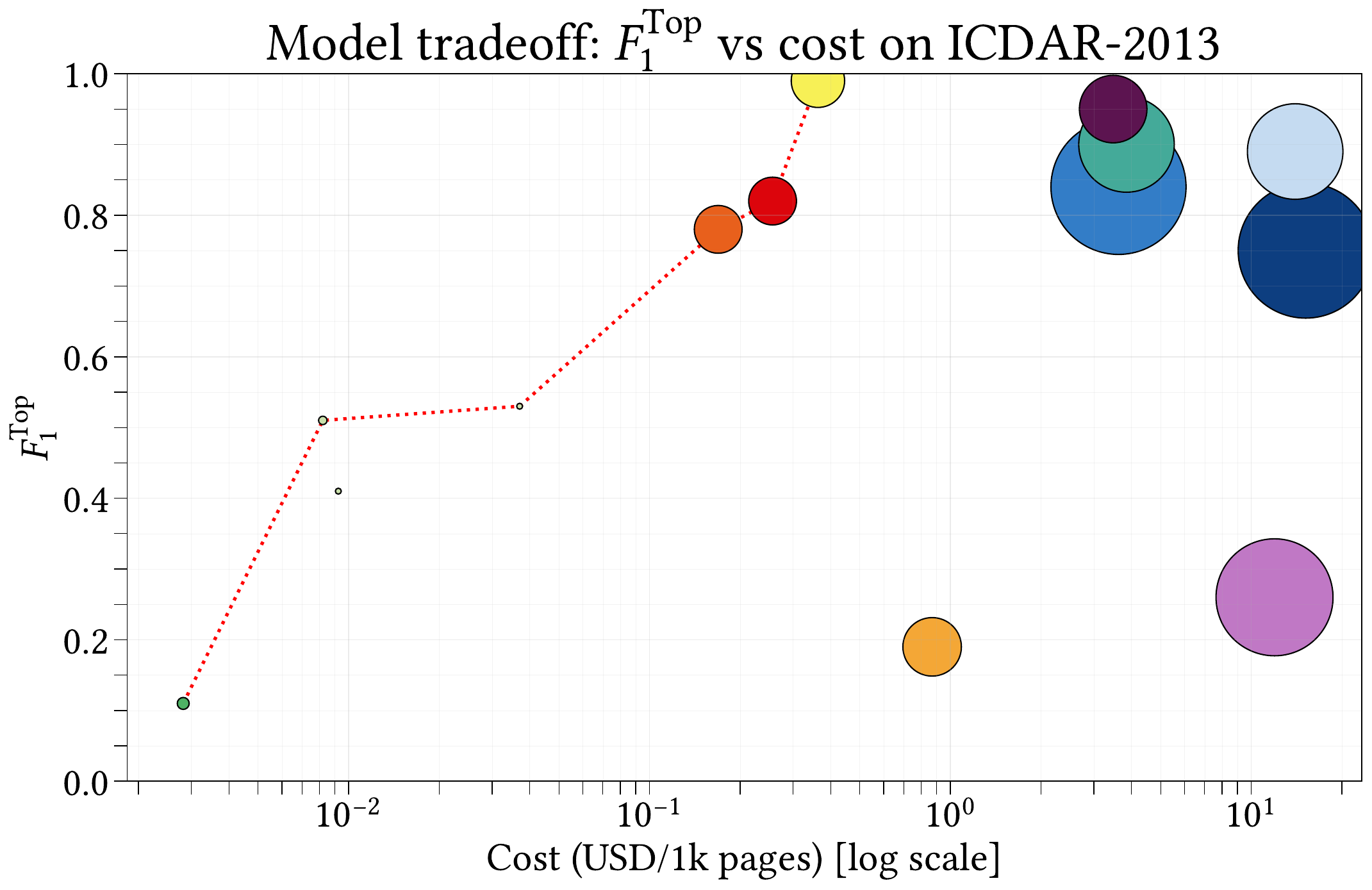}
    \caption{\FTop{} on ICDAR-2013}
  \end{subfigure}
  \hfill
  \begin{subfigure}[b]{0.3\textwidth}
    \centering
    \includegraphics[width=\linewidth]{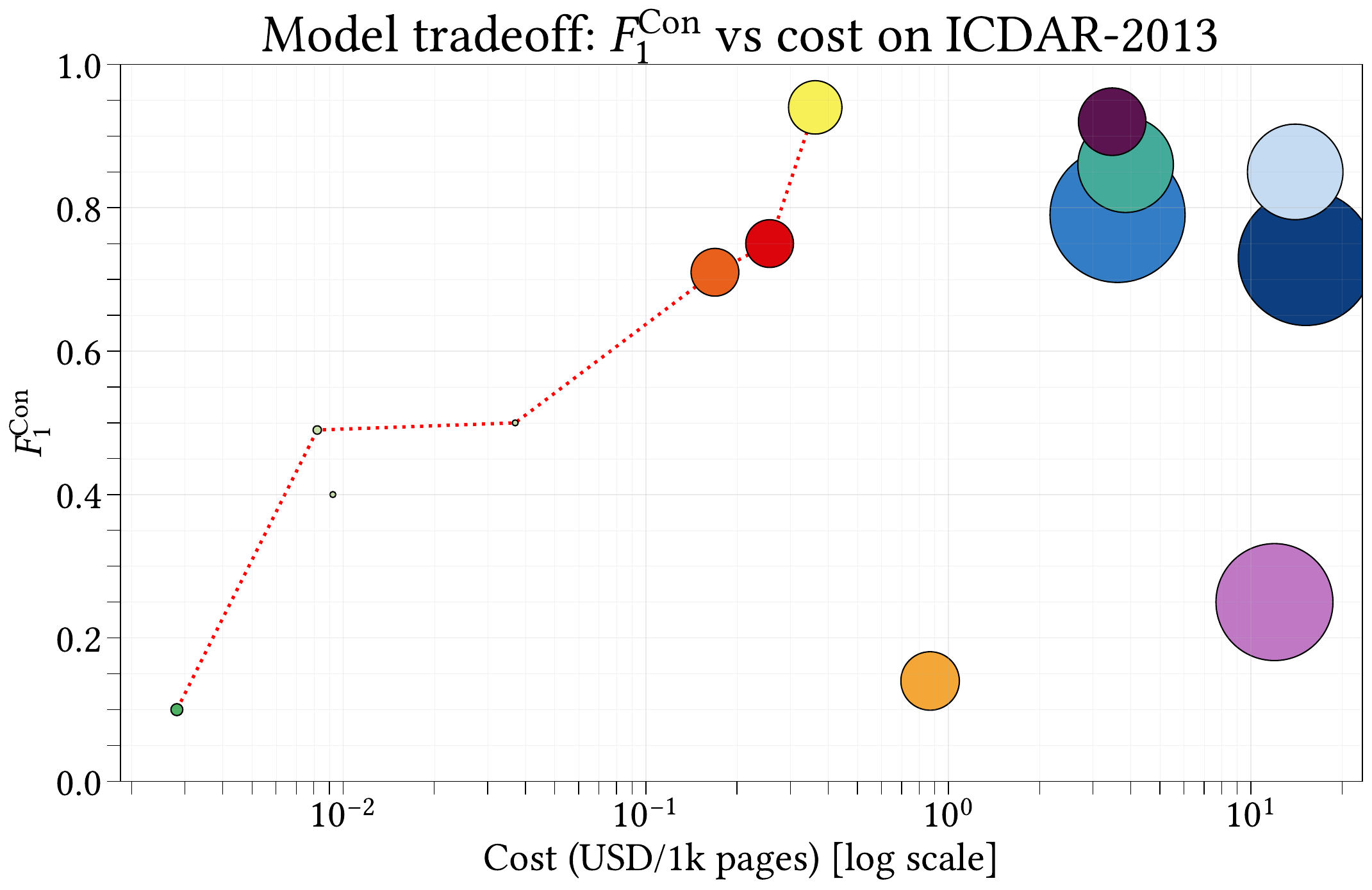}
    \caption{\FCon{} on ICDAR-2013}
  \end{subfigure}
  \hfill
  \begin{subfigure}[b]{0.3\textwidth}
    \centering
    \includegraphics[width=\linewidth]{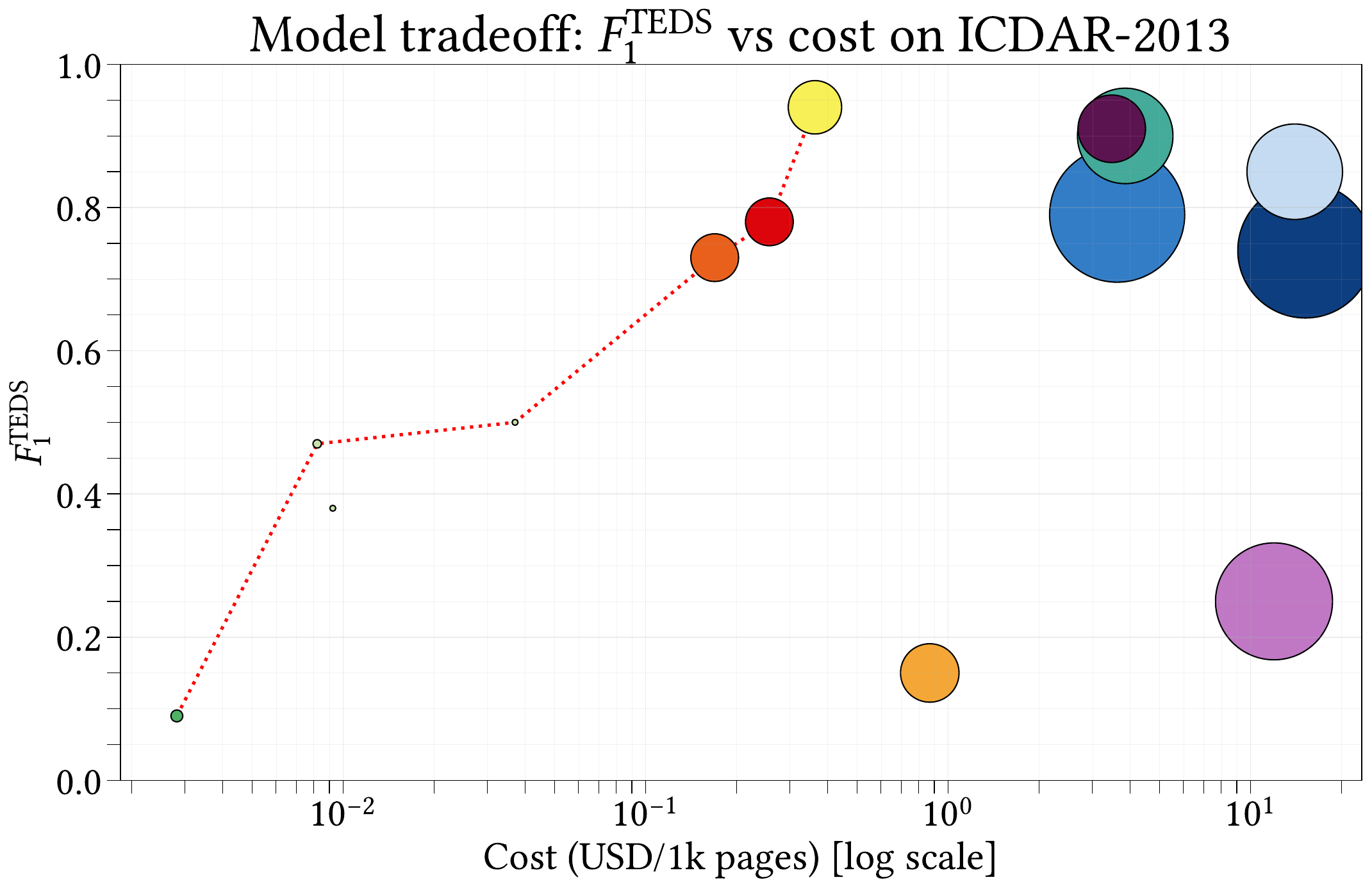}
    \caption{\FTEDS{} on ICDAR-2013}
  \end{subfigure}
  \caption{Tradeoff: TE metrics vs. computational cost for each \emph{txt} TE metric across datasets.
    Bubble size is roughly proportional to the logarithm number of parameters (see \autoref{fig:tradeoff}).
    Colors are the same as in \autoref{fig:tradeoff} and \autoref{tab:all-model}.}
  \label{fig:tradeoff-all-txt}
\end{figure}

\expandafter\def\csname list_brgm\endcsname{RP-56431-FR_Lycee_Dembeni_30, RP-61170-FR_Tsoundzou_39, RP-61170-FR_Kwale_51, RP-68294-FR_envoiClient_118}%
\expandafter\def\csname list_icdar\endcsname{us-015_3, us-024_4, us-025_2, us-035a_2}
\expandafter\def\csname list_arx\endcsname{0801-0002_8, 0801-0134_25, 0801-0287_21, 1401-0480_8}
\expandafter\def\csname list_pub\endcsname{PMC4534043_7, PMC4879640_2, PMC5831713_3, PMC6147433_4}
\def\myModels{"cam_plum_pymu_grobid", "tatr_xy_vgt_doc", "monkey_pedia", "gemini_gpt_qwen"}
\newpage
\clearpage
\section{Visualization of Results}
\foreach \dataset/\datasetDisplay in {icdar/ICDAR-2013, brgm/Table-BRGM, arx/Table-arXiv, pub/PubTables} {
    \edef\currentImageList{\csname list_\dataset\endcsname}
    \foreach \img in \currentImageList {
        \begin{figure}[htbp]
            \centering
            \foreach \mod [count=\i] in \myModels {
                \begin{subfigure}[b]{0.24\textwidth}
                    \centering
                        \fbox{\includegraphics[height=0.18\textheight, width=\linewidth, keepaspectratio]{figure/visu/\dataset/\mod\img.jpg}}
                    \caption{
                        \ifcase\i\or Baseline\or Computer Vision\or Expert VLMs\or General VLMs\fi
                    }
                \end{subfigure}
                \ifnum\i<4 \hfill \fi
            }
            \caption{Results for \texttt{\detokenize\expandafter{\img}.jpg} from \datasetDisplay}
        \end{figure}
    }
}

\begin{figure}[htbp]
  \centering
  \begin{subfigure}[b]{0.4\columnwidth}
    \centering
    \fbox{\includegraphics[height=0.4\textheight]{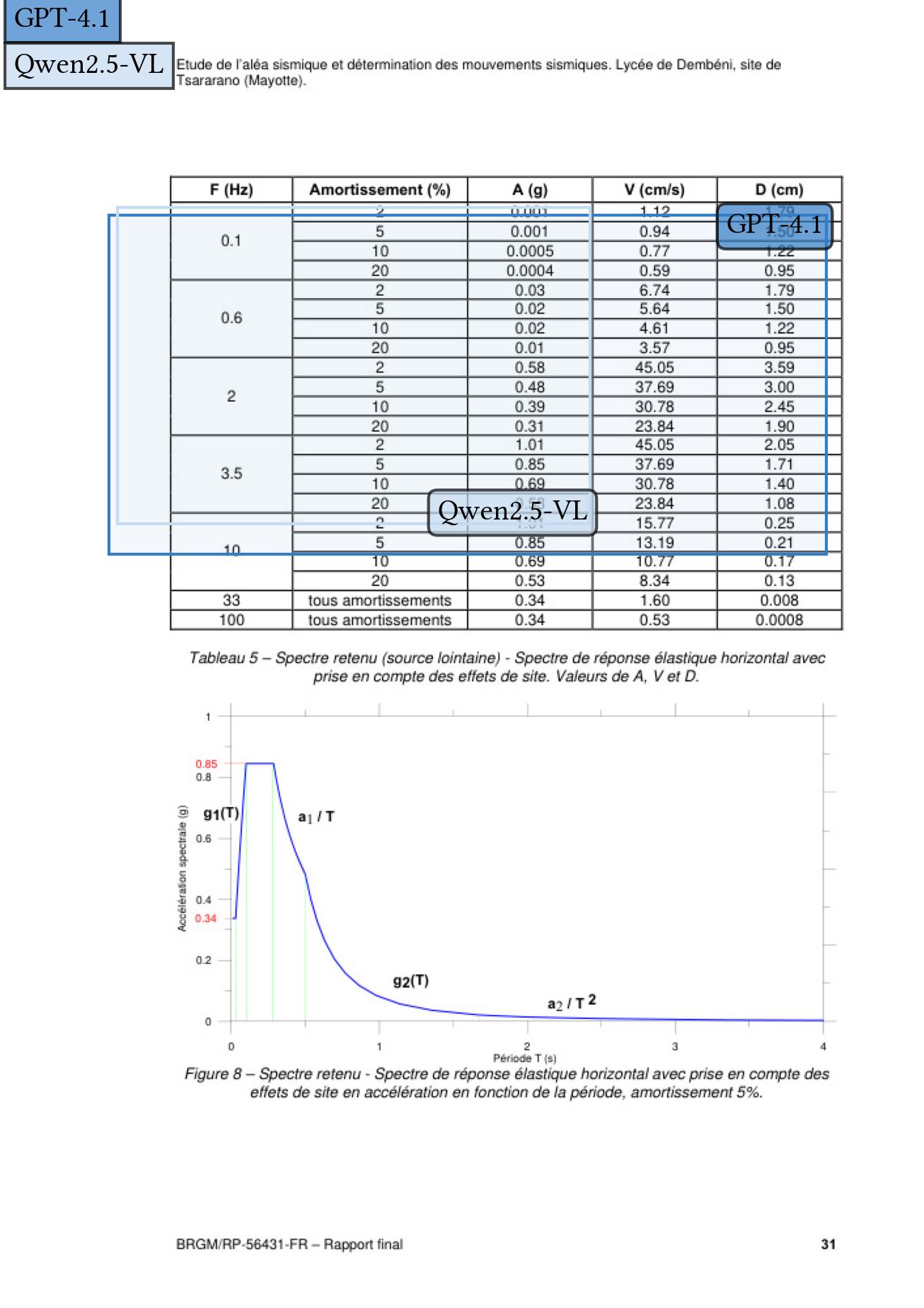}}
    \caption{Bounding box output by \gpt.}\label{fig:gpt-bbox}
  \end{subfigure}\hfill
  \begin{subfigure}[b]{0.4\columnwidth}
    \centering
    \resizebox{\linewidth}{!}{\begin{tabular}{|l|l|l|l|l|}
        \hline
        F (Hz)                               & Amortissement (\%)  & A (g)  & V (cm/s)              & D (cm)                \\ \hline
        \multirow{4}{*}{0.1}                 & 2                   & 0.001  & 1.12                  & 1.79                  \\ \cline{2-5}
                                             & 5                   & 0.001  & 0.94                  & 1.50                  \\ \cline{2-5}
                                             & 10                  & 0.0005 & 0.77                  & 1.22                  \\ \cline{2-5}
                                             & 20                  & 0.0004 & 0.59                  & 0.95                  \\ \hline
        \multirow{4}{*}{0.6}                 & 2                   & 0.03   & 6.74                  & 1.79                  \\ \cline{2-5}
                                             & 5                   & 0.02   & 5.64                  & 1.50                  \\ \cline{2-5}
                                             & 10                  & 0.02   & 4.61                  & 1.22                  \\ \cline{2-5}
                                             & 20                  & 0.01   & 3.57                  & 0.95                  \\ \hline
        \multirow{4}{*}{\colorbox{red}{3.5}} & 2                   & 0.58   & 45.05                 & \colorbox{red}{3.00}  \\ \cline{2-5}
                                             & 5                   & 0.48   & 37.69                 & 3.00                  \\ \cline{2-5}
                                             & 10                  & 0.39   & 30.78                 & 2.45                  \\ \cline{2-5}
                                             & 20                  & 0.31   & 23.84                 & 1.90                  \\ \hline
        \multirow{4}{*}{\colorbox{red}{10}}  & 2                   & 1.01   & \colorbox{red}{41.35} & \colorbox{red}{1.71}  \\ \cline{2-5}
                                             & 5                   & 0.85   & 37.09                 & 1.71                  \\ \cline{2-5}
                                             & 10                  & 0.69   & \colorbox{red}{32.83} & 1.40                  \\ \cline{2-5}
                                             & 20                  & 0.53   & \colorbox{red}{28.57} & 1.08                  \\ \hline
        \multirow{4}{*}{\colorbox{red}{33}}  & 2                   & 1.01   & \colorbox{red}{18.13} & 0.37                  \\ \cline{2-5}
                                             & 5                   & 0.85   & \colorbox{red}{16.01} & 0.37                  \\ \cline{2-5}
                                             & 10                  & 0.69   & \colorbox{red}{13.89} & 0.29                  \\ \cline{2-5}
                                             & 20                  & 0.53   & \colorbox{red}{11.76} & \colorbox{red}{0.22}  \\ \hline
        \multirow{2}{*}{\colorbox{red}{100}} & tous amortissements & 0.34   & \colorbox{red}{1.64}  & 0.008                 \\ \cline{2-5}
                                             & tous amortissements & 0.34   & \colorbox{red}{1.64}  & \colorbox{red}{0.008} \\ \hline
      \end{tabular}}
    \caption{HTML output by \gpt.}\label{fig:gpt-html}
  \end{subfigure}
  \caption{GPT-4.1 hallucination example.
    First, in \autoref{fig:gpt-bbox}, the bounding box is approximate (IoU=0.66) resulting from poor visual/spatial understanding.
    However, it does not prevent GPT-4.1 from extracting the whole table (see \autoref{fig:gpt-html}).
    Second, the table structure (topology, \autoref{fig:gpt-html}) has been correctly detected,
    but GPT-4.1 \colorbox{red}{hallucinates on the table content}.}\label{fig:gpt-hallu}
\end{figure}

\else \fi
\end{document}